\documentclass{emulateapj}

\usepackage{psfig}
\usepackage{amsmath}
\usepackage{amssymb}
\usepackage{graphicx}
\usepackage{appendix}

\newcommand{\lSect}[1]{{\label{sec:#1}}}
\newcommand{\lFig}[1]{{\label{fig:#1}}}
\newcommand{\lEq}[1]{{\label{eq:#1}}}

\def\gtaprx {\lower .1ex\hbox{\rlap{\raise .6ex\hbox{\hskip .3ex
	{\ifmmode{\scriptscriptstyle >}\else
		{$\scriptscriptstyle >$}\fi}}}
	\kern -.4ex{\ifmmode{\scriptscriptstyle \sim}\else
		{$\scriptscriptstyle\sim$}\fi}}}
\def\ltaprx {\lower .1ex\hbox{\rlap{\raise .6ex\hbox{\hskip .3ex
	{\ifmmode{\scriptscriptstyle <}\else
		{$\scriptscriptstyle <$}\fi}}}
	\kern -.4ex{\ifmmode{\scriptscriptstyle \sim}\else
		{$\scriptscriptstyle\sim$}\fi}}}
\newcommand{\FIGFF}[2]{{\ref{fig:#2}{#1}}}

\newcommand{\FIG}[2]{{Fig.~\FIGFF{#1}{#2}}}
\newcommand{\Fig}[1]{{\FIG{}{#1}}}

\newcommand{\Figures}[1]{{Figures~\FIGFF{}{#1}}}
\newcommand{\Sectff}[1]{{\ref{sec:#1}}}
\newcommand{\Sect}[1]{{\S~\Sectff{#1}}}

\newcommand{\Eqref}[1]{{\ref{eq:#1}}}

\newcommand{\eqff}[1]{{\Eqref{#1}}}

\newcommand{\eq}[1]{{eq.~\eqff{#1}}}

\begin{document}


\title{Type~Ia Supernova: Calculations of Turbulent Flames Using the
  Linear Eddy Model}

\author{S. E. Woosley\altaffilmark{1}, A. R. Kerstein\altaffilmark{2},
  V. Sankaran\altaffilmark{2}, and F. R\"opke\altaffilmark{3}}

\altaffiltext{1}{Department of Astronomy and Astrophysics, University
  of California, Santa Cruz, CA 95064; woosley@ucolick.org}
\altaffiltext{2}{Combustion Research Facility, Sandia National Laboratory,
  Livermore, CA; arkerst@sandia.gov} 
\altaffiltext{3}{Max Planck Institut f\"ur Astrophysik, Garching,
  Germany; fritz@mpa-Garching.mpg.de}

\begin{abstract} 
The nature of carbon burning flames in Type Ia supernovae is explored
as they interact with Kolmogorov turbulence. One-dimensional
calculations using the Linear Eddy Model of \citet{Ker91}
elucidate three regimes of turbulent burning. In the simplest case,
large scale turbulence folds and deforms thin laminar flamelets to
produce a flame brush with a total burning rate given approximately by
the speed of turbulent fluctuations on the integral scale, $U_L$. This
is the regime where the supernova explosion begins and where most of
its pre-detonation burning occurs. As the density declines,
turbulence starts to tear the individual flamelets, making broader
structures that move faster. For a brief time, these turbulent
flamelets are still narrow compared to their spacing and the concept
of a flame brush moving with an overall speed of $U_L$ remains
valid. However, the typical width of the individual flamelets, which
is given by the condition that their turnover time equals their
burning time, continues to increase as the density
declines. Eventually, mixed regions almost as large as the integral
scale itself are transiently formed. At that point, a transition to
detonation can occur. The conditions for such a transition are
explored numerically and it is estimated that the transition will
occur for densities near $1 \times 10^7$ g cm$^{-3}$, provided the
turbulent speed on the integral scale exceeds about 15\% sonic. An
example calculation shows the details of a detonation actually
developing.
\end{abstract}


\section{INTRODUCTION}
\lSect{intro}

\citet{Dam40} first discussed multiple regimes of turbulent chemical
combustion and gave scaling relations for each. In modern terms, the
two regimes can be distinguished by their Karlovitz number, Ka,
\citep[e.g.,][]{Pet00},
\begin{equation}
\frac{U_L}{S_{\rm lam}} = {\rm Ka}^{2/3} \left( \frac{L}{\delta_{\rm
    lam}} \right)^{1/3},
\end{equation}
or equivalently,
\begin{equation}
{\rm Ka} = \left( \frac{\delta_{\rm lam}}{l_{\rm G}} \right)^{1/2}.
\lEq{Karlovitz}
\end{equation}
Here $U_L$ is the {\sl rms} velocity of turbulent fluctuations on an
integral scale, $L$; $S_{\rm lam}$ is the laminar conductive speed;
$\delta_{\rm lam}$ is the width of the laminar flame; and $l_G$ is the
Gibson length.  For isotropic Kolmogorov turbulence (assumed
throughout this paper), the turbulent speed on the scale of the flame
thickness is
\begin{equation}
v_{\rm turb}(\delta_{\rm lam}) = \left(\frac{\delta_{\rm lam}}{L}\right)^{1/3}
U_L,
\end{equation}
and the Gibson scale, the size of the eddy that turns over in a
(laminar) flame crossing time, is
\begin{equation}
l_G = \left( \frac{S_{\rm lam}}{U_L} \right)^3 L.
\lEq{Gibson}
\end{equation}

For Ka $\ltaprx 1$, individual laminar flames are moved around by the
largest turbulent eddies while smaller eddies have little effect.  The
overall burning progresses at a speed determined by turbulence
properties and is independent of the burning rate on small scales.
This regime has been extensively explored in the astrophysical context
\citep{Nie95a,Nie95b,Nie97} and its properties are reflected in the
Munich group's subgrid model for flame propagation
\citep{Sch06a,Sch06b}.

The condition Ka $\gg 1$, on the other hand, implies that turbulence
can penetrate into the flame and transport heat, and possibly fuel,
faster than laminar burning crosses a flame width. An equivalent
condition is that the Gibson scale is much less than the flame
thickness.  This regime too has been discussed in the astrophysical
literature \citep{Nie97,Nie97a,Kho97,Lis00b}. It is generally agreed
that if spontaneous detonation is to occur, it requires Ka $> 1$ and
probably Ka $> 10$ so that the burning region itself is disrupted, not
just the preheat zone.

It is also known in the chemical combustion community
\citep{Pet86,Pet00,Ker01} that the region Ka $\gg 1$ can be further
divided based on the value of the Damk\"ohler number, Da = $L/(U_L
\tau_{\rm nuc})$. Here $\tau_{nuc}$ is the characteristic burning time
scale, appropriately modified by turbulence. For Da $<$ 1, the eddy
turnover time on the integral scale is short compared with the nuclear
time; for Da $>$ 1, it is longer.  In the literature, the term
``distributed reaction zone(s)'' has been used with reference to the
Da $<$ 1 regime, the Da $>$ 1 regime, or both lumped together (i.e.,
all flames with Ka $\gg$ 1).  Therefore we avoid this terminology,
choosing instead to follow \citet{Pet86} in referring to Da $<$ 1 as
the ``well-stirred reactor regime'' (WSR regime), and to follow
\citet{Ker01} in referring to Da $>$ 1 as the ``stirred flame regime''
(SF regime).

In the WSR regime, there is only one flame. It has a width broader
than the integral scale and a speed slower than the turbulent speed on
the integral scale. This sort of flame is similar to the usual laminar
flame, except that the turbulent diffusion coefficient ($D_{\rm turb}
\sim U_L L$) now substitutes for conduction. Within the SF regime, on
the other hand, there can be multiple burning regions, but the idea of
a flame brush composed of individual flamelets with well-defined local
properties is no longer valid.  The overall burning continues with an
average rate given by the turbulent speed on the integral scale, but
the flamelets do not have a uniform width and their number and
individual speeds are quite variable.

In this paper, we explore these three regimes of turbulent nuclear
combustion, Ka $< 1$, WSR, and SF, in the context of a Type Ia
supernova using a numerical tool, the Linear Eddy Model
(\Sect{LEM}). If a transition to detonation is to occur, we conclude
that it must happen in the SF regime, specifically where Da $\sim 10$
and in the presence of a high degree of turbulence, $U_L \gtaprx 0.15
\, c_{\rm sound}$. An example of a successful spontaneous transition to
detonation is given.

\section{Special Conditions in a Type Ia Supernova}
\lSect{special}

The conditions characterizing turbulent (nuclear) combustion in a Type
Ia supernova are novel and have no direct analogue on earth. This
makes the supernova an interesting environment for testing new
physics, but it also means that our terrestrial intuition regarding
flames can be misleading. For example, it is thought that laboratory
flames in the Ka $\gg 1$ regime simply go out \citep{Pet00} because
they are unable to maintain their heat in the presence of so much
turbulence. But the flame in a supernovae can never ``go out'' until
the star comes apart and, in terms, of local flame variables, that
takes a very long time. The relevant time scale is the hydrodynamic
time scale for the whole star, $\sim$1 s, not the shorter local
turbulent time scale, $U_L/L$. Also the star is very large, $\gtaprx
10^8$ laminar flame thicknesses and 100 integral scales.  Rare events
have many opportunities for realization. There are also other
conditions that are vastly different and, in some cases, make
detonation more easily achievable.

\subsection{High Reynolds Number}

The Reynolds number in a Type Ia supernova is orders of magnitude
greater than achieved thus far in any terrestrial experiment or
in any numerical simulation. The source of viscosity is electron-ion
interactions in a fully ionized plasma \citep{Nan84}
\begin{equation}
\begin{split}
\nu &= \frac{1.9 \times 10^6}{Z} \left(\frac{(\rho_6/\mu_e)^{5/3}}{1 \ +
  \ 1.02 \, (\rho_6/\mu_e)^{2/3}}\right) \ \frac{1}{I_2} \\ &\sim 2 \times
10^6 \ {\rm gm \ cm^{-1} \ s^{-1}}
\end{split}
\end{equation}
for Z = 7, $\mu_e$ = 2, $\rho_6 = 10$, and $I_2$ = 0.5. The Reynolds
number is then
\begin{equation}
Re = \frac{\rho U_L L}{\nu} \sim 5 \times 10^{13},
\end{equation}
for $\rho =$ 10$^7$ g cm$^{-3}$, $U_L$ = 100 km s$^{-1}$, and $L$ = 10
km.  This large value of Re implies a tiny Kolmogorov scale, much
smaller than any laminar flame thickness under consideration here, and
irresolvably small in most numerical simulations,
\begin{equation}
\eta \ = \ Re^{-3/4} L \ \sim \  10^{-4} \ {\rm cm}.
\end{equation}
Note also that this implies ten orders of magnitude in length scale
where the turbulence is (assumed to be) Kolmogorov and isotropic with
constant energy dissipation, $U_L^3/L$.

On earth, the Kolmogorov scale is usually not so small compared with
the flame thickness. This makes achieving the SF and WSR regimes more
difficult on earth, though certainly not impossible.

\subsection{Temperature-Dependent Heat Capacity}
\lSect{cp}

The heat capacity in the supernova at a relevant density, $\rho \sim
10^7$ g cm$^{-3}$, is due to a combination of semi-degenerate
electrons, ions and radiation. Radiation is an important component of
the heat capacity at these low densities and hence the heat capacity
is a rapidly increasing function of the temperature. At a density of
$10^7$ g cm$^{-3}$, for 50\% C and 50\% O, the heat capacity at
constant pressure is 3.5, 6.0, 9.0, 12.7, 18.2,, and 27.3 $\times
10^7$ erg gm$^{-1}$ K$^{-1}$ for a temperature of 0.5, 1.0, 1.5, 2.0,
2.5, and 3.0 $\times 10^9$ K respectively. The power of T upon which
$C_p$ depends varies from 0.70 to 2.70 in the same temperature range.

This means that when the fuel is cold, a small amount of burning
raises the temperature a lot. Given the high power of temperature upon
which the burning depends, burning just a little fuel dramatically
shortens the nuclear time scale.

\subsection{Temperature-Dependent Reaction Rate}
\lSect{enuct}

The most important reaction rate in the regime where detonation might
occur is $^{12}$C + $^{12}$C. The rate for this reaction is
proportional to $\rho X^2(^{12}C) T^n$ with $n \sim 19 - 27$ for
temperatures in the range 1 - 3 $\times 10^9$ K (more sensitive at
lower temperature).  This high temperature sensitivity coupled with
the temperature-dependent heat capacity means that once about half of
the carbon has burned, the remaining increase in the temperature
happens very rapidly. As we shall see later this leads to small but
very rapid increases in the local pressure that can help initiate a
detonation.

\subsection{Strong Turbulence}

As burning plumes of ash float due to the Rayleigh-Taylor instability,
they create shear and turbulence on their boundaries. Since the plumes
are large, the speed at which buoyancy balances drag is high. A speed of
order 10 - 30\% sonic is necessary to burn a large fraction of the
star before it comes apart.  Typical turbulent speeds on an integral
scale of 10 km are about 150 km s$^{-1}$ \citep{Roe07a}, but speeds as
great as 1000 km s$^{-1}$ may occasionally occur. The sound speed in
the star at a density of 10$^7$ g cm$^{-3}$ is 3500 km s$^{-1}$, so
the fastest turbulence is not terribly subsonic. Fluctuations in
burning rate do not have to accelerate the burning by orders of
magnitude in order to make a detonation happen.

\subsection{Large Lewis Number}
\lSect{turb}

The Lewis number, which is the ratio of thermal diffusivity to mass
diffusivity, is very large in the supernova. The ionic diffusion
coefficient for a carbon-oxygen plasma is \citep{Han75,Bil01}
\begin{equation}
D_{\rm ion} \ = \ 3 \omega_p \alpha^2 \Gamma^{-4/3} \sim 0.1 \ {\rm cm^2 \ s^{-1}}, 
\end{equation}
where $\omega_p = (4 \pi n_i (Ze)^2/A m_p)^{1/2}$ is the plasma frequency,
$\alpha = (3/(4 \pi n_i))^{1/3}$, with $n_i$, the ion density, and
$\Gamma = (Ze)^2/\alpha k T$. The thermal diffusion coefficient is
\begin{equation}
D_{\rm rad} \ = \ \frac{4 a c T^3}{3 \rho^2 C_P \kappa} \ \sim \ 10^4.
\lEq{drad}
\end{equation}
Here representative conditions and opacities have been assumed: $T
\approx 1 - 2 \times 10^9$ K, $\rho = 10^7$ g cm$^{-3}$, and $\kappa
\approx 0.02$ cm$^2$ gm$^{-1}$ \citep{Tim00a}. Combining these two
equations, $Le = D_{\rm rad}/D_{\rm ion}\sim 10^5$.

Terrestrial Lewis numbers are close to unity. The large Lewis number
in the star has an effect on the laminar speed (\Sect{laminar}), but
this dependence is greatly mitigated in the turbulent regime
(\Sect{distrib}) where the turbulent diffusion coefficient exceeds
$D_{\rm rad}$. In that case, the effective Lewis number approaches
unity since both ions and heat are transported with equal efficiency by
the turbulent eddies \citep{Asp08}.

\section{The Linear Eddy Model}

\subsection{The LEM Code}
\lSect{LEM}

The range of scales that must be resolved to address the flame
propagation problem in a Type Ia supernova is very large, $\sim
10^{-1} - 10^6$ cm even if the Kolmogorov scale is not resolved. This
range exceeds the current or anticipated capabilities of 3D
simulations, or even 2D simulations.  This conundrum arises for many
turbulent flow environments, and motivated the development of the
Linear Eddy Model (LEM), a 1D simulation tool \citep{Ker91}.

LEM simulates the evolution of scalar properties on a 1D spatial
domain.  This can be interpreted as property profile evolution along a
1D line of sight through 3D turbulent flow.  The 1D domain is treated
as a closed system with respect to enforcement of conservation laws.
In this and other respects, LEM is not fully consistent with the
evolution observed along a line of sight, yet it captures the salient
features and provides useful results.

The physical processes that are time advanced on the LEM domain are
diffusive (e.g., Fickian) transport (in the present context
representing species transport, radiation transport or subgrid
turbulent transport), chemical (or nuclear) reactions, and turbulent
eddy motions.  For combustion applications, including the present 
application, LEM includes an equation of state, an energy equation, 
and thermal expansion, using a zero-Mach-number (constant-pressure) 
formulation \citep{Smi97}.  The novelty of the LEM approach is the 
representation of eddy motions.

On a 1D domain, advection in the usual sense cannot reorder the fluid
elements along the domain, hence cannot emulate the folding of
material surfaces that is an essential feature of turbulent stirring.
A model of eddy motions is introduced that is formulated to capture
this folding effect and other properties of turbulent stirring.
Namely, a turbulent eddy is represented as an instantaneous map,
called the triplet map, that is applied to a designated interval of
the 1D domain.

The triplet map is applied by first compressing all property profiles
in the interval by a factor of three.  The property profiles in the
interval are then replaced by three side-by-side copies (`images') of
the compressed profiles.  The middle copy is then flipped.

This procedure preserves the continuity of spatial property profiles.
It also leaves the total linear measure (model analog of volume) of
fluid corresponding to a given state or set of states (based on the
property values) unchanged.  This is the 1D analog of the solenoidal
condition.  Flipping of the middle copy introduces fluid-element
reordering analogous to the folding of material surfaces.  The
three-fold compression emulates gradient amplification by compressive
strain.  The dispersive effect of extensive strain is represented in
LEM by the mapping of neighboring fluid elements to different images.

Computationally, the triplet map is implemented as a permutation of
the cells of a uniform discretization of the 1D domain.  This and
other details of LEM are explained elsewhere (Kerstein 1991).  Here,
features relevant to what follows are described.  Some symbols that
are used have different meanings than in later sections.

An LEM simulation time advances processes other than advection until
it is time to implement a triplet map.  After the map, advancement of
the other processes resumes until it is time for the next map.

Model inputs are the initial and boundary conditions, the transport
coefficients and rate constants governing diffusive and chemical
advancement, and parameters controlling the time sequence of triplet
maps.  Map size $l$ is sampled from a probability density function
(pdf) $f(l)$ that is designed to reproduce relevant features of the
inertial-range turbulent cascade.  One key feature is the turbulent
diffusivity $D_{\rm turb}$ associated with maps of size $l<S$.  The 
inertial-range scaling $D_{\rm turb} \propto S^{4/3}$ is enforced.  
In \citet{Ker91}, it is shown that this implies $f(l)=Al^{-8/3}$, 
where $A$ is a normalization factor.

Map size $l$ is restricted to the range $[\eta, L]$, where $\eta$ and
$L$ are the model analogs of the Kolmogorov scale and the integral
scale, respectively.  Then as shown in the Appendix, the total
turbulent diffusivity is given by $D_{\rm turb}= \frac{1}{18} \Lambda
A(L^{4/3}-\eta^{4/3})$, where $\Lambda$ is the total frequency of maps
of all sizes per unit domain length.  Here, homogeneous turbulence is
assumed, so none of the model parameters depend on location, and map
location is sampled uniformly within the notionally infinite 1D domain.

The parameters $\eta$, $L$, and $D_{\rm turb}$ are model inputs and 
$\Lambda$ is determined from them.  $D_{\rm turb}$ is not
typically known for turbulent flows, but rather is inferred from the
relation $D_{\rm turb} = U_LL/C$, where $C$ is an empirical coefficient.  Here,
LEM results are compared to 3D simulations of turbulent premixed
combustion for which $U_L$ and $L$ are known, but $D_{\rm turb}$ is not, so $C$
must be specified in order to evaluate the LEM input parameter $D_{\rm turb}$.
\citet{Smi97} calibrated $C$ by comparing LEM results for turbulent
premixed methane-air combustion (using a simplified chemical
mechanism) to turbulent burning velocity measurements.  They chose the
value $C=15$ for some cases and $C=3.5$ for others, reflecting the
wide variation of turbulent burning velocity results obtained in
different experiments.

The experimental set-ups did not correspond to the idealized case of
flames freely propagating through stationary, homogeneous turbulence.
Here, LEM results are compared to 3D simulation results more closely
analogous to the LEM flow configuration.  The $C$ value inferred from
this comparison is close to the lower of the two $C$ values reported
by Smith and Menon.

\begin{figure*}
\begin{center}
\includegraphics[width=0.475\textwidth]{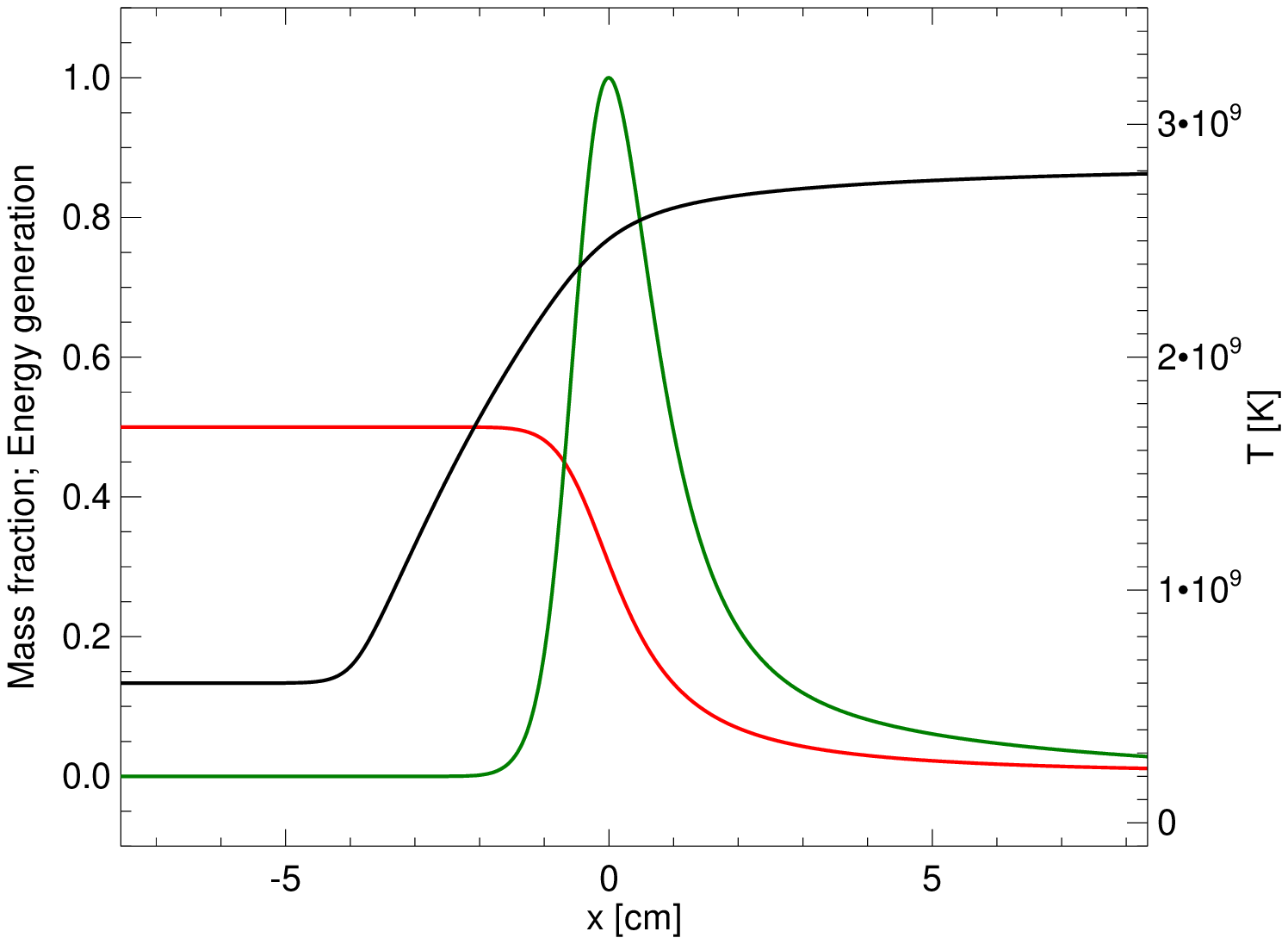}
\hfill
\includegraphics[width=0.475\textwidth]{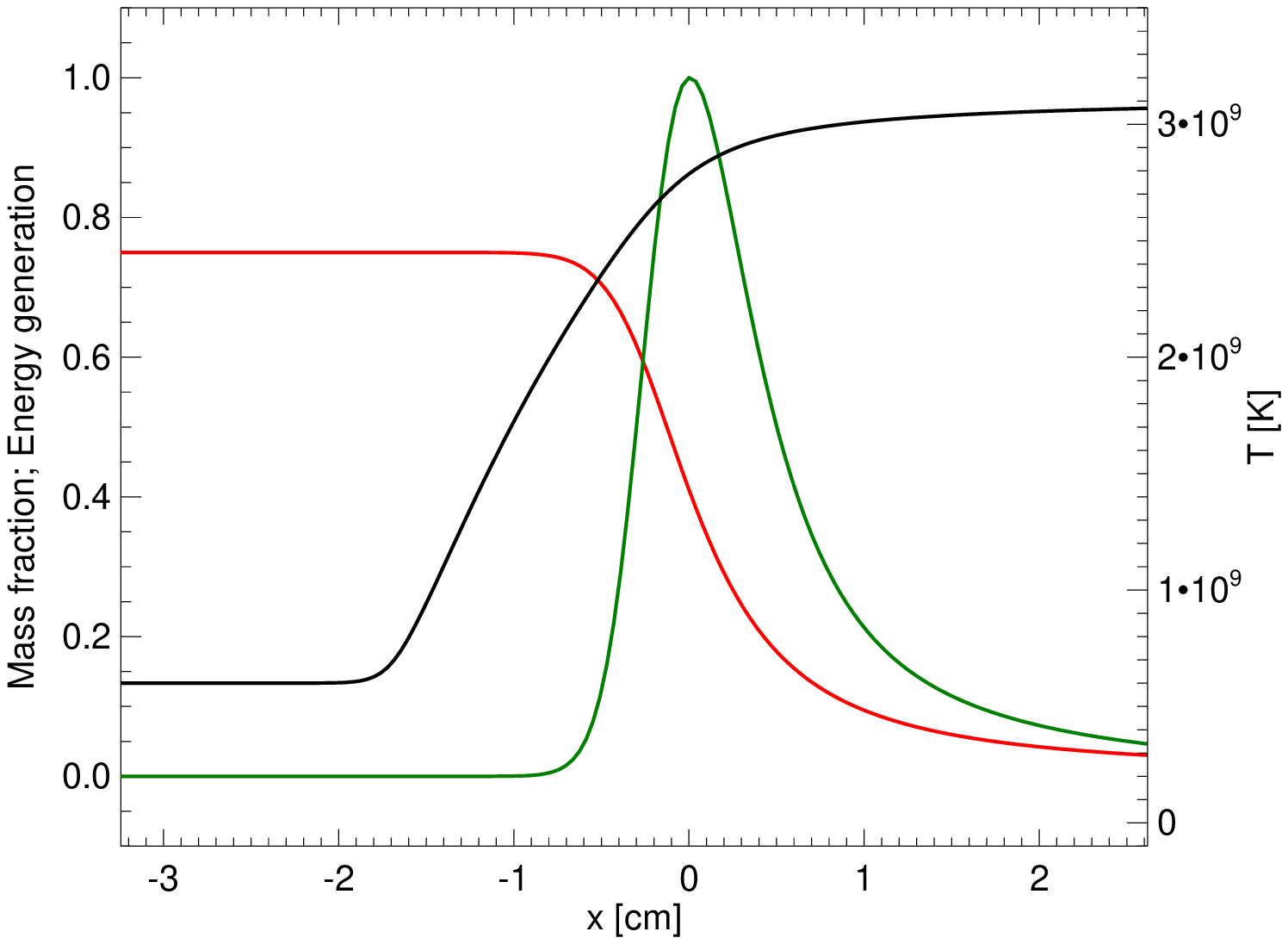}
\caption{Laminar flame structure for initial carbon mass fractions of
  0.50 (left frame) and 0.75 (right frame) and a density of 1.0
  $\times 10^7$ g cm$^{-3}$. Carbon mass fraction is given in red and
  the temperature, in units of 10$^9$ K, is in black. The energy
  generation (green) has been normalized to a maximum value of 7.47
  $\times 10^{20}$ erg g$^{-1}$ s$^{-1}$ ($X^0_{12} = 0.50$) and $8.40
  \times 10^{21}$ erg g$^{-1}$ s$^{-1}$ ($X^0_{12} = 0.75$).  The
  flame is propagating to the left with a speed of $3.2 \times 10^3$
  cm s$^{-1}$ ($X^0_{12} = 0.50$) and $1.1 \times 10^4$ cm s$^{-1}$
  ($X^0_{12} = 0.75$). The assumed Lewis Number in both cases was
  1000. These calculations with the LEM code used 2048 zones and show
  its ability to resolve and physically calculate laminar flames where
  radiation transport is dominant. \lFig{flame}}
\end{center}
\end{figure*}

\subsection{Modifications for the SN Ia Problem}
\lSect{modifications}

The standard (i.e., terrestrial chemical combustion) version of LEM
was modified to use an equation of state, opacities, and a nuclear
reaction network appropriate to the supernova. The equation of state
\citep[the ``Helmholtz EOS'';][]{Tim00b} included pressure and energy
contributions from radiation, ions (treated as an ideal gas), and
electrons and pairs of arbitrary speed and degeneracy.  The opacity
routine \citep{Tim00a} included heat transport by radiative diffusion
and conduction. Energy generation and composition changes were
followed using a nuclear reaction network with 7 species: $^{4}$He,
$^{12}$C, $^{16}$O, $^{20}$Ne, $^{24}$Mg, $^{28}$Si, and $^{56}$Ni
(though no appreciable $^{56}$Ni was ever produced in this study).
These species were connected together by a chain of $(\alpha,\gamma)$
reactions as well as the heavy ion reactions 3$\alpha$, $^{12}$C +
$^{12}$C, $^{16}$O + $^{16}$O, and $^{12}$C + $^{16}$O.  Though it is
an unpublished version written by the authors, the network was similar to
that of \citet{Tim00c}. The network did not include electron screening
corrections to the reaction rates which were, in general,
small at the low densities considered here. Testing the fast 7 isotope
network against a larger 19 isotope network that included screening
and about 85 reactions \citep{Wea78} showed that the small network
gave an energy generation that was typically 20 - 40\% smaller. Since
this is equivalent to a small error in the temperature, the results of
the 7 isotope, unscreened network were deemed sufficiently accurate.

The code was then checked and calibrated against several other studies
of flame propagation. For simple laminar flames (\Sect{laminar}), the
turbulence parameter, $C$ (\Sect{LEM}), doesn't enter.  For turbulent
flames in the flamelet regime, the default setting $C = 15$ was used
(\Sect{flamelet}). However, for flames in the WSR regime
(\Sect{distrib}), good agreement with prior 3D studies required a
smaller value of $C = 5$. This value is within the range anticipated
from terrestrial experiments (\Sect{LEM}) and was used for all
calculations in the WSR and SF regimes (i.e., all studies with Ka $>
1$) unless otherwise noted. Future 3D studies, especially of the SF
regime, are encouraged in order to gain better confidence and
understanding of $C$ for flames of this sort.

\section{Single Laminar Flames}
\lSect{laminar}

In the absence of turbulence, a flame has a simple structure in
which a self-similar profile of temperature and fuel concentration
propagates into the fuel with a well-defined width and speed. Heat is
transported by a combination of conduction and radiative diffusion,
and this heat raises the temperature to the point where fuel burns
on a diffusion time scale.  Such flames are well understood
\citep{Lan59}, even in the supernova context \citep{Tim92}. In
multiple dimensions, laminar flames may become deformed and take on a
cusp-like appearance due to the Landau-Darrieus instability
\citep{Nie95b}, but this instability does not lead to a major
destabilization of the overall burning \citep{Roe04}, nor does it
greatly affect the propagation speed \citep{Bel04a}.

As a test of the LEM implementation and to provide a calibration point
for more complex studies to follow, the one-dimensional laminar speed
was calculated for carbon-burning flames at a variety of densities for
two initial carbon mass fractions, $X^0_{12}$ = 0.50 and 0.75. The
results are given in Table 1 and \Fig{flame}. The speed of the flame
with carbon mass fraction 0.50 and density 10$^7$ g cm$^{-3}$ was $3.2
\times 10^3$ cm s$^{-1}$, in good agreement with the $3.5 \times 10^3$
cm s$^{-1}$ found by \citet{Asp08}, who used somewhat different
nuclear physics.  The ash temperatures for the two compositions
considered here were also in good agreement with the values calculated
by \citet{Woo07}. The flame speeds for other densities were in
agreement with previous studies by \citet{Bel04b}, and \citet{Tim92}.

In these calculations, heat was transported by conduction and
radiation and the laminar scale was well resolved. Zoning for the case
of 10$^7$ g cm$^{-3}$, $X^0_{12}$ = 0.50 was 0.0244 cm and the flame
width, several cm.  A recalculation of the flame speed with coarser
grids gave essentially the same answer. For resolutions of 0.196 cm
and 0.0978 cm the calculated flame speed was 3.2 and 3.1 $\times 10^3$
cm s$^{-1}$, respectively, indicating that our calculation and that of
\citet{Asp08} were both converged and did not differ because of
resolution.

Some dependence on Lewis Number was noted. Two calculations of the
speed for $X^0_{12}$ = 0.50 and Le = 100 and Le = 1000 gave speeds of
$3.0 \times 10^4$ cm s$^{-1}$ and $3.2 \times 10^4$ cm s$^{-1}$
respectively.

\begin{figure*}
\begin{center}
\includegraphics[width=0.475\textwidth]{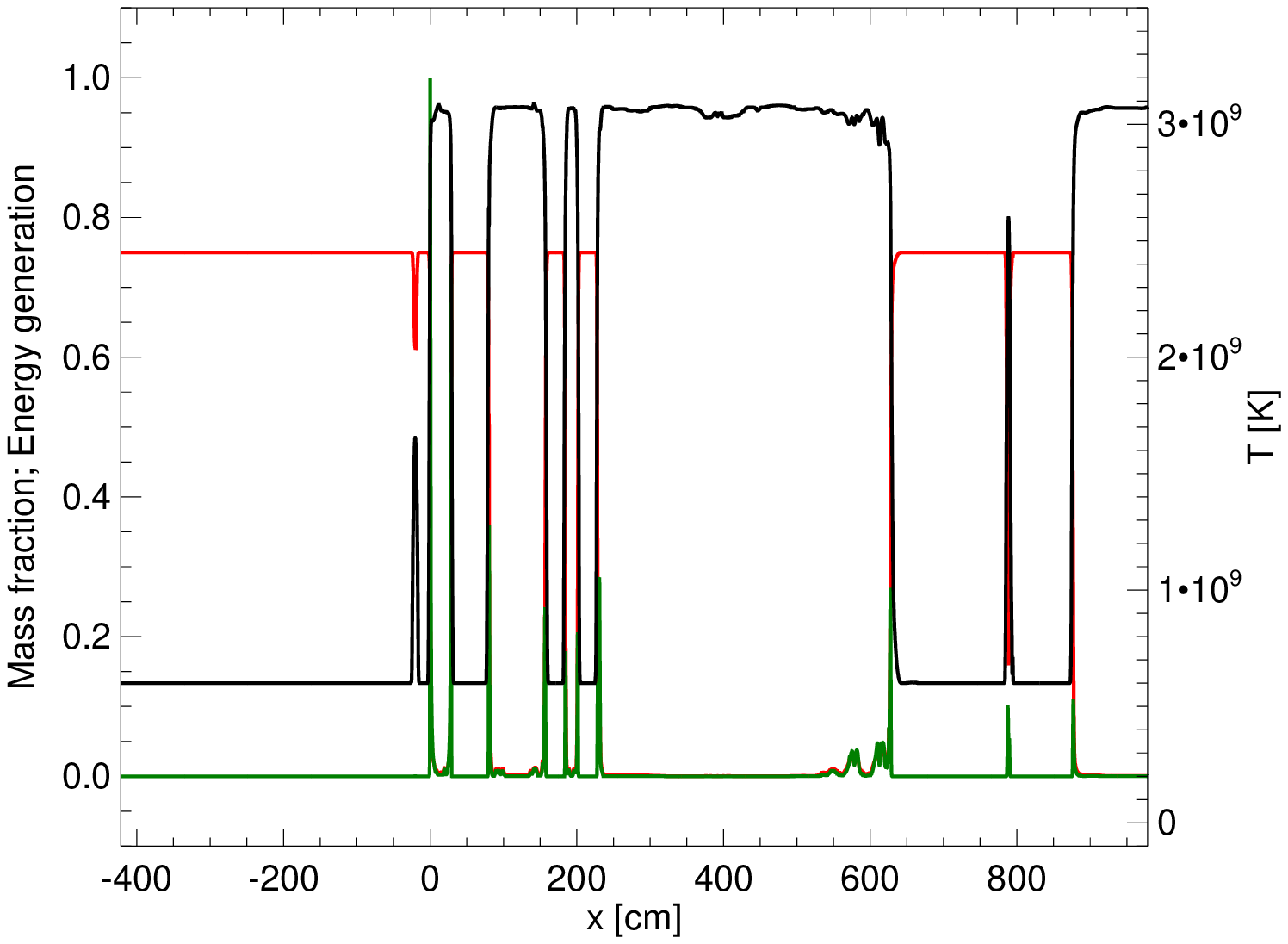}
\hfill
\includegraphics[width=0.475\textwidth]{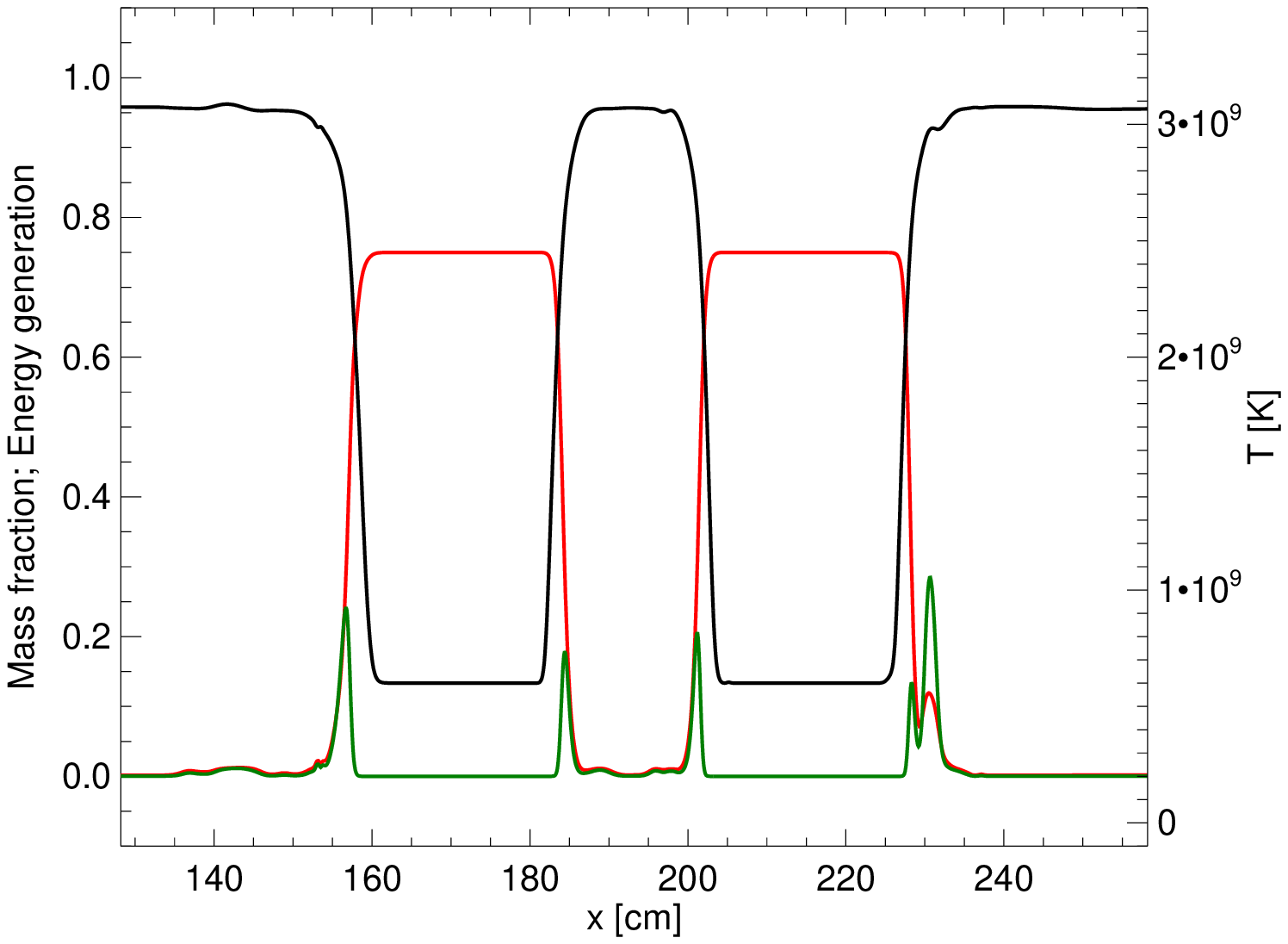}
\caption{Multiple laminar flames for a carbon mass fraction of 0.75,
  density, 1.0 $\times 10^7$ g cm$^{-3}$, integral length scale, 10 m,
  and turbulent speed on that scale, 400 m $^{-1}$. The left frame
  shows the entire flame brush, while the right frame shows greater
  detail for a few of the flamelets.  The entire collection of
  flamelets is moving to the left at an average rate of 500 m s$^{-1}$
  (see \Fig{flameletspd}).The time here is 0.015 s into the
  calculation and the instantaneous rate of burning corresponds to a
  speed of 980 m s$^{-1}$.  This calculation with the LEM code used
  40000 zones and Le = 100.  \lFig{flamelet}}
\end{center}
\end{figure*}

\section{Three Regimes of Turbulent Burning}
\lSect{regime}

\subsection {Large Scale Turbulence and Flame Brushes (Flamelet Regime)}
\lSect{flamelet}

The interior of a Type Ia supernova is turbulent, first because of the
convection that precedes the explosion, which gives a lower bound for
$U_L \sim$ 50 - 100 km s$^{-1}$ \citep{Woo04,Kuh06}, and second
because of the Rayleigh-Taylor and Kelvin-Helmholtz instabilities
associated with the flame itself \citep{Hil00}.  Because these speeds
are so much greater than the laminar speed (Table 1), especially at
low density, it is expected that turbulence has a major effect on
flame propagation. Initially, however, the flame is very thin compared
to the Gibson length and is just carried around by the eddies
\citep{Dam40}. Overall the burning rate is independent of the speed of
each little flamelet, and is governed instead by the speed of the
largest turbulent eddies. This is Damk\"ohler's ``large scale
turbulence'' limit.

To illustrate burning in this regime, we considered a laminar flame
similar to that in \Fig{flame} with a carbon mass fraction of 0.75 and
a fuel density $1.0 \times 10^7$ g cm$^{-3}$, but embedded in a
turbulent background with characteristic speed 400 m s$^{-1}$ on an
integral scale of 10 m. On this length scale in a supernova, one would
actually expect much larger turbulent speeds, 10 to 100 km s$^{-1}$,
cascading down from still higher values on larger scales. For the time
being such large, highly turbulent regimes with narrow flames remain
out of computational reach, even in 1D.  Conditions chosen here were
thus artificial, but illustrative.  The Gibson scale here
(\eq{Gibson}) is 21 cm, considerably larger than the laminar flame
width, which is $\sim$1 cm (\Fig{flame}), but still about still 50
times less than the integral scale.  The results are given in
\Fig{flamelet}.  Sometimes only a single flame was observed, but at
the particular moment sampled here, there were eleven (counting the
small blip of entrained ash at 800 cm).

\Fig{flameletspd} shows the overall propagation speed of the flame
brush. Here, as in the rest of the paper, the burning ``speed'' is
defined by an integral over the grid of the burning rate, 
\begin{equation}
v_f \ = \ \frac{\int S_{\rm nuc} dx}{q},
\end{equation}
where $S_{\rm nuc}$ is the nuclear energy generation rate on the grid
(erg g$^{-1}$ s$^{-1}$), and $q$ is the energy released by burning a
gram of fuel. For fuel with 50\% carbon, $q$ is $3.2 \times 10^{17}$
erg g$^{-1}$ and for 75\% carbon it is $4.5 \times 10^{17}$ erg
g$^{-1}$.

For the case shown in \Fig{flameletspd}, the average burning speed is
500 m s$^{-1}$, close to the turbulent {\sl rms} speed on the integral
scale, 400 m s$^{-1}$. In LEM, one expects for the flamelet regime that
$v_{\rm turb}$ = 18 $D_{\rm turb}/L$ (see Appendix). Combining this
with $D_{\rm turb} = U_L L/C$ (\Sect{LEM}) and using the same value
for $C$ as the simulation ($C = 15$), one has $v_{\rm turb} = (18/15) \,
U_L$, which is excellent agreement.

\Fig{flameletspd} also shows that the burning rate is
far from regular. Speeds over three times the average are sometimes
seen. At other times, the flame briefly almost goes out.
A similar behavior would be expected in the actual supernova early on
when the density is higher and the turbulent speed is not that much
greater than the laminar one. Then, as here, there would be just a few
flames within the integral scale and the burning rate would be highly
irregular. For example, at a density of $10^9$ g cm$^{-3}$ with a
carbon mass fraction of 0.50, the laminar flame speed is 36 km
s$^{-1}$ \citep{Tim92} and the ratio $L/l_{\rm Gib} \sim (U_L/S_{\rm
  lam})^3$ is, for $U_L = 130$ km s$^{-1}$, also about 50.  The
difference at these high densities is that the flame would be
irresolvably small, even in the present version of LEM, for a
calculation that carried the entire integral scale.

\subsection{Transition to Stirred Flames}

The flames at such high density are individually very thin and the
laminar speed far below sonic, so detonation at early times is
avoided. As the density declines, the Gibson length shrinks and the
flame brush contains an increasingly large number of flamelets. Just
before entering the SF regime in the supernova, there
are roughly a thousand flamelets within the flame brush ($U_L \sim
100$ km s$^{-1}$; $S_{\rm lam} \sim 100$ m s$^{-1}$).  The statistical
variations in overall speed are therefore small.

\begin{figure}
\includegraphics[angle=90,width=0.475\textwidth]{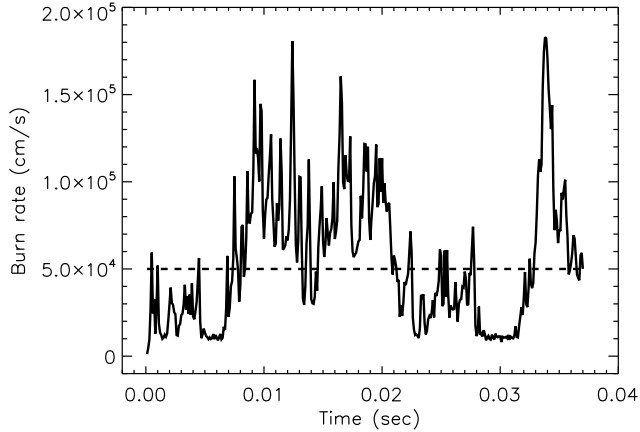}
\caption{Speed of the flame brush as a function of time for the same
  calculation shown in \Fig{flamelet}. The speed, as measured by the
  overall energy generation on the grid, is highly variable ranging
  from a lower bound given by the laminar speed of a single flamelet
  ($1.1 \times 10^4$ cm s$^{-1}$) to several times the mean value. The
  dashed line indicates the average during the time interval studied,
  $5.5 \times 10^4$ cm s$^{-1}$, which is close to the {\sl rms}
  turbulent speed on the integral scale, $4 \times 10^4$ cm s$^{-1}$.
  The turnover time on the integral scale is 0.025 s and one sees the
  effect of the occasional large eddy in accelerating the burning.
  \lFig{flameletspd}}
\end{figure}

As the density in the star declines below several times 10$^7$ g
cm$^{-3}$, the Gibson length becomes less than the laminar flame width
and the transport of fuel and heat by turbulent advection starts to
become important on the scale of the flame. By the time the density
reaches $1.0 \times 10^7$ g cm$^{-3}$, for typical turbulent speeds,
not only is the preheat region of the flame being mixed by turbulence,
but the burning zone itself has been disrupted and combustion has a
qualitatively different character \citep{Asp08,Pet00}. In the star,
this transition is brought about by the slowing and thickening of the
laminar flame with decreasing density in a background of turbulence
with nearly constant energy dissipation rate and integral length
scale.

\subsection{The Well-Stirred Reactor (WSR Regime)}
\lSect{distrib}

Before discussing the stirred flame regime further though
(\Sect{stirred}), it is helpful to consider a limiting case where the
turbulence thickens a single flame to a dimension greater than the
integral scale, i.e., $\delta_{\rm turb} > l$, where $l$ is some
integral scale to be varied subject to the condition $v'^3(l)/l =
\epsilon$ = constant, where $v'$ is the velocity on the integral
scale, $l$.  While probably not realized in the supernova because of
the large integral scale, this is the regime most easily accessible to
multi-dimensional simulations and corresponds to what is commonly
meant by the ``well-stirred reactor'' \citep[e.g., Fig. 7
  of][]{Pet86}. The solutions here also obey scaling relations
predicted by \citet{Dam40}.

\begin{figure}
\includegraphics[width=0.475\textwidth]{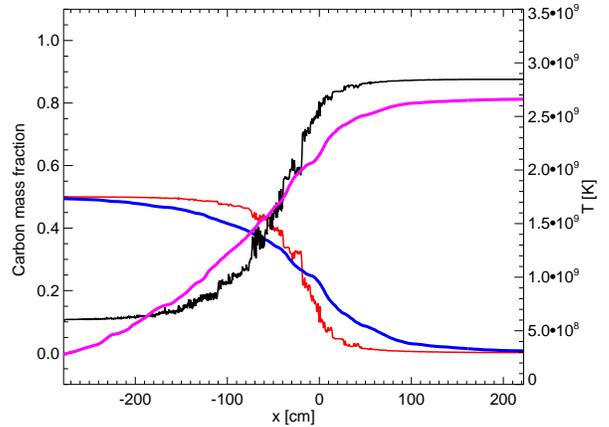}
\caption{Comparison of carbon mass fraction and temperature for two
  flames calculated at a density of 10$^{7}$ g cm$^{-3}$, integral
  scale, $l = 15$ cm, and turbulent speed, $2.45 \times 10^5$ cm
  s$^{-1}$. The purple and blue curves are the average temperature and
  carbon mass fraction from the 3D study of \citet{Asp08}. The black
  and red curves are from the corresponding study using LEM and $C =
  5$ (Table 2). The speed of the flame calculated in 3D was $1.8
  \times 10^4$ cm s$^{-1}$. With LEM it was $1.5 \times 10^4$ cm
  s$^{-1}$, about five times the laminar flame speed for these
  conditions. The flame widths calculated in the two studies are quite
  similar. Differences in the temperature are attributable, in part,
  to the different fuel temperature employed in the two studies - $1
  \times 10^8$ K for the 3D study and $6 \times 10^8$ K for the LEM
  study and the smaller network used for the 3D study (see text).
  \lFig{andywidth}}
\end{figure}

\subsubsection{An example compared with previous 3D simulations}
\lSect{Aspden}

\citet{Asp08} recently carried out 3D simulations of flame propagation
in the WSR regime for a density and composition appropriate to Type Ia
supernovae. We focus here on their calculation at 10$^{7}$ g
cm$^{-3}$, $X_{12}^0$ = 0.50, and $\epsilon = U_L^3/L = 10^{15}$ erg g
s$^{-1}$.  \Figures{andywidth}, \ref{fig:andy}, \ref{fig:andyspd}, and
Table 2 show the results for an LEM calculation at the same values of
integral scale, $l$ = 15 cm, and $v'(l)$ = 2.47 km s$^{-1}$ that they
employed. A single broad flame propagates at a steady speed
about 5 times faster than the laminar value. The width
of the flame, as measured by the full width at half maximum of the
energy generation curve is about 50 cm, or 20 times the laminar width
(\Fig{flame}). 

The most noticeable differences with the 3D calculation
(\Fig{andywidth}) are a consequence of differing assumptions regarding
the background (fuel) temperature in the supernova. Aspden et al. used
$1 \times 10^8$ K; here we use $6 \times 10^8$ K. The lower
temperature would characterize a region of low turbulence following
some expansion prior to a detonation transition. The latter is
characteristic of the convection zone in the supernova at the time of
runaway. For a non-turbulent medium, the former is more
physical. However, anticipating that we will be interested here in
turbulent energies $U_L^3/L \sim 10^{17} - 10^{18}$ erg g$^{-1}$
s$^{-1}$ and that at least a few turnovers on the integral scale are
expected prior to a detonation, the latter is also a reasonable
assumption. For $U_L \sim 10^8$ cm s$^{-1}$ and $L = 10$ km, 10$^{16}$
erg g$^{-1}$ would be deposited in one turnover time. At a density of
10$^7$ g cm$^{-3}$, this corresponds to a temperature increase from $1
\times 10^8$ K (or essentially zero) to $6 \times 10^8$ K. Because of
the temperature sensitivity of the heat capacity (\Sect{cp}), we
expect our results to be insensitive to the assumed fuel
temperature. However we do note that, if a fuel temperature of $1
\times 10^8$ is assumed in LEM, better agreement with the width is
obtained than in \Fig{andywidth}, but the best value of $C$ is lowered
to $C = 2$.

While lower values of $C$ are actually more favorable to detonation, we
prefer to use the larger value because it is more consistent with the
range seen in terrestrial experiments (\Sect{LEM}) and because a
hotter fuel temperature is both physically justifiable and more stable
numerically.

\begin{figure}
\begin{center}
\includegraphics[width=0.475\textwidth]{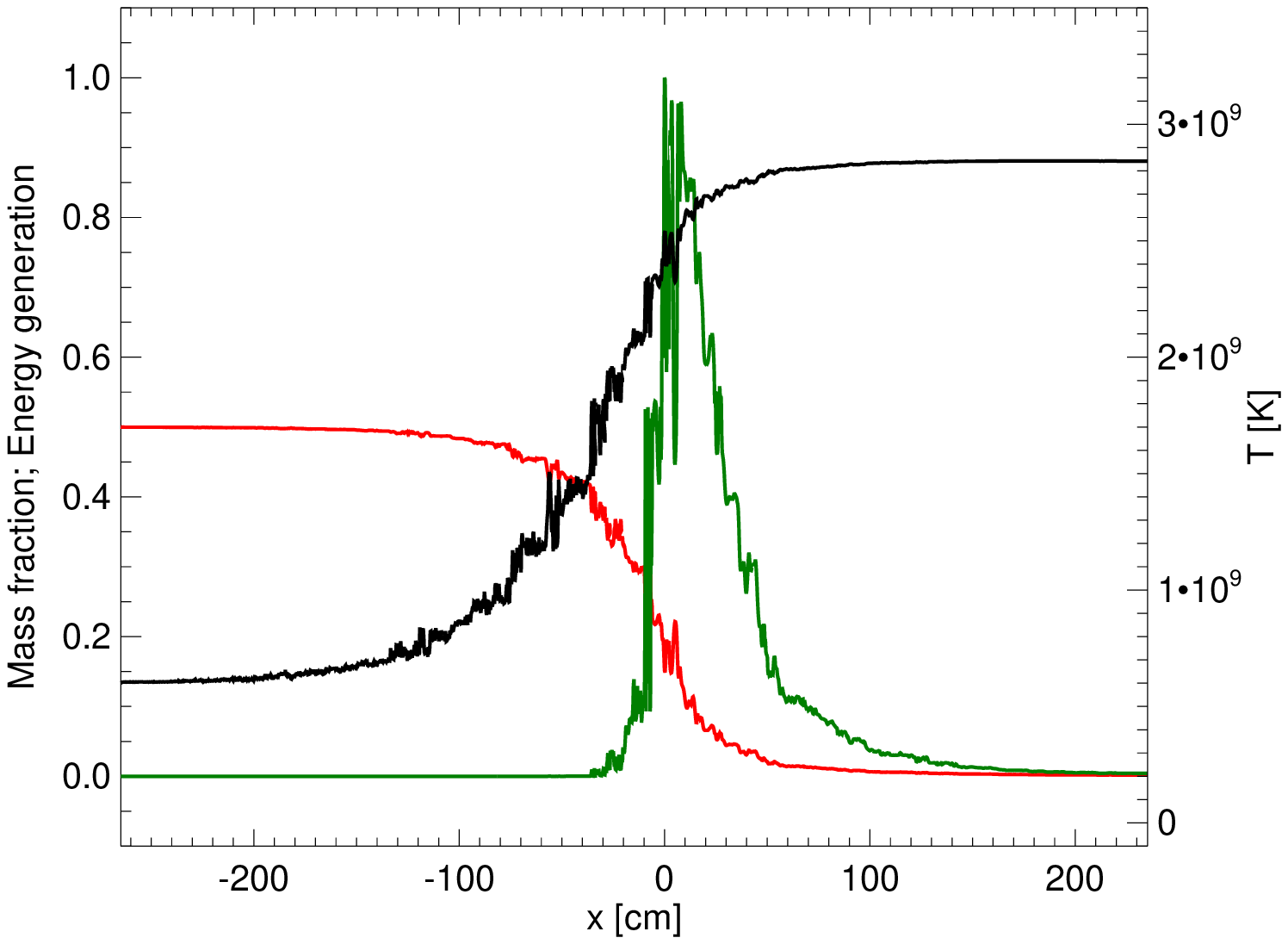}
\hfill
\includegraphics[width=0.475\textwidth]{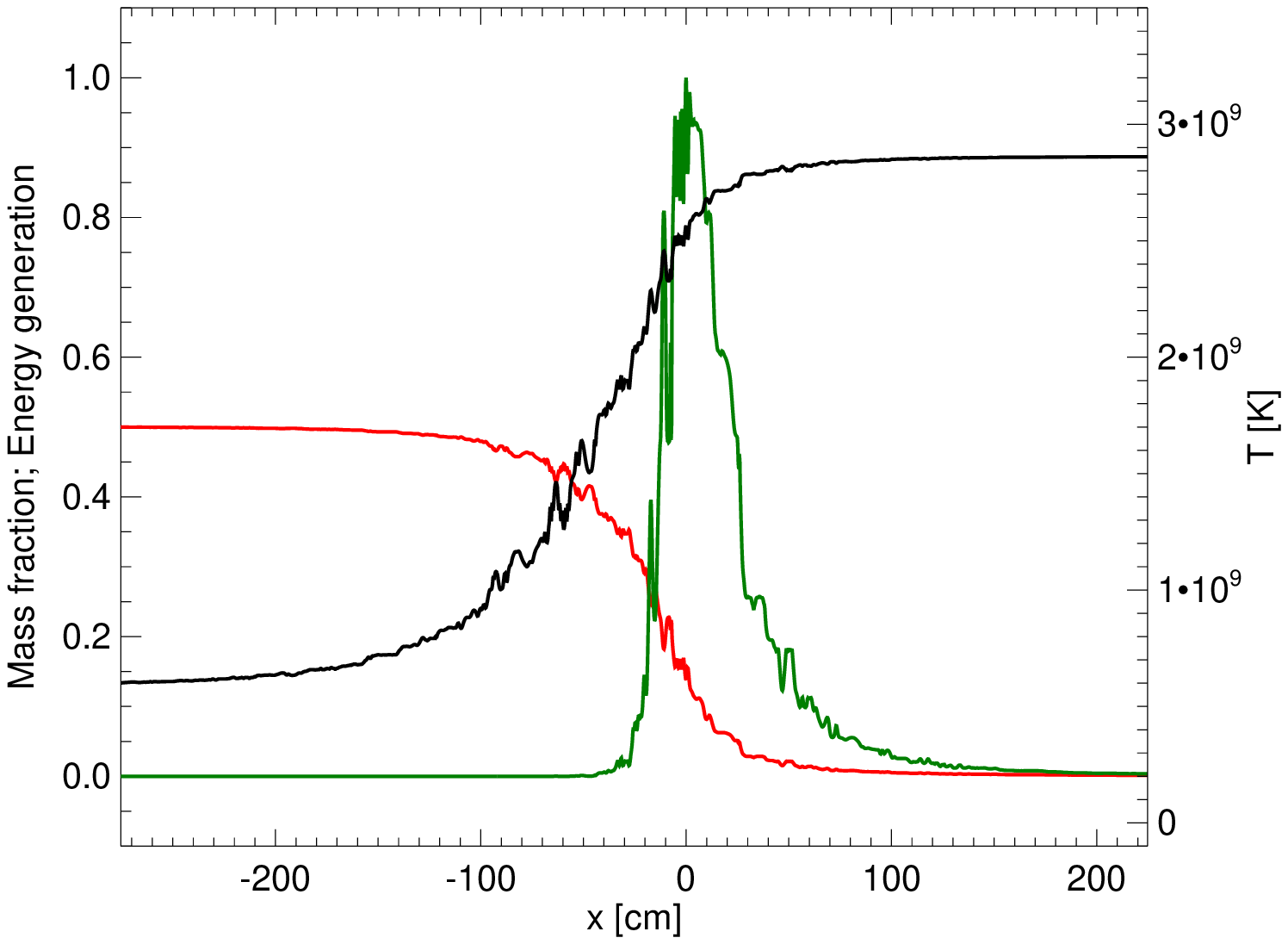}
\caption{Two recomputations of the turbulently broadened flame in
  Fig. 4.  Both calculations used the LEM code with a zoning of 2048
  zones ($\Delta x$ = 0.147 cm). The maximum energy generation in both
  plots is normalized to $2.3 \times 10^{20}$ erg g$^{-1}$
  s$^{-1}$. The calculation in the first frame used a diffusive energy
  transport coefficient based on radiation and conduction. The one in
  the second frame, virtually identical in result, used the subgrid
  model discussed in the text.
  \lFig{andy}}
\end{center}
\end{figure}

\begin{figure}
\includegraphics[angle=90,width=0.475\textwidth]{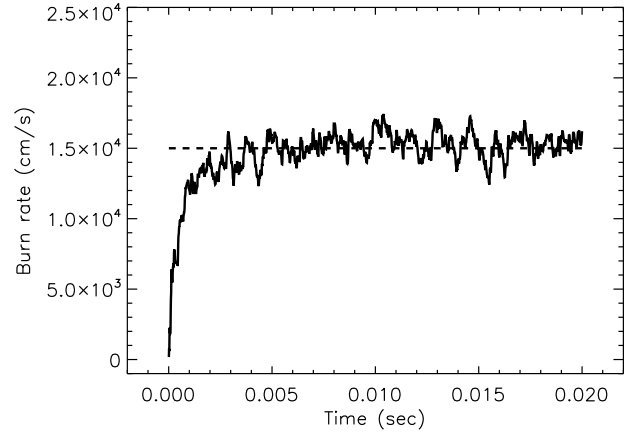}
\caption{Speed of the turbulently broadened flame shown in \Fig{andy}
  as a function of time. The turnover time on the integral scale is
  $6.1 \times 10^{-5}$ s. It takes many turnover times to reach the
  terminal speed, but then that speed is maintained rather precisely.
  The dashed line shows the average speed, $1.5 \times 10^4$ cm
  s$^{-1}$, or 5 times the laminar value. \lFig{andyspd}}
\end{figure}

\begin{figure*}
\begin{center}
\includegraphics[width=0.475\textwidth]{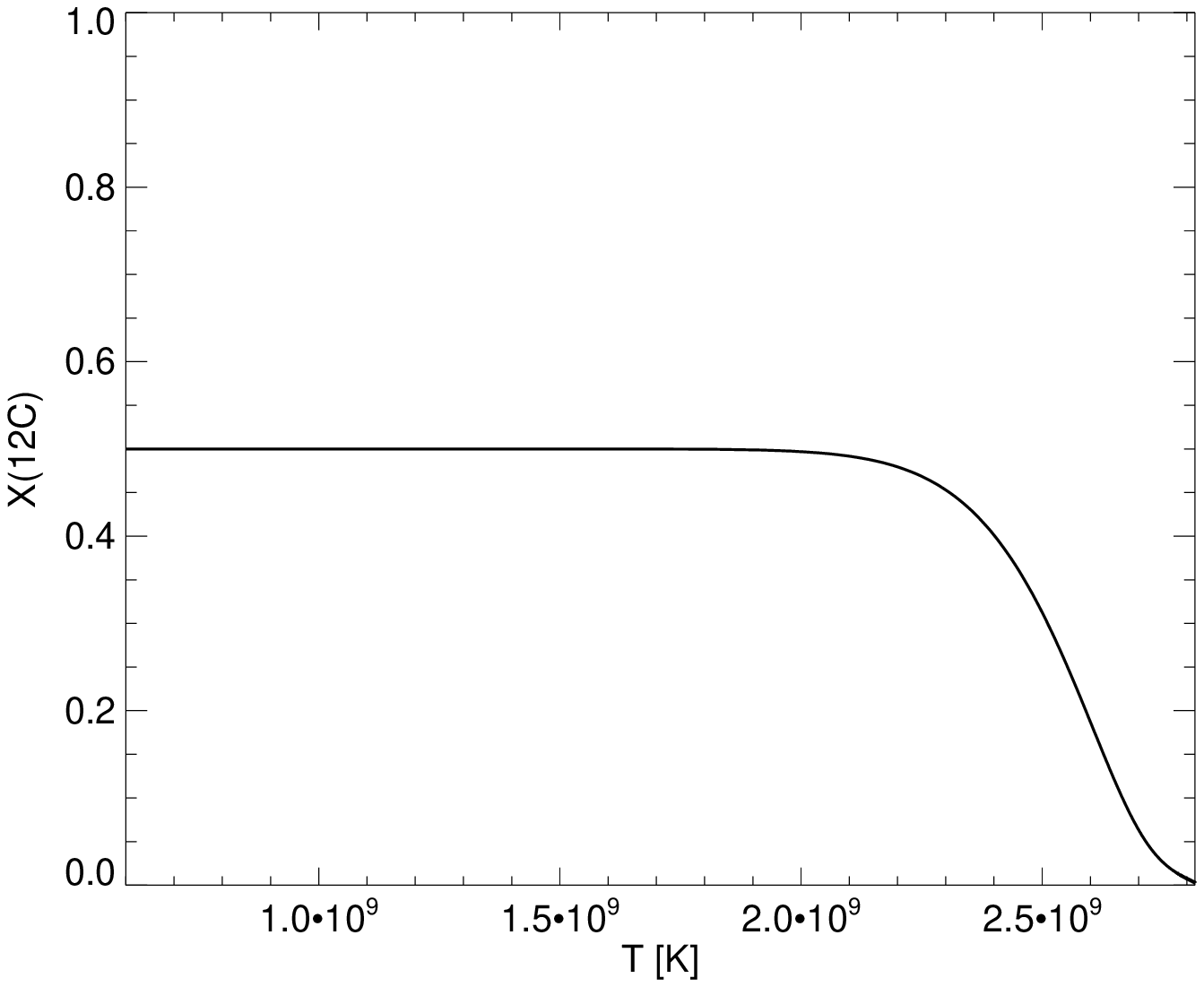}
\hfill
\includegraphics[width=0.475\textwidth]{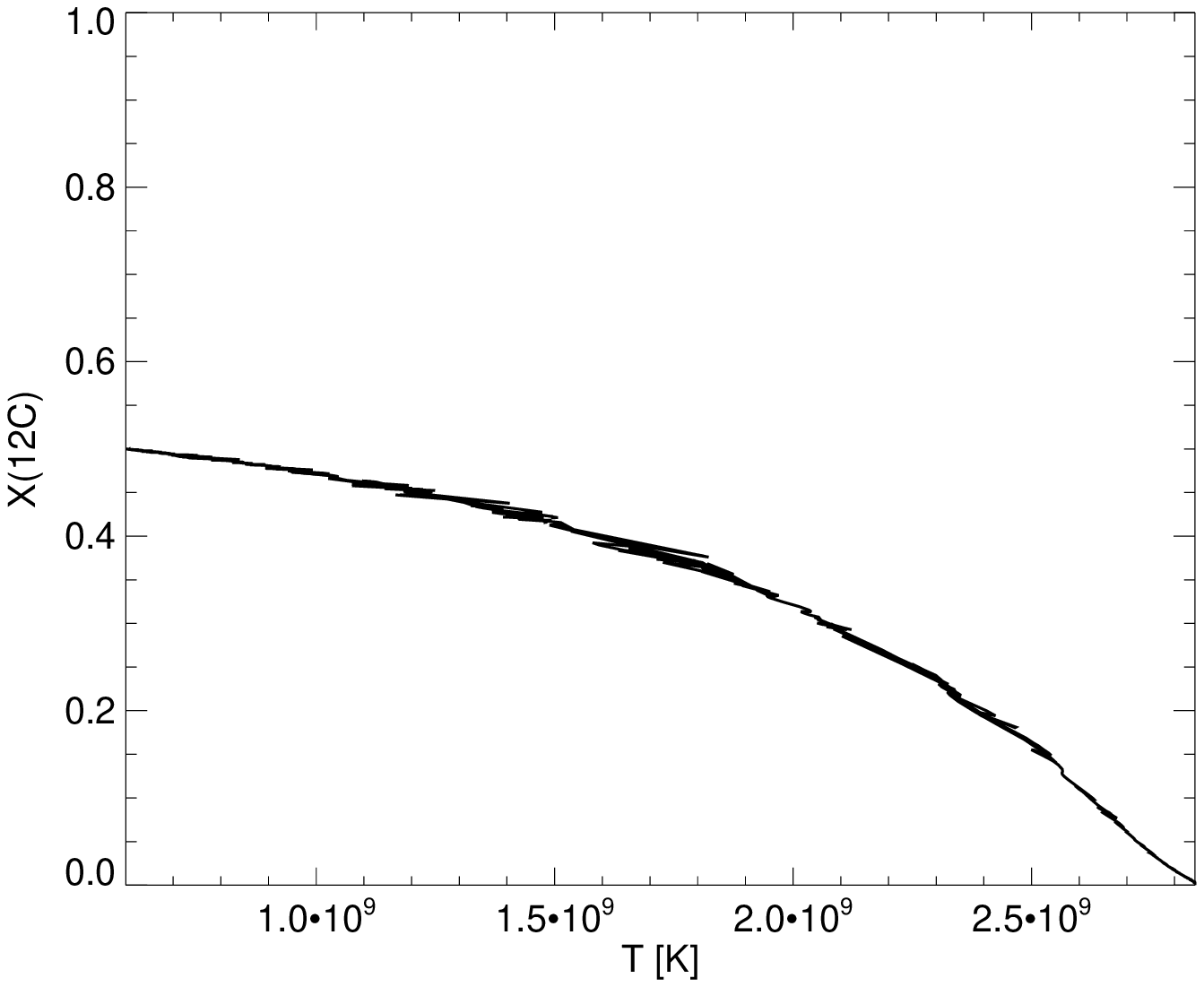}
\caption{Carbon mass fraction as a function of temperature for all
  grid points for the calculation of the laminar flame (left; see also
  \Fig{flame}) and the turbulently broadened flame (right; see also
  \Fig{andy}). The initial density and carbon mass fraction in both
  cases were $X^0_{12} = 0.5$ and $1.0 \times 10^7$ g
  cm$^{-3}$. Burning in the laminar flame occurs chiefly at a higher
  carbon mass fraction due to preheating by radiation. Burning in the
  turbulent flame occurs as hot fuel and cold ash are mixed with
  negligible radiation transport. Because burning a certain fraction
  of carbon releases a certain amount of energy and, at constant
  pressure, gives a unique temperature, all points lie on a
  well-defined line. A mixture of radiation transport and advection
  would have given points in between these two lines (see also
  \citet{Asp08}).  \lFig{c12oft}}
\end{center}
\end{figure*}

For the calculation shown in the first frame of \Fig{andy}, energy
transport by conduction and radiative diffusion was included, just as
in the laminar flame cases, and zoning was sufficient to resolve a
laminar flame had one been present. However, the transport here is
really dominated by the turbulence. The radiative diffusion
coefficient (\eq{drad}) varies from 730 cm$^2$ s$^{-1}$ in the fuel to
$1.8 \times 10^4$ cm$^2$ s$^{-1}$ in the ash while the turbulent
diffusion coefficient  
(\Sect{LEM}; $D_{\rm turb} = v' l /C$ with $v' = 2.47$ km s$^{-1}$, $l$ =
15 cm, and $C = 5$) is $7.4 \times 10^5$ cm$^2$ s$^{-1}$.

\Fig{c12oft} shows the distribution of carbon with temperature in the
case of a laminar flame (\Fig{flame}) and the turbulent flame
(\Fig{andy}). Both plots show a single-valued, monotonic relation. For
the laminar flame this is not surprising. At each temperature in the
self-similar front there is a unique carbon abundance. When eddies,
coupled with diffusion move heat and carbon around in a way that, in
LEM at least, is discontinuous, the result is perhaps more
surprising. It is also important that the curves are substantially
different in the two cases.

The curve for the turbulent case in \Fig{andy} is what one would
result were carbon to burn at constant pressure {\sl with no transport
  whatsoever} \citep{Asp08}. This means the process that is
transporting heat is transporting composition with equal efficiency,
i.e., the effective Lewis number is one. One expects this sort of
behavior when turbulence is dominating in the transport of both. The
Lewis number in the calculation was still 1000, but small scale eddies
kept the mixture so well stirred that even a small amount of ionic
diffusion at the grid scale gave the same result as if Le = 1. In
fact, the actual turbulent transport would have been even more
efficient had the resolution been higher. Recall that the Kolmogorov
scale is not resolved here.

\subsubsection{A subgrid model for turbulent transport}

If the turbulent diffusion coefficient, $D_{\rm turb} \sim v' l$, is so much
greater than conduction and radiation, then the results should be
independent of $D_{\rm rad}$. The second frame of \Fig{andy} shows
that this is indeed the case. Here the calculation sets conduction and
radiative transport to zero, but instead uses a diffusion coefficient
coupling individual zones of
\begin{equation}
D_{\rm SG} = v(\Delta x) \, \Delta x \, \frac{15}{C}
\end{equation}
where $\Delta x$ is the grid resolution (constant in this paper), and 
\begin{equation}
v(\Delta x) = \left(\frac{\Delta x}{l}\right)^3 v'.
\end{equation}
The subgrid diffusion coefficient $D_{\rm SG}$ represents the mixing
effects of unresolved eddies smaller than the grid scale.  The results
for $C = 5$ are identical.

The exact value of $D_{\rm SG}$ is not very important so long as it is
derived from a length scale that is very much less than the
(turbulently broadened) flame thickness. There is a characteristic
eddy size that is chiefly responsible for the diffusion (Appendix A)
and so long as that scale is well resolved, the results are
similar. However there is a choice for $D_{\rm SG}$ that works best as
the resolution becomes coarser. \Fig{subgrid} shows that a diffusion
coefficient derived from the turbulent speed on the grid scale has
much better scaling properties than one based on e.g., $6 \Delta x$,
even though the smallest eddy in LEM is $6 \Delta x$. The figure also
shows that the diffusive properties of a temperature discontinuity
remain unaltered as the resolution is decreased by a factor of 50.

\begin{figure*}
\begin{center}
\includegraphics[width=0.475\textwidth]{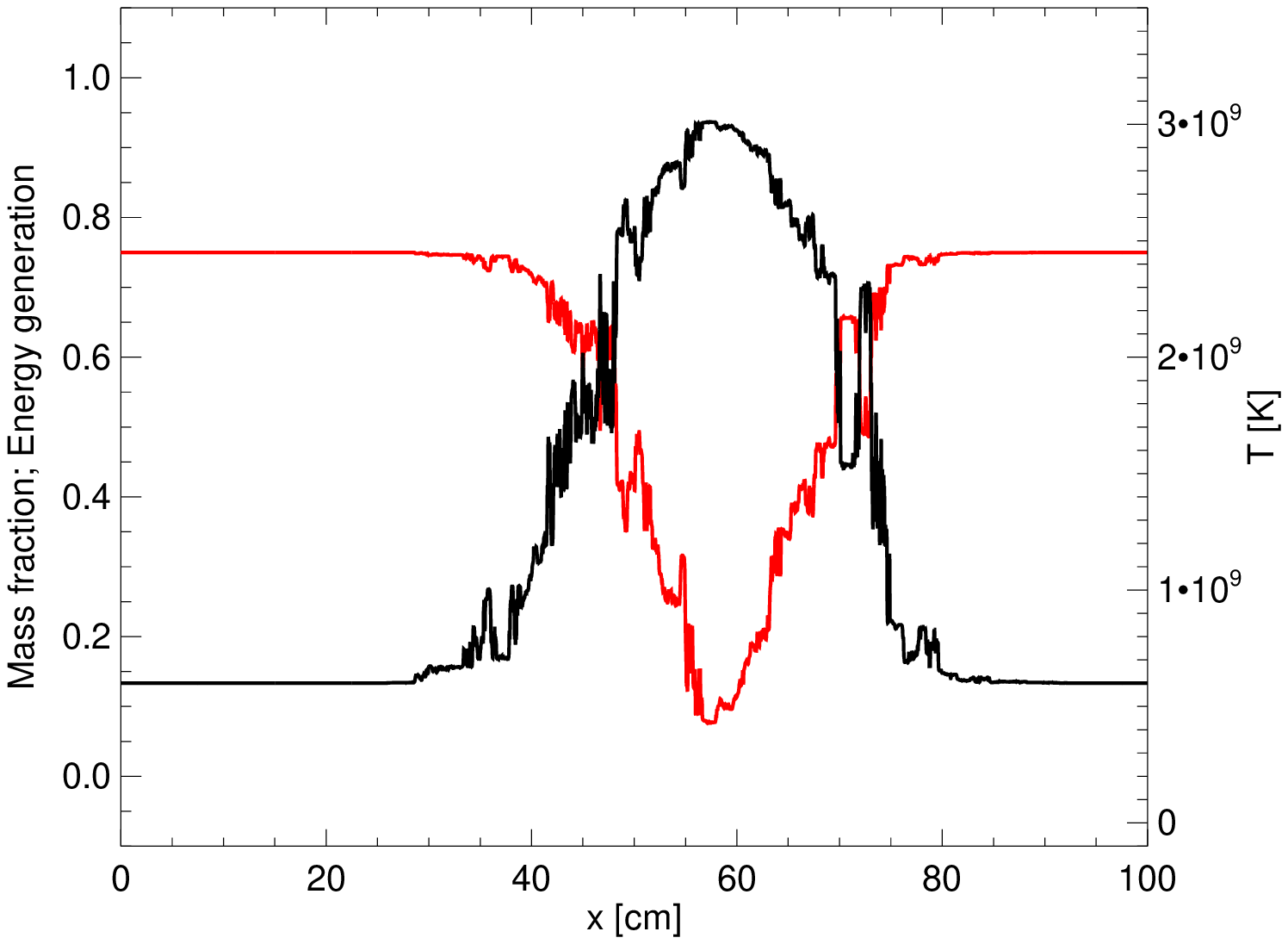}
\hfill
\includegraphics[width=0.475\textwidth]{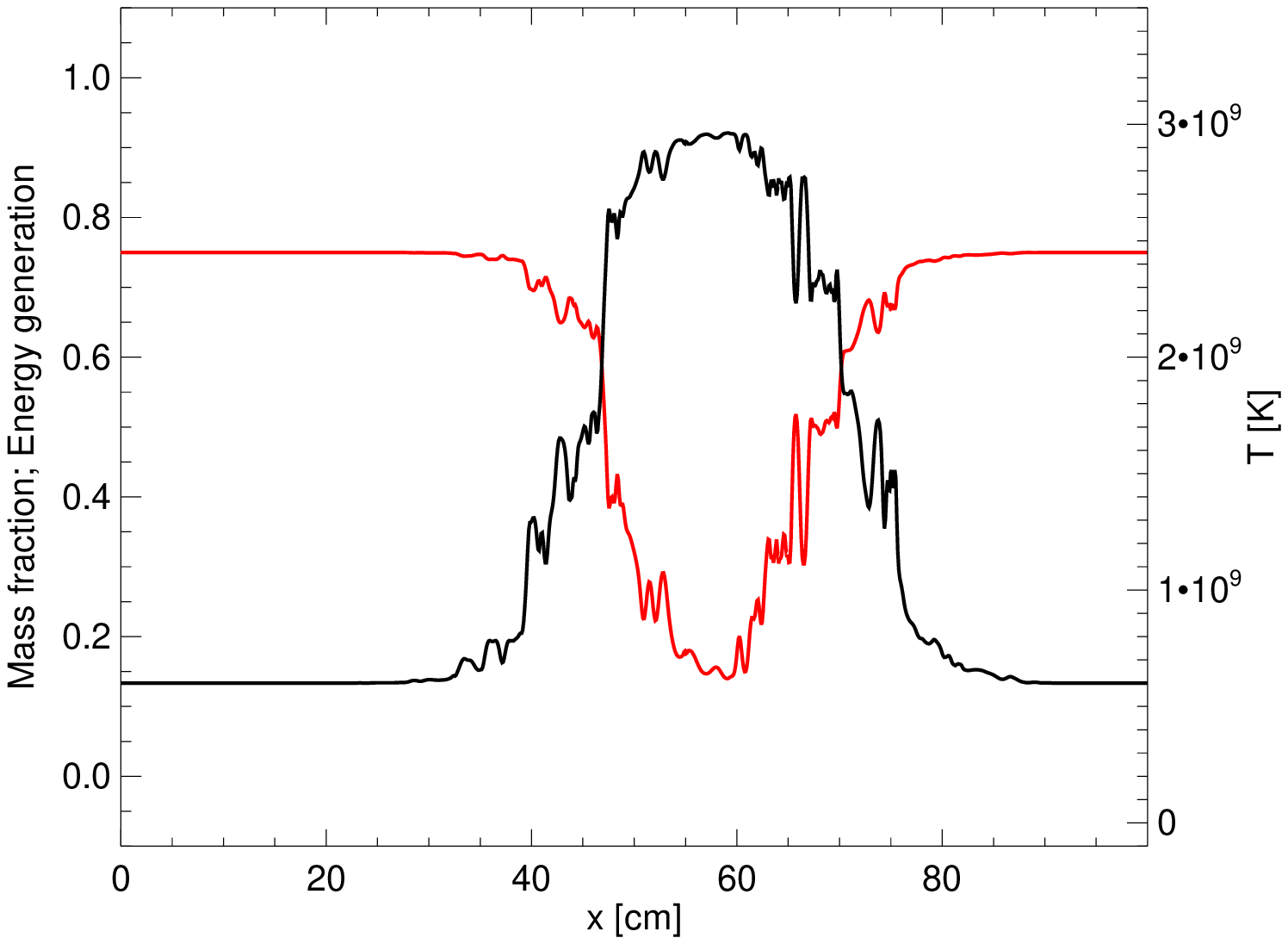}
\hfill
\includegraphics[width=0.475\textwidth]{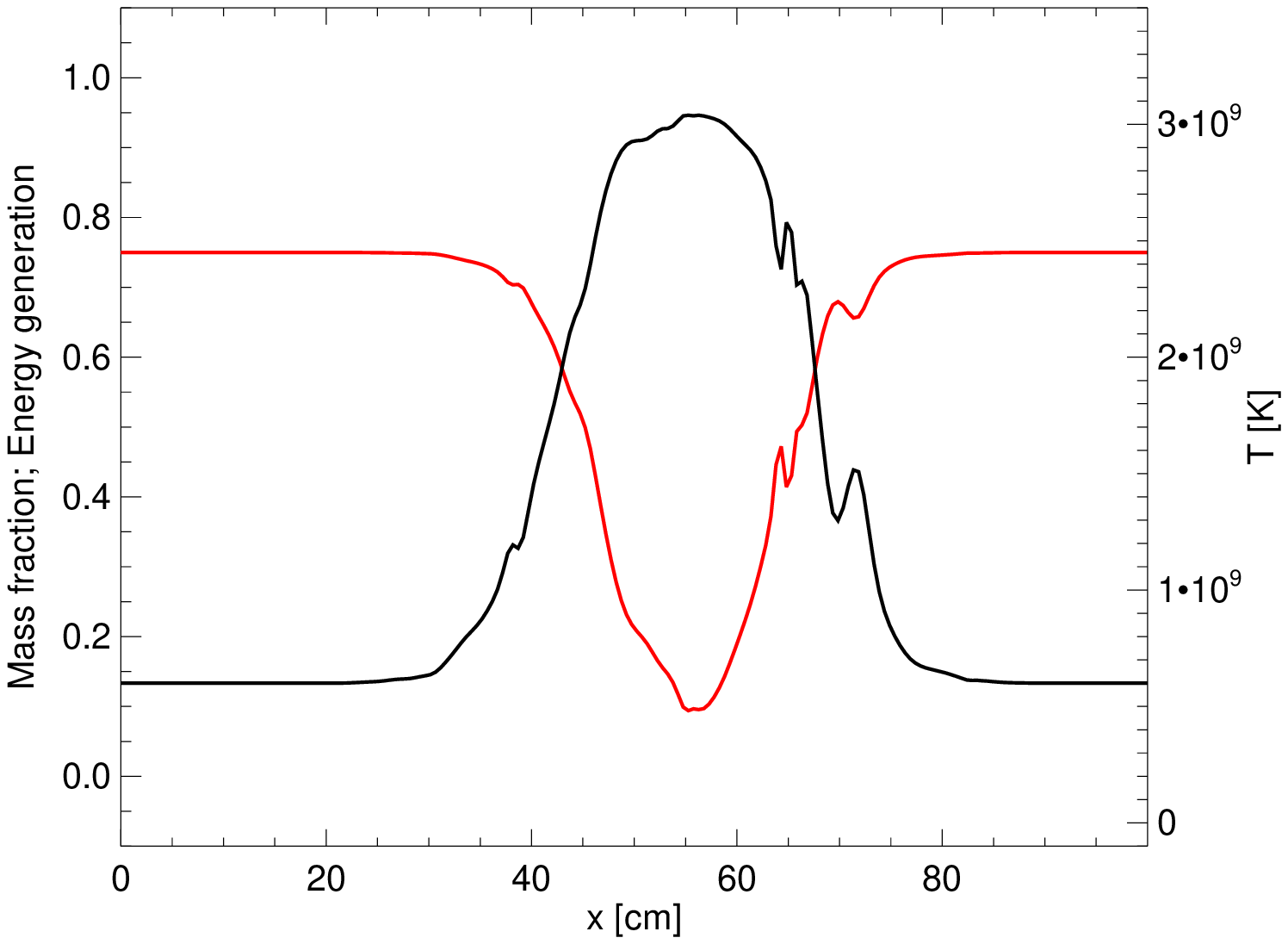}
\hfill
\includegraphics[width=0.475\textwidth]{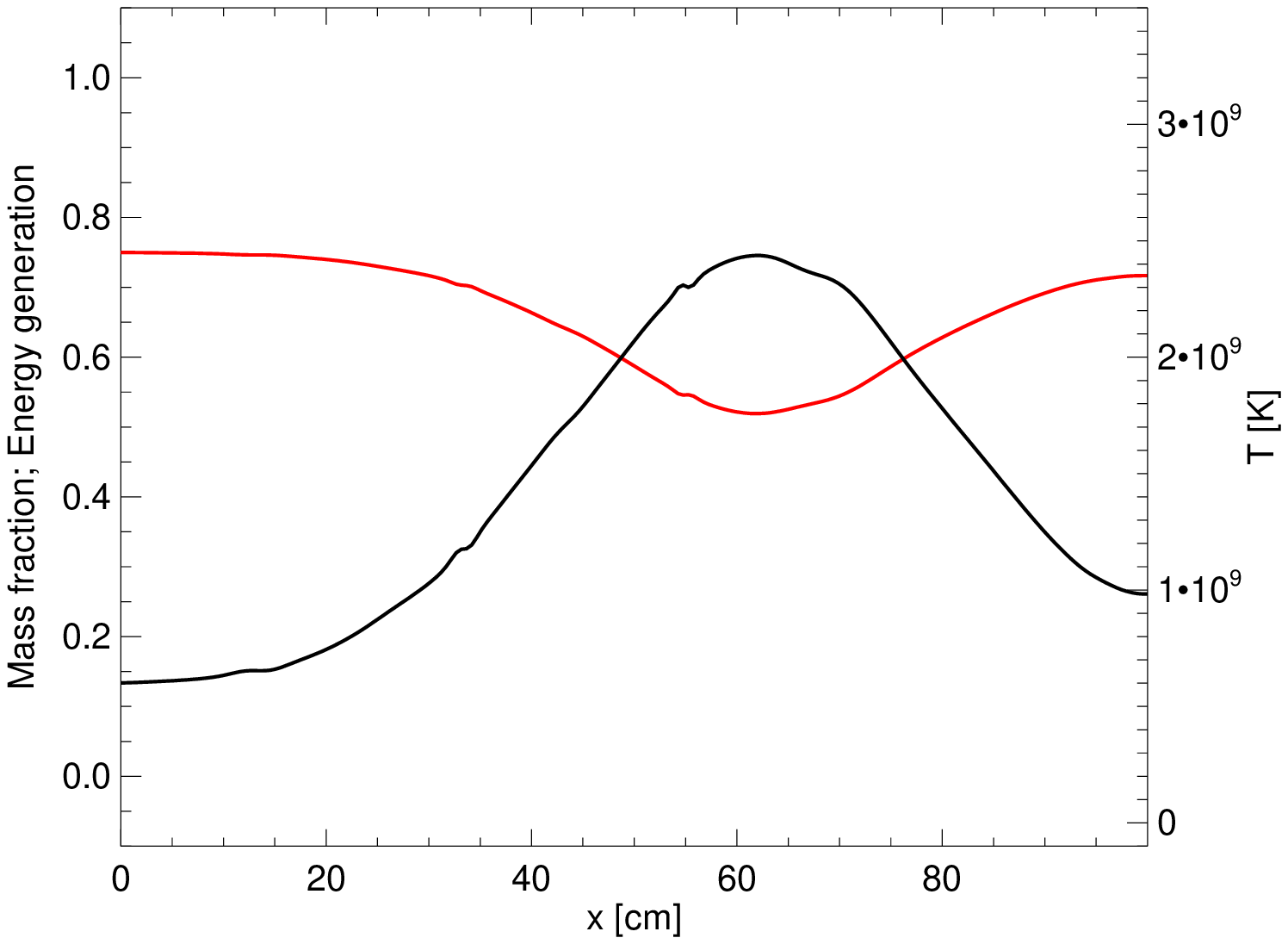}
\caption{Resolution study of the subgrid model. Nuclear burning and
  conduction were turned off so that all transport was by turbulent
  eddies. Turbulence with a speed of 10 km s$^{-1}$ was assumed on an
  integral scale of 5 cm. An island of ash 20 cm wide and initially
  centered at 50 cm was inserted in a background of fuel. The ensuing
  diffusive spreading was calculated using a grid of 10,000 (first
  frame), 1000 (second frame), and 200 (third frame) zones. The extent
  to which the fuel diffuses after $5 \times 10^{-5}$ s is very
  similar in all three runs. Larger subgrid diffusion coefficients on
  the other hand gave a resolution-dependent spreading that was more
  extensive for lower resolution. The fourth frame shows the result
  for $D_{\rm turb} = v(6 \Delta x) \, 6 \Delta x$ instead of $D_{\rm turb} = v(\Delta
  x) \, \Delta x$ (see text). In that case, the image for 10000 zones
  (not shown) is virtually unchanged. (Note that the grid moves to
  keep the point $2 \times 10^9$ K centered.) \lFig{subgrid}}
\end{center}
\end{figure*}

\begin{figure}
\begin{center}
\includegraphics[width=0.475\textwidth]{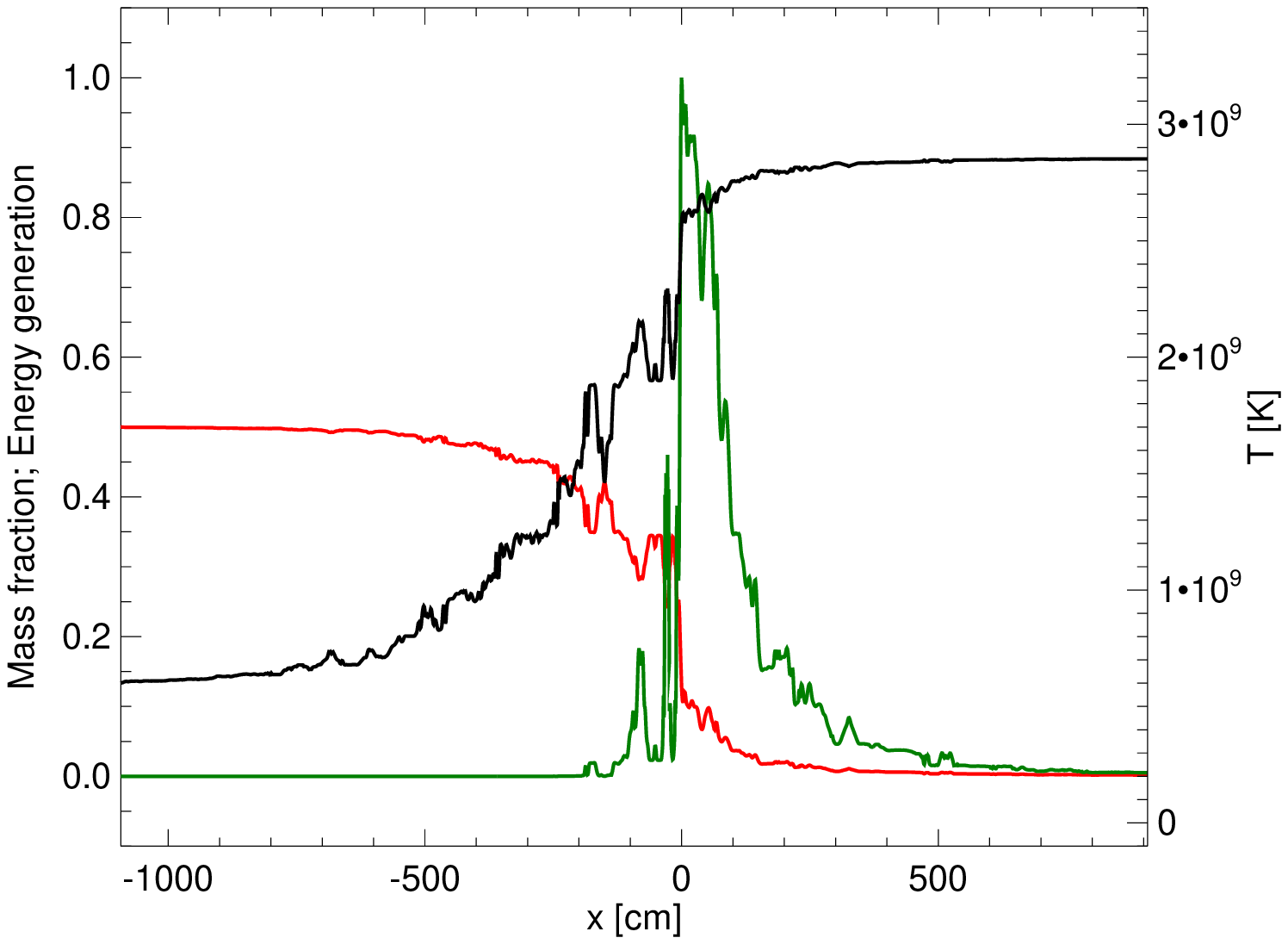}
\hfill
\includegraphics[width=0.475\textwidth]{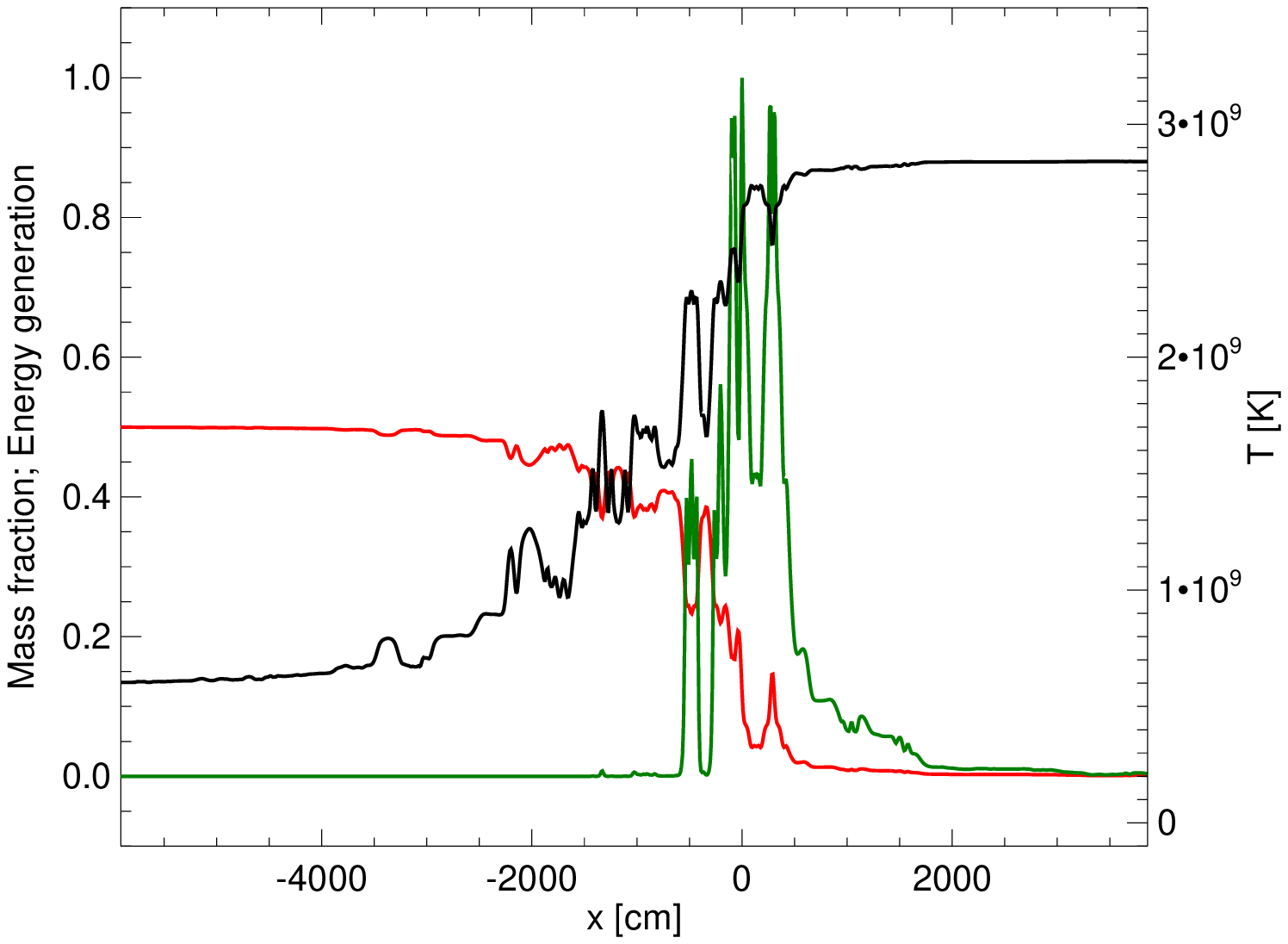}
\caption{Flames for larger integral scales than \Fig{andy}, but still
  in the WSR regime. The integral scales are 120 cm (first frame) and
  960 cm (second frame), and the corresponding {\sl rms} turbulent
  speeds on those scales, 4.93 km s$^{-1}$ and 9.86 km s$^{-1}$. The
  average propagation speeds are 0.61 km s$^{-1}$ and and 2.3 km
  s$^{-1}$. Note that both the flame speeds and widths scale as
  $l^{2/3}$ in this regime, both in this figure and as compared with
  \Fig{andy}. In the flame with the larger integral scale (right
  frame), one begins to see some structure in the flame and small
  ``ledges'' of mixture with nearly constant fuel abundance and
  temperature (\Sect{ledges}).  \lFig{andy2}}
\end{center}
\end{figure}

\subsubsection{The effect of increasing the integral scale}
\lSect{scaling}

Encouraged by these results, which imply that one need not resolve the
laminar scale in this regime in order to obtain accurate results, we
proceeded to explore flame properties for a large range of $l$
consistent with the condition $\epsilon_t = v'^3/l = 10^{15}$ erg
g$^{-1}$ s$^{-1}$ (Table 2).

In general, the speed of a flame propagated by diffusion should obey
the scaling relation
\begin{equation}
v_f \ \sim \ \sqrt{\frac{D}{\tau_{\rm nuc}}}.
\lEq{scale}
\end{equation}
where $D_{\rm turb} \sim v' \, l$.  \Fig{andy2} shows the effect on
the flame of increasing $l$ at constant turbulent energy dissipation
$\epsilon$, mass density, and carbon mass fraction. For awhile, the
profiles follow the scaling implied by \eq{scale}.  If the burning time
is independent of $l$,
\begin{equation}
\begin{split}
v_{\rm turb} \ \propto (v' l)^{1/2} \ \propto \ l^{2/3}, \\
\delta_{\rm turb} \ \propto \ v_{\rm turb} \ \propto \ l^{2/3}.
\end{split}
\end{equation}
For given, $l$, the flame speed also remains roughly steady.

However, this scaling also predicts its own breakdown.  Since
$v_{turb}$ increases as $l^{2/3}$ while $v'$ only increases as
$l^{1/3}$ there must be a critical integral scale $\lambda$ for which 
the two become equal.  $\lambda$ is determined by the relation 
$\lambda/v' = \tau_{\rm nuc}$, where $v'$ is the velocity fluctuation 
at scale $\lambda$.  Rewriting this as 
$\lambda = \left( v'^3 \tau_{\rm nuc}^3/\lambda \right)^{1/2}$ gives
\begin{equation}
\lambda = \left(\epsilon \tau_{\rm nuc}^3\right)^{1/2}.
\lEq{lambda}
\end{equation}

At $l = \lambda$, the flame width and integral scale are the same and
the burning speed is approximately equal to the turbulent speed on
that scale \citep{Ker01,Woo07}.  This is also the size of an eddy that
burns in one turnover time. The quantity, $\epsilon = U_L^3/L$, is a
constant by assumption.  Approximate values of $\lambda$ have been
tabulated for various densities, turbulent energies and carbon mass
fractions by \citet{Woo07}.  Better values for the cases considered
here can be inferred from the data in Table 2 which uses our reaction
network. Once $l > \lambda$, the flamelets no longer has the 
appearance of a diffusively broadened structure, though the 
individual flamelets can occasionally coalesce into
larger structures. This is the SF regime of the next section.  As
\Fig{andy3} and \Fig{andyvofl} show, the flame structure and speed
both become highly variable as the integral scale approaches this
value.

\begin{figure*}
\begin{center}
\includegraphics[width=0.475\textwidth]{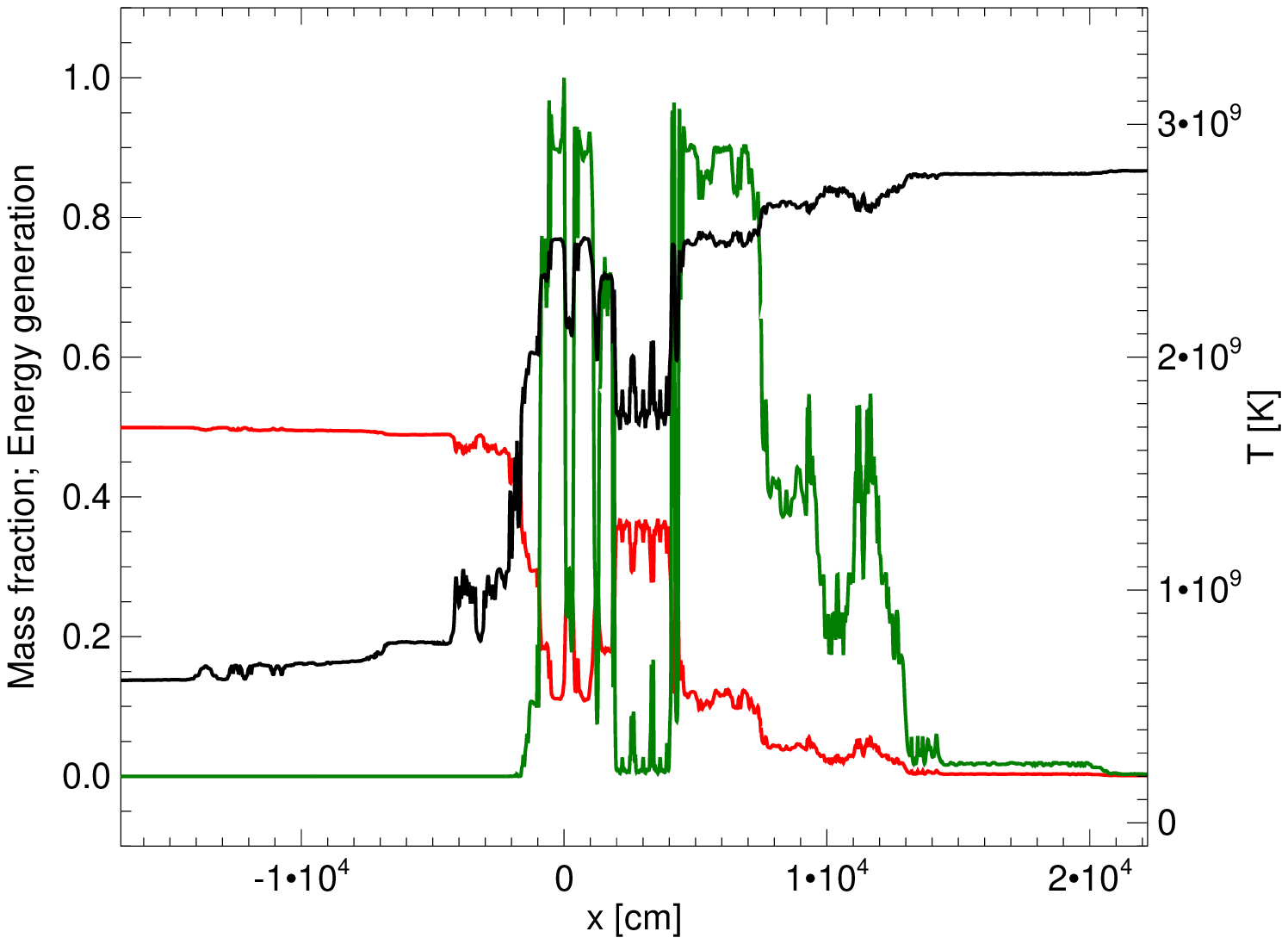}
\hfill
\includegraphics[width=0.475\textwidth]{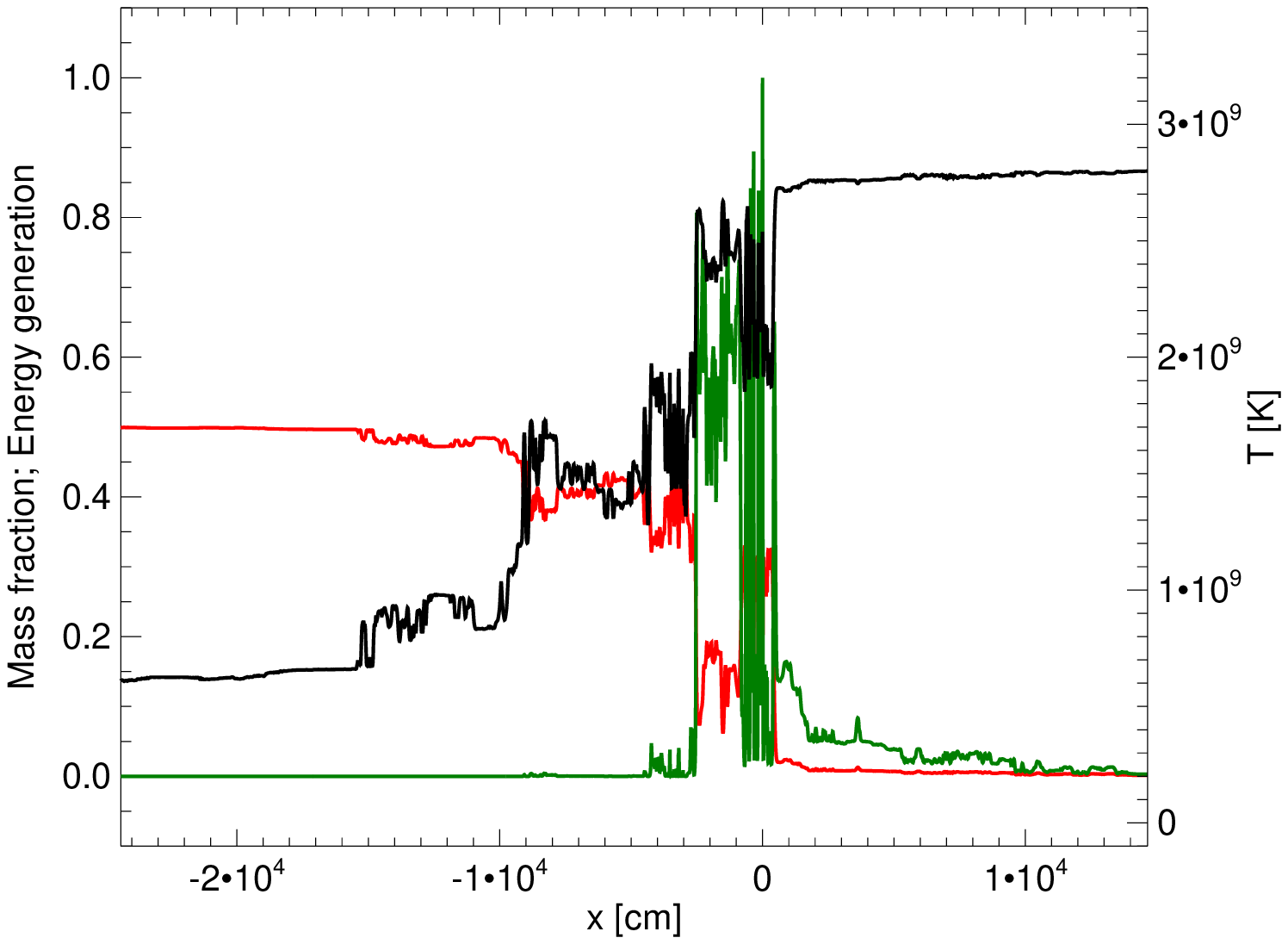}
\includegraphics[width=0.475\textwidth]{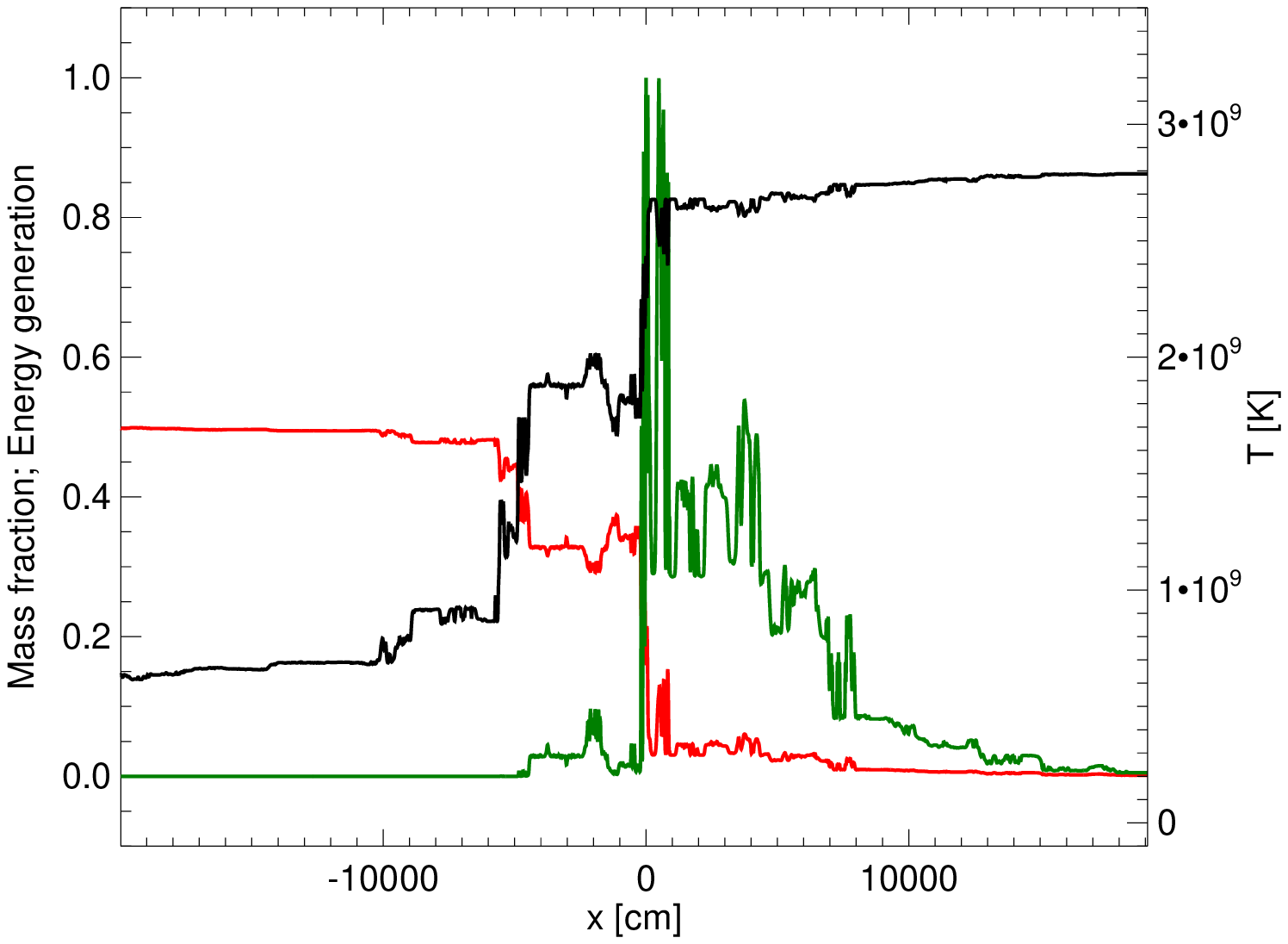}
\hfill
\includegraphics[width=0.475\textwidth]{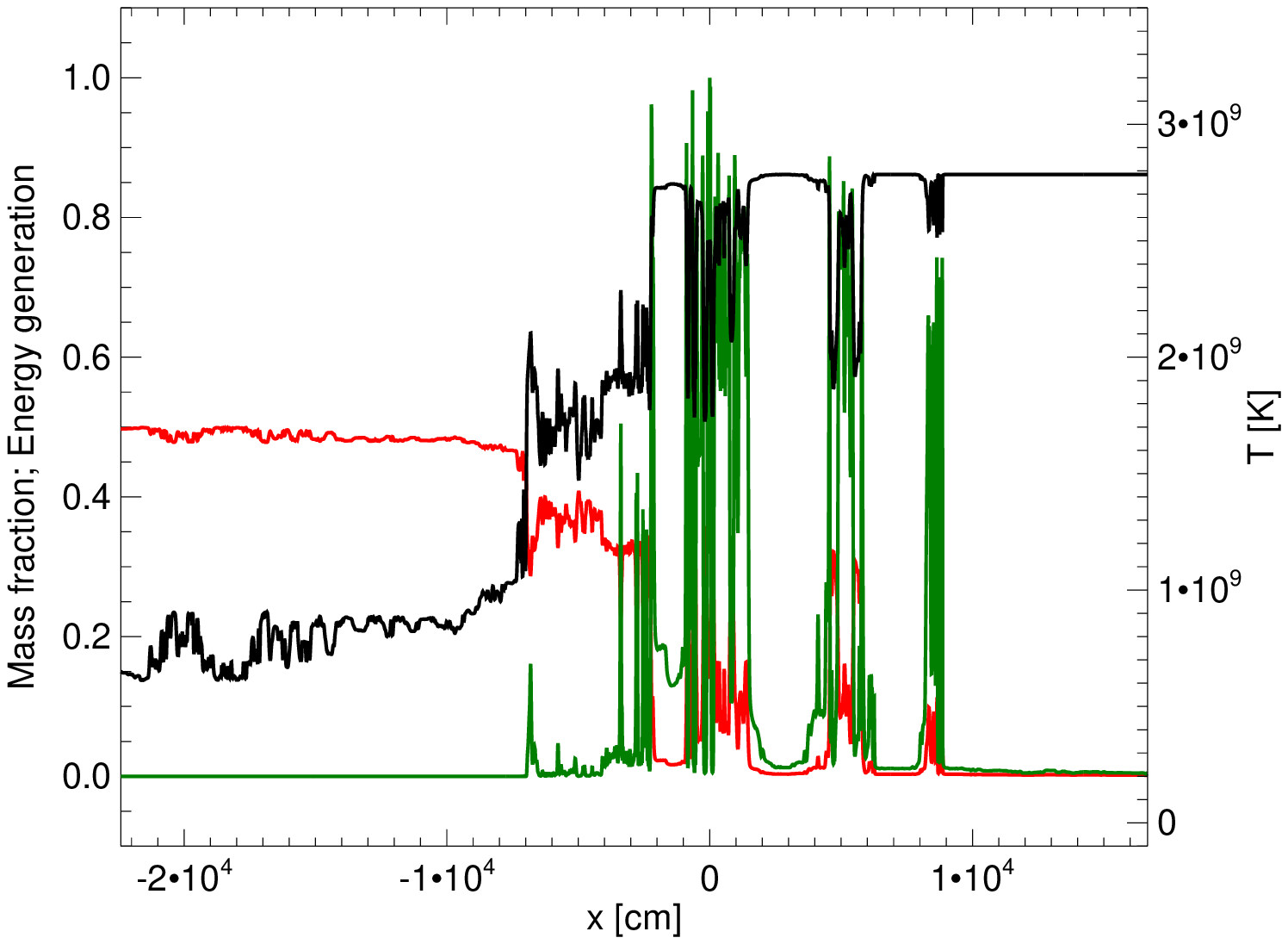}
\caption{A turbulent flame calculated for a still larger integral
  scale than in \Fig{andy2} and sampled at four different times.  The
  density and carbon mass fraction continue to be $1.0 \times 10^7$ g
  cm$^{-3}$ and $X^0_{12}$ = 0.50, and $\epsilon = 10^{15}$ erg
  g$^{-1}$ s$^{-1}$. The integral scale here is 76.8 m and the flame's
  average speed is 8.3 km s$^{-1}$ compared with an {\rm rms}
  turbulent speed on the integral scale of 19.7 km s$^{-1}$.  No
  well-organized, self-similar solution is visible at the four
  different times (contrast to \Fig{andy} and \Fig{andy2}). At some
  times the flame resembles the multiple structures in the flamelet
  regime (compare the lower right with \Fig{flamelet}). At others,
  sizable islands of well-mixed fuel and ash have nearly constant
  temperature. Some of these (lower left) are almost as large as the
  integral scale itself. The average width of the flame is still
  somewhat broader than the integral scale by about a factor of two.
  \lFig{andy3}}
\end{center}
\end{figure*}

\begin{figure*}
\begin{center}
\includegraphics[angle=90,width=0.475\textwidth]{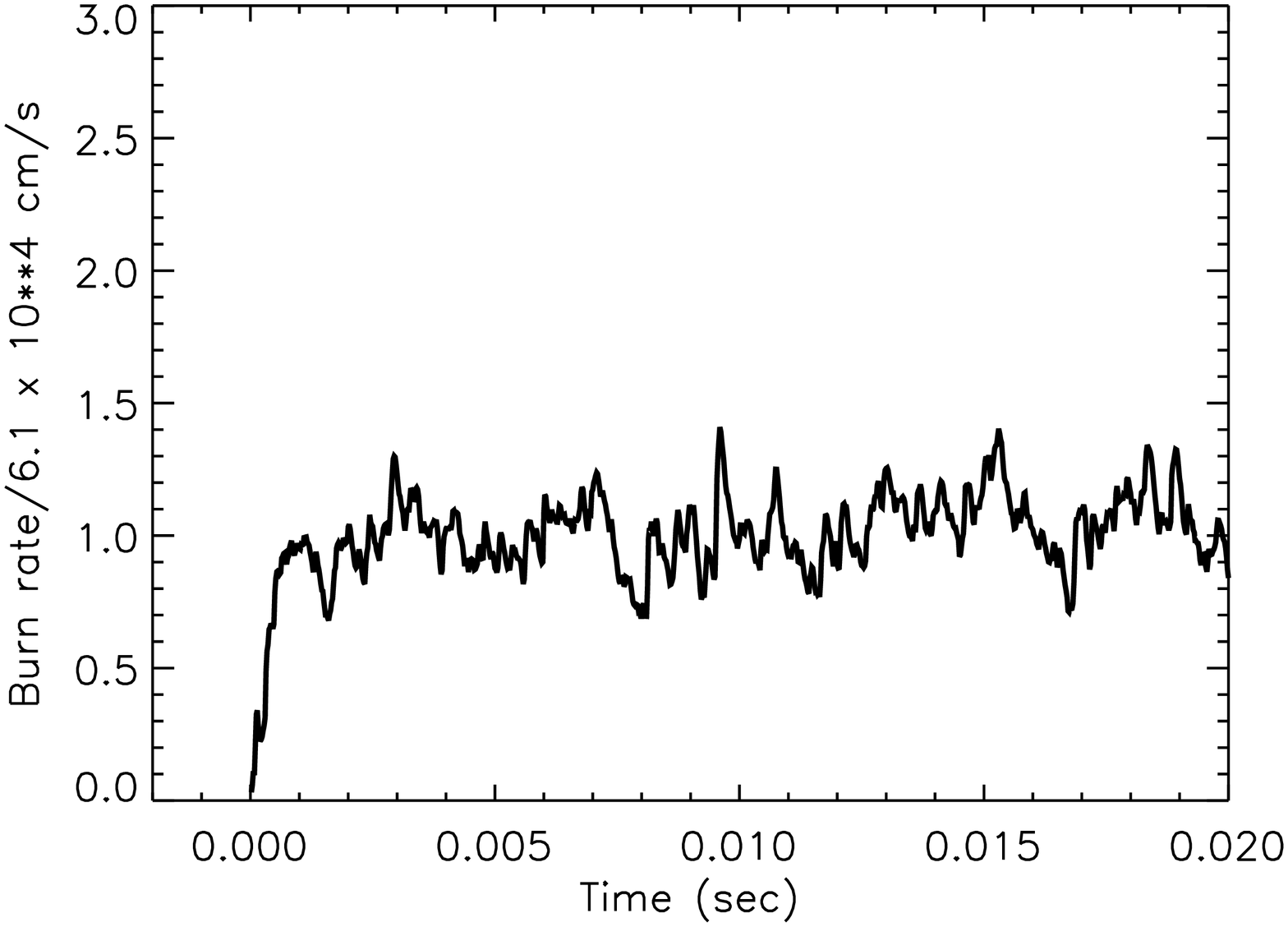}
\hfill
\includegraphics[angle=90,width=0.475\textwidth]{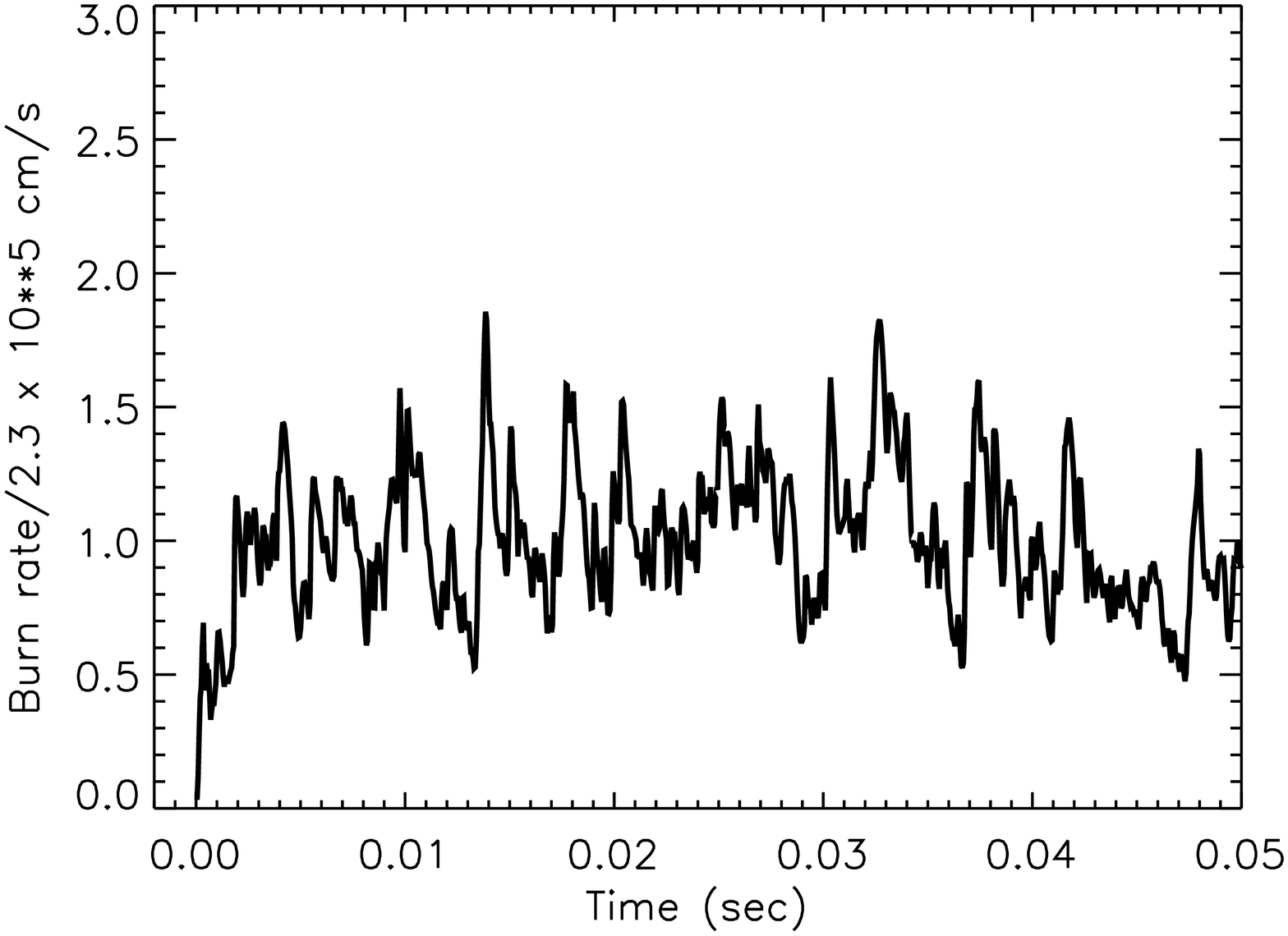}
\includegraphics[angle=90,width=0.475\textwidth]{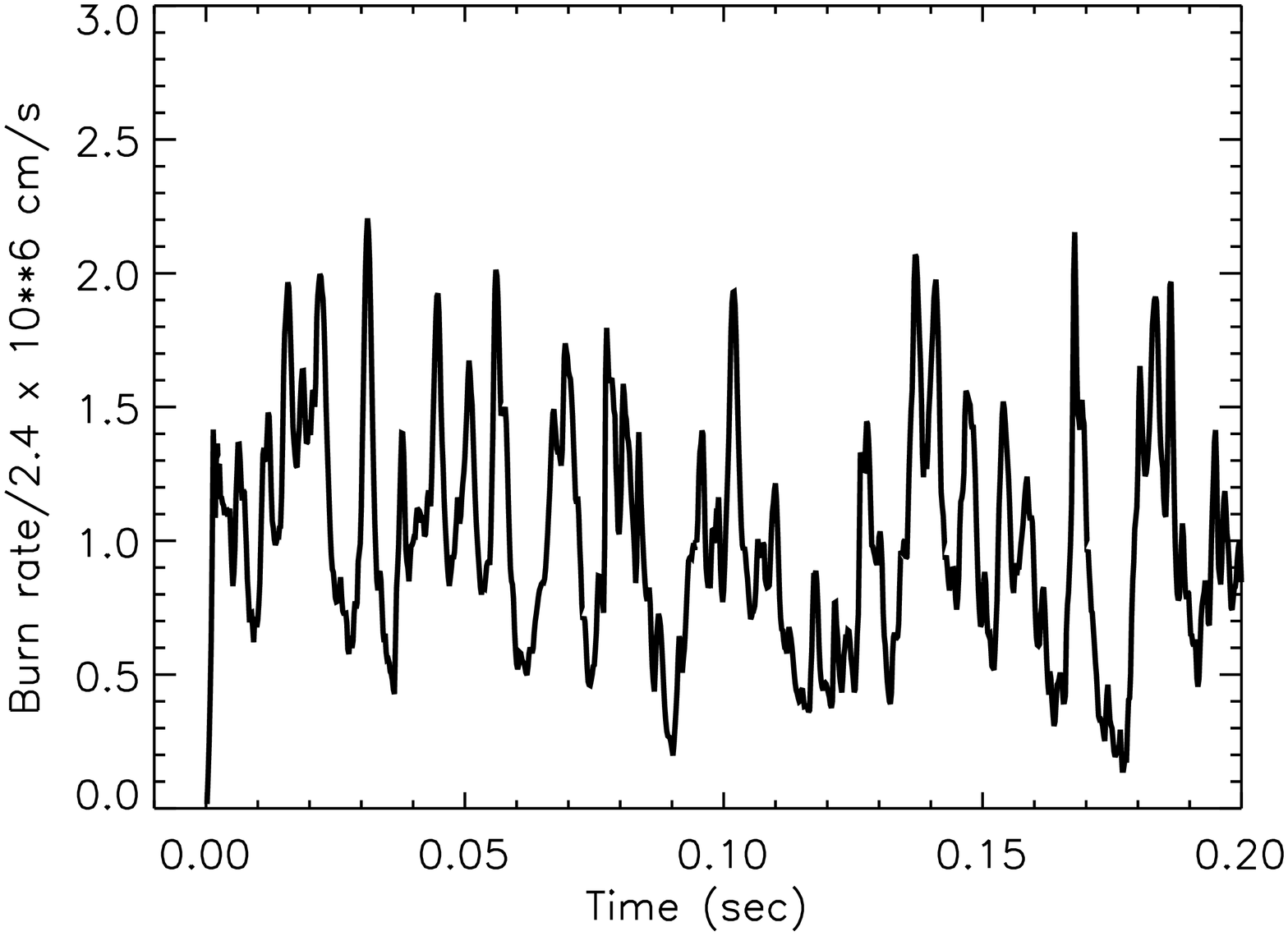}
\hfill
\includegraphics[angle=90,width=0.475\textwidth]{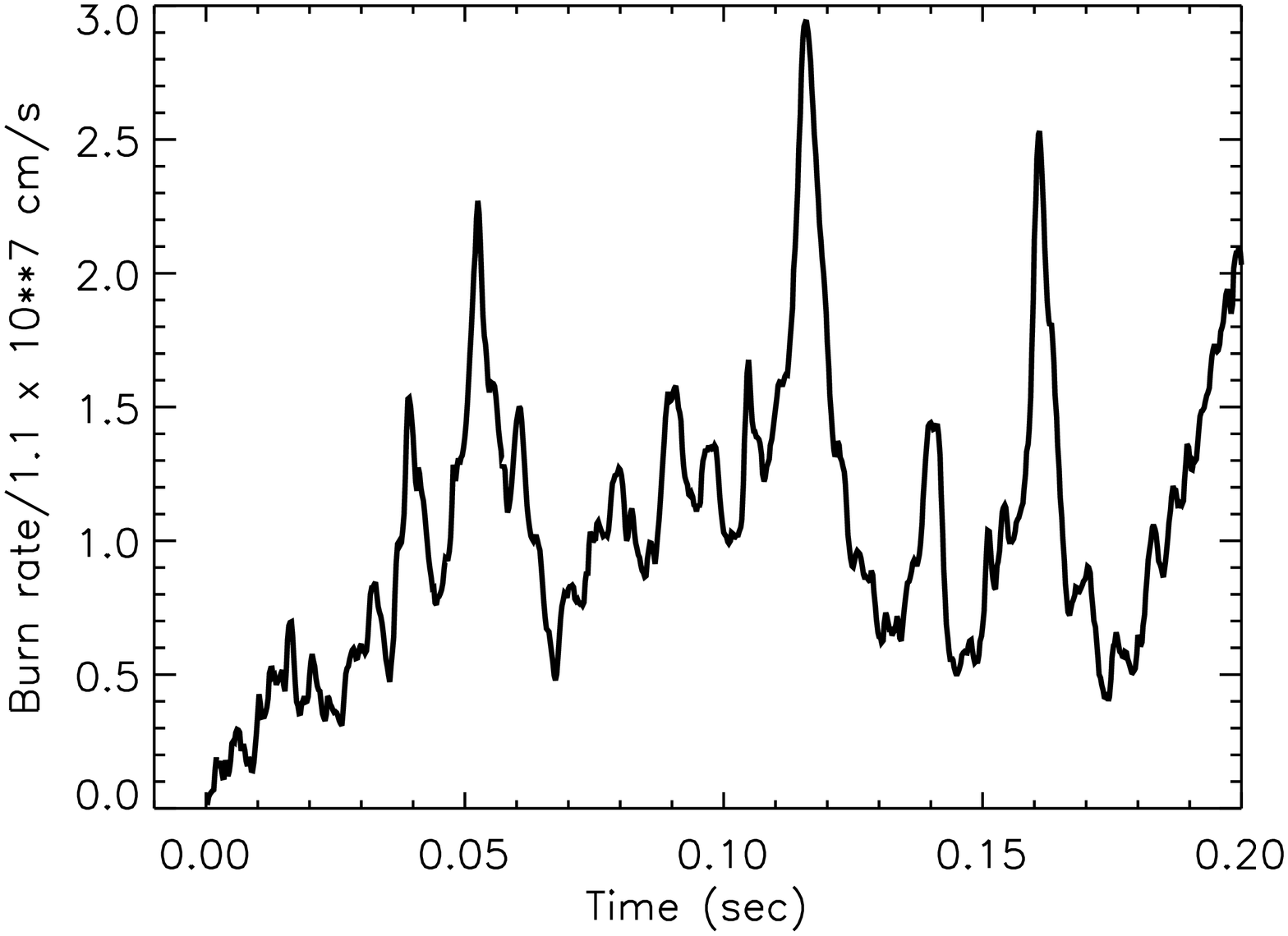}
\caption{Instantaneous flame speed divided by average speed for four
  different choices of $l$ and $v'$ at a density of 10$^7$ g
  cm$^{-3}$, fuel carbon concentration 50\%, and turbulent energy
  dissipation rate 10$^{15}$ erg g$^{-1}$ s$^{-1}$. The four choices
  correspond to the four integral scales in Table 2, $l$ = 120 cm, 960
  cm, 614 m, and 4.92 km. For the smallest scale (top left), $l <
  \lambda = 64$ m, and the flame speed (and width) are nearly
  constant. Going to larger length scales one sees the effect of
  individual large eddies and unsteady burning. In the last frame
  (lower right), the flame width and speed are approximately the same
  as turbulence on the integral scale and large burning rates are seen
  approximately every $l/v'$ seconds. The speed plotted here is the
  rate at which a single flame would move into the fuel with an
  burning rate equivalent to that on the entire grid. The actual
  burning here actually happens mostly in regions with much smaller
  carbon concentrations than in the unmixed fuel.  \lFig{andyvofl}}
\end{center}
\end{figure*}

\begin{figure*}
\begin{center}
\includegraphics[angle=90,width=0.475\textwidth]{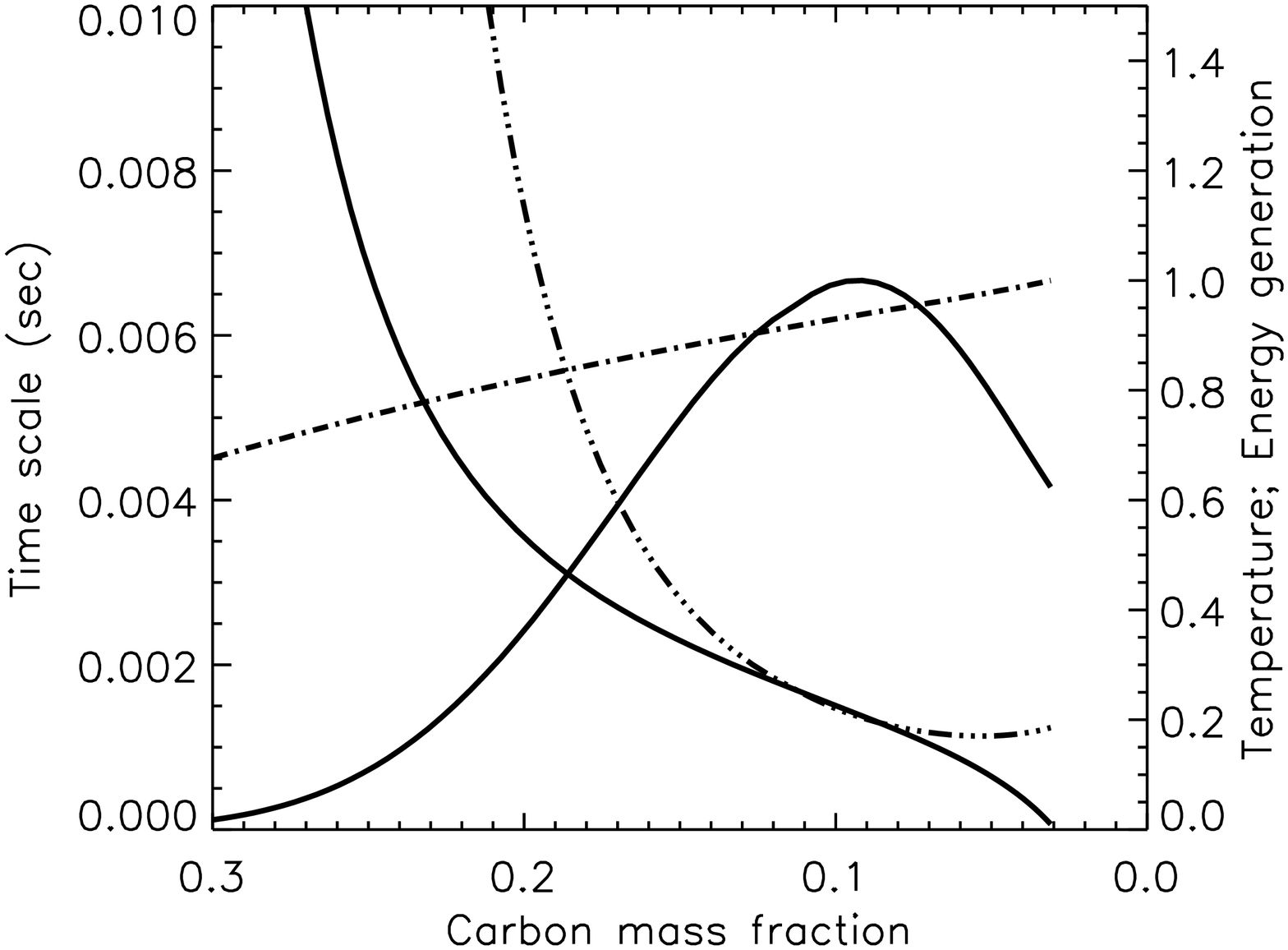}
\hfill
\includegraphics[angle=90,width=0.475\textwidth]{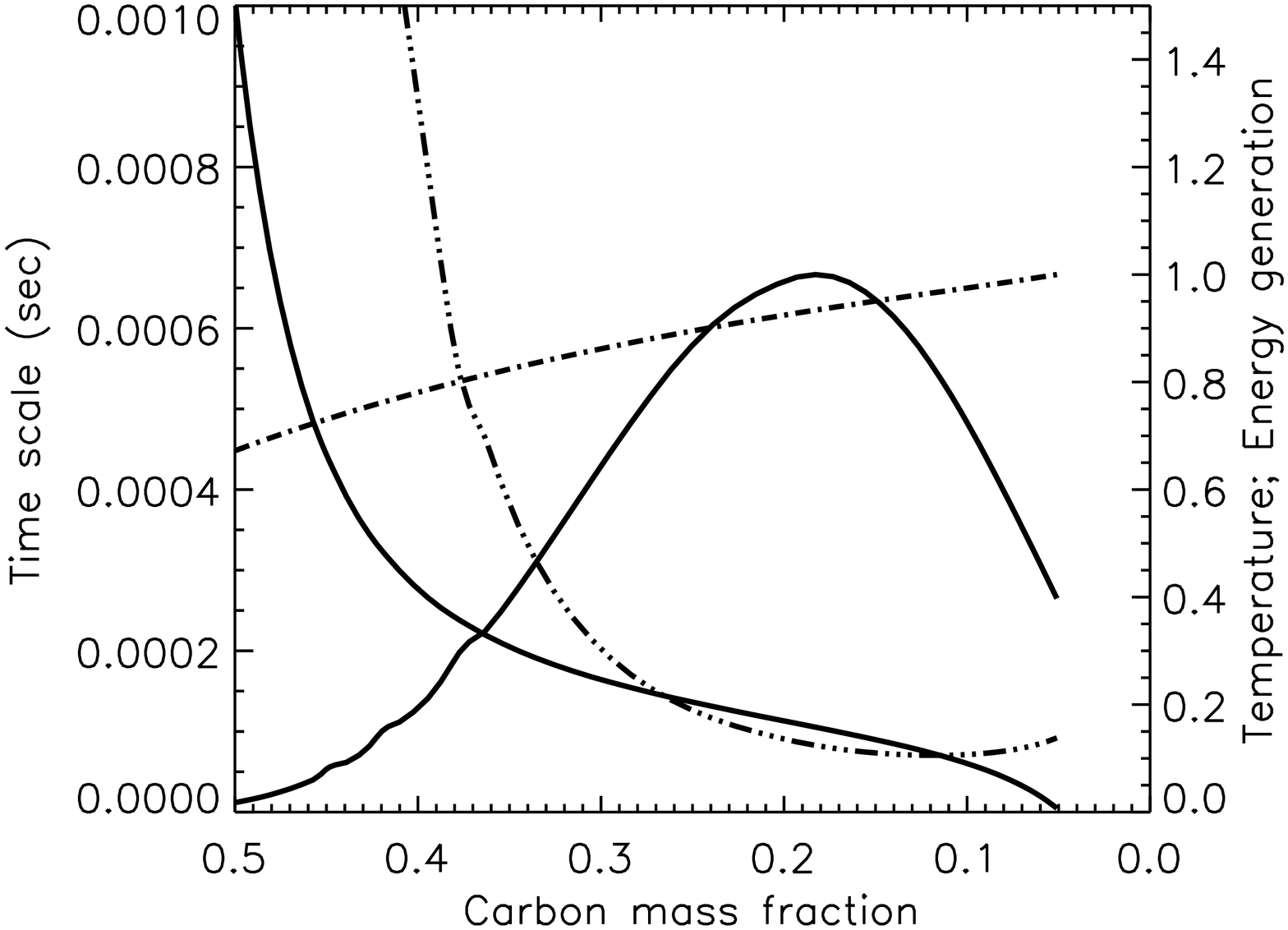}
\caption{Nuclear time scale and energy generation as a function of
  carbon mass fraction for carbon burning under isobaric conditions.
  The initial carbon mass fraction in the first frame was 0.50, though
  only the evolution below 0.30 is shown. Similarly in the second
  frame the initial carbon mass fraction was 0.75. Solid lines show
  the induction time scale (the time remaining until most of the
  carbon is consumed) and the nuclear energy generation rate divided
  by the maximum value achieved in the evolution. Dashed lines show
  the nuclear time scale, $[(1/X_{12})(dX_{12}/dt)]^{-1}$, and the
  temperature divided by the ash temperature. For the 50\% carbon run
  the maximum energy generation and ash temperature were $3.89 \times
  10^{19}$ erg g$^{-1}$ s$^{-1}$, and $2.59 \times 10^9$ K. For the
  75\% run, the corresponding values were $1.29 \times 10^{21}$ erg
  g$^{-1}$ s$^{-1}$, and $2.95 \times 10^9$ K. The induction time was
  arbitrarily defined to reach zero when the remaining carbon fraction
  was 0.03 for the 50\% carbon case and 0.05 for the 75\% carbon
  case. Energy released beyond these points is negligible and the time
  for the carbon to go to precisely zero is arbitrarily long. For high
  remaining carbon abundance, the induction time scale is much shorter
  than the instantaneous nuclear time scale because much of the
  remaining carbon will burn at a higher temperature. Note the
  relatively small change in induction time over an interesting range
  of carbon mass fraction, 0.2 to 0.1 (left) and 0.4 to 0.1
  (right). \lFig{taunuc}}
\end{center}
\end{figure*}

\subsubsection{The nuclear time scale in the WSR regime}

Burning in the WSR, $l < \lambda$, produces flames that, for a given
constant density and fuel composition, have a well-defined nuclear
time scale, $\tau_{\rm nuc}$.  \Fig{taunuc} shows the temperature, and
nuclear time scale as function of carbon mass fraction in the fuel for
two initial carbon concentrations (50\% and 75\% by mass) and fuel
density 10$^7$ g cm$^{-3}$. The burning is assumed to be isobaric and
two time scales are computed. One time scale reflects the
instantaneous rate of carbon consumption, $\tau = X_{12}/(dX_{12}/dt)$
(the dash-dotted line). More relevant to the flame speed, however, is
the {\sl induction} time scale, which is the time required to consume
most of the remaining fuel. In a situation where burning increases the
temperature, and therefore the reaction rate, the induction time can
be much shorter than the instantaneous burning time scale (defined by the 
current abundance divided by the current burning rate). Dividing the
turbulent flame width by the turbulent flame speed for those cases in
Table 2 where both quantities are well defined gives an approximately
constant value $\tau_{\rm nuc} \approx 0.003$ s for $\epsilon =
10^{15}$ erg g$^{-1}$ s$^{-1}$ and $X_{12}^0$ = 0.50.  This
corresponds to a carbon mass fraction in \Fig{taunuc} of $X_{12}
\approx 0.20$ where the energy generation is about 1/$e$ of its
maximum.

Defining $\tau_{\rm nuc}$ in this fashion as the induction time scale
from the point where the energy generation reaches $1/e$ of its
eventual maximum allows the computation of $\lambda = (\epsilon
\tau_{\rm nuc}^3)^{1/2}$. For the case of carbon mass fraction in the
fuel = 75\%, the induction time scale was calculated based on a
starting carbon mass fraction of 37\%. The values of $\lambda$ so
determined are given in Table 3. These values are about a factor of
two less than given in \citet{Woo07} because they refer to a time
scale derived from the width of the energy generating region. If one
instead uses the larger width based upon temperature or carbon mass
fraction, the derived values for $\lambda$ are about twice as large,
consistent with \citet{Woo07}.

\subsection{Stirred Flames (SF Regime)}
\lSect{stirred}

The stirred flame (SF) regime, which is characterized by Ka $\gg 1$
and Da = $L/(U_L \tau) = (L/\lambda)^{2/3} > 1$, is the most complex
of the three regimes of turbulent burning. In this case, there is no
well-determined scale for the flame width. Structures of size $\sim
\lambda$ persist to some extent, but since the flame experiences a new
eddy of length $\lambda$, in approximately the same time it takes to
burn a distance $\lambda$, it is subject to continual disruption. At
times there may be almost no burning; at others, nearly simultaneous
burning happens on scales much larger than $\lambda$.  The fact that
the flame has no persistent steady state also reflects a poorly
defined nuclear time scale. Each temperature has a different burning
time scale and thus the characteristic widths of mixtures prepared by
the turbulence is very temperature sensitive. Cooler mixtures have
larger length scales.  In general though, the average time scales are
longer and the flame structures larger than expected from our studies
of the WSR.

Because of the large integral scale in the supernova, the SF regime is
encountered as soon as the Karlovitz Number exceeds about 10. The
Damk\"ohler number at that point is already much greater than unity and
the WSR regime (Da $< 1$) is encountered much later, if ever.

\subsubsection{The transition from the flamelet to the SF regime}
\lSect{transition}

Just before the laminar flame is disrupted at Ka $\sim$ 10, there are
$\sim U_L/S_{\rm lam}$ flame surfaces folded in the flame brush. The
average spacing between these flamelets is $d \sim (L/U_L) S_{\rm
  lam}$ (Table 3) and the thickness of each is $\delta_{\rm lam}$. A
short time later, as Ka continues to increase, these flamelets are
smeared out to make a smaller number of broader, faster structures
with characteristic average thickness $\lambda$.  Their number then is
$\sim \sqrt{\rm Da} = (L/\lambda)^{1/3}$, and their spacing is $\sim
L/\sqrt{{\rm Da}} = L^{2/3} \lambda^{1/3}$.  So long as Da $\gg 1$, as
it is at the transition, the new broadened structures will, on the
average, still be separated by distances much greater than their size
and will not coalesce into one large mixture.

The quantity $\lambda$ (\eq{lambda}) that plays a such a critical role
in this discussion can be thought of as a generalization of the Gibson
length (\eq{Gibson}). For the flamelet regime, the Gibson scale is the
size of the eddy that would be crossed by a laminar flame with speed
$S_{\rm lam}$ on an eddy turnover time. It is smaller for larger
turbulent energy. The quantity $\lambda$, on the other hand, is also
the size of an eddy that turns over on a nuclear time scale, but its
size depends only on turbulence properties, not the radiative
diffusion that sets $S_{\rm lam}$. It is larger for larger turbulent
energy. As the supernova expands, the Gibson scale shrinks as $S_{\rm
  lam}$ declines. Eventually when Ka $= 1$, $l_G = \delta_{\rm lam}$,
but except for differences due to a changing nuclear time scale
(\Sect{scaling}), at Ka = 1, $\lambda$ is also approximately equal to
$l_G$ and $\delta_{\rm lam}$. Thereafter, as the density decreases
more, $l_G$ ceases to have much meaning since laminar flame speeds are
no longer relevant. The new relevant scale, $\lambda = (\epsilon
\tau^3)^{1/2}$, on the other hand, grows rapidly at lower density due
to the increasing nuclear time scale.

What this means is that the transition from laminar burning to the SF
regime is probably smooth and uneventful. Flamelets will grow
gradually in speed and width as the density declines, not
discontinuously. Because the nuclear time, and thus $\lambda$, lack
precise definitions in the SF regime, and because of intermittency,
one cannot completely rule out large scale transient mixing at the
laminar-SF transition, especially if the turbulent energy is very
large, but a detonation here seems unlikely. If it did happen at such
high density, a very bright supernova would result due to the near
complete incineration of the star.

\subsubsection{Complex structure}
\lSect{complex}

As the density continues to decrease, $\lambda$ increases and there
are fewer, thicker, faster turbulent flamelets in the flame brush
(\Fig{andy3} and \Fig{andyvofl}). The limit of one flame of size
$\lambda = L$ is reached when Da = 1. This is the condition suggested by
\citet{Woo07} as likely for detonation.  The present study shows,
however, that mixed regions larger than $\lambda$ can exist
transiently even for Da $\gg 1$. Since the critical mass for detonation
decreases rapidly with increasing density, this makes detonation
easier.

Consider the case of 50\% carbon, a turbulent dissipation rate of
10$^{15}$ erg g$^{-1}$ s$^{-1}$, density 10$^7$ g cm$^{-3}$, and
integral scale $l = 4.92$ km (Table 2). For these conditions,
$\lambda$ = 64 m (Table 3) and Ka = $(4.92 \times 10^5/6.4 \times
10^3)^{2/3} = 18$. The average number of flame surfaces is $\sqrt{\rm
  Ka} \sim 4$. \Fig{andy5} shows that mixed regions as big as L (and
much bigger than $\lambda$) occasionally exist. The extent of mixing
and hence the flame speed is highly variable (\Fig{andyvofl}). Most of
the time, less mixing is seen than in \Fig{andy5} and only a few
disjoint regions of burning are present, but structures like these can
exist for an eddy turnover time or so. Characteristics of these
regions include a carbon mass fraction much less than in the fuel,
typically on the rapidly rising part of the energy generation curve in
\Fig{taunuc}, and temperatures that are a substantial fraction of the
ash temperature. The relatively high temperatures are a consequence of
the temperature dependent nature of the heat capacity (\Sect{cp}).

\begin{figure*}
\begin{center}
\includegraphics[width=0.475\textwidth]{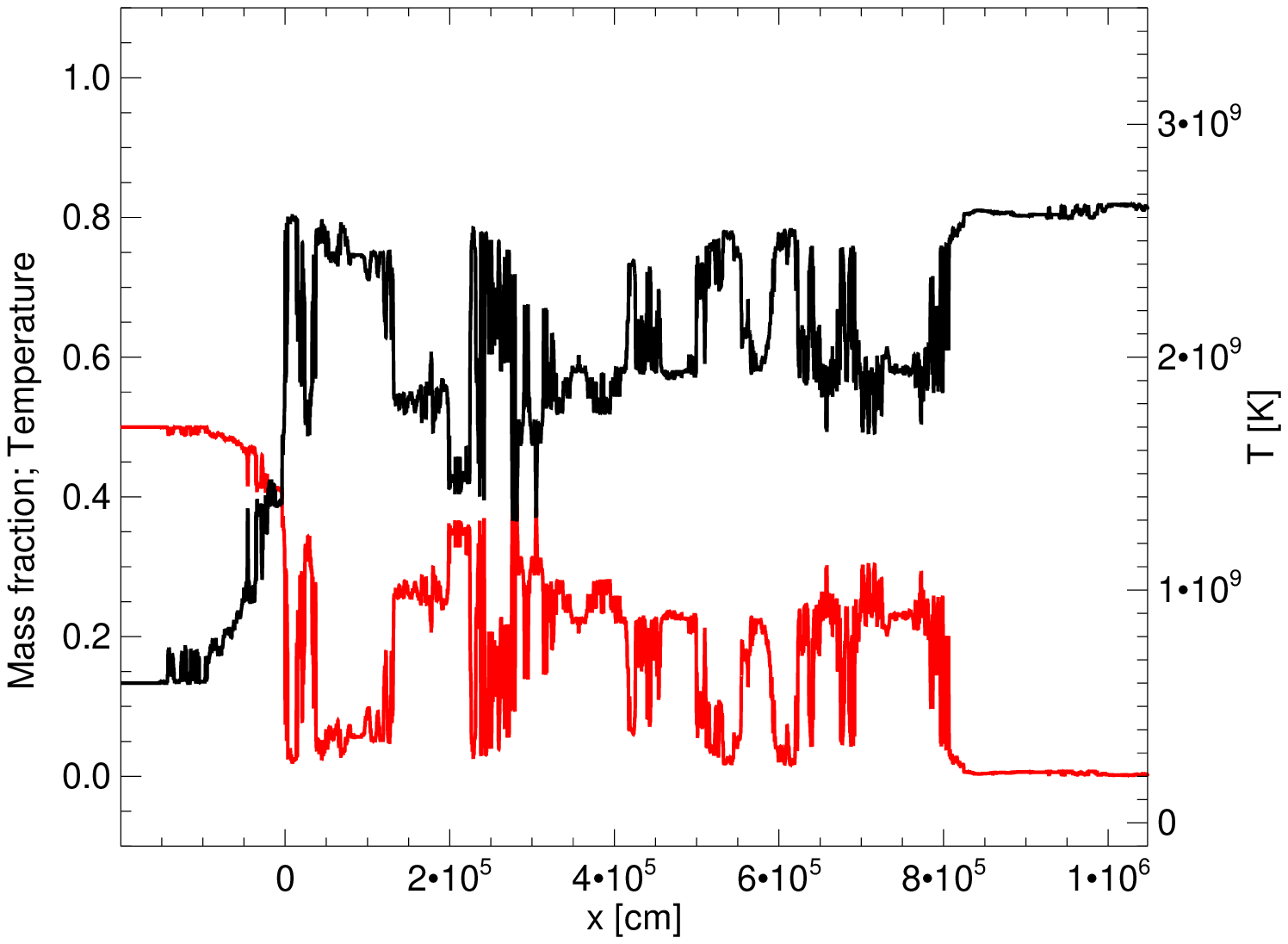} \hfill
\includegraphics[width=0.475\textwidth]{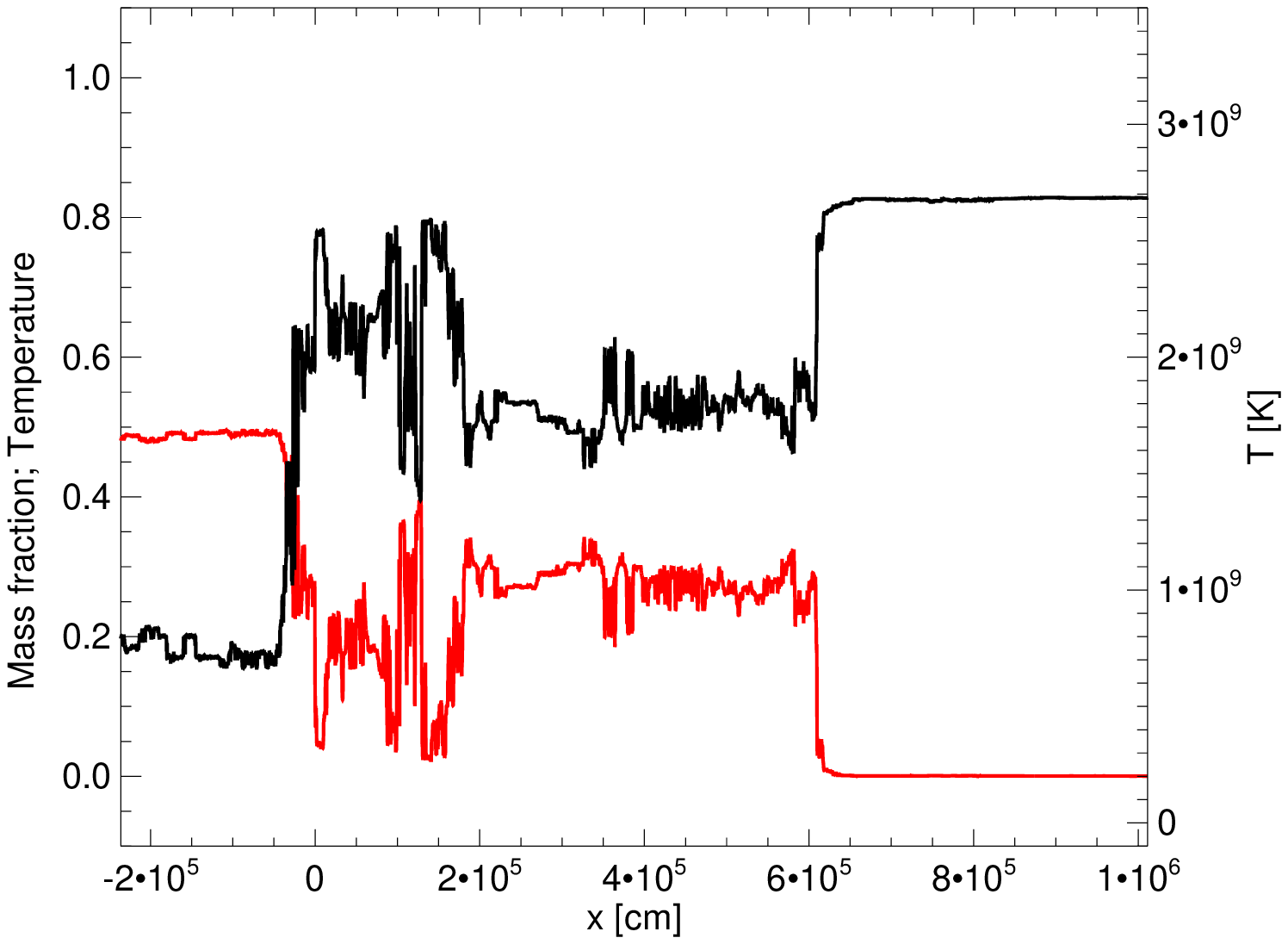} 
\caption{A single flame in the stirred flame regime for an integral
  scale of 4.92 km. The turbulent speed on that scale is 78.9 km
  s$^{-1}$ and the average flame speed and width are also close to
  these values. Many complex and folded structures are seen like in
  \Fig{andy3}, but the above two snapshots show that occasionally
  quite regular, well-mixed regions of fuel and ash exist. Sometimes
  several of these structures add together to create a mixed region
  about as large, or even larger than the integral scale.
  \lFig{andy5}}
\end{center}
\end{figure*}

\subsubsection{Ledges}
\lSect{ledges}

The ledges seen in \Fig{andy5} play an important role in promoting
detonation and it is thus worth spending a moment to discuss their
credibility. They have not been seen previously in any combustion
simulation or experiment.  For accessible terrestrial conditions, it
is believed that chemical flames in this regime would extinguish
\citep{Pet00}.  Numerical exploration of this regime, other than with
LEM, has so far been impractical.

Nevertheless, it has long been known that property fields in
turbulence exhibit intermittent behavior, including the occurrence of
transient well-mixed regions separated by cliffs (sharp property
changes).  This structure is seen, for instance, in 1D profile data
from 3D numerical simulations \citep{Wat06}.  A quantitative signature
of this structure is the saturation of high-order structure-function
exponents, reflecting the dominant contribution of the cliff regions
to high-order intermittency statistics \citep{Cel00}.  Measurements by
\citet{Moi01} show clear evidence of the saturation of the scalar
structure-function exponents.  LEM structure-function exponents
exhibit a somewhat slower roll-off to saturation \citep{Ker91},
suggesting that the prevalence and duration of well-mixed regions in
LEM might be lower than actually occurs in turbulence.  This is
plausible because maps in LEM are statistically independent events,
but the eddies that they represent actually occur in bunches because
each eddy breakdown is an energy source for subsequent eddy breakdown.
This bunching contributes to the intermittency of turbulence.

There is some limited understanding of why turbulence exhibits these
properties \citep{Cel00}.  Further clarification would provide
insight into a mechanism that appears to play a key role in the timely
occurrence of detonations in supernovae.

\begin{figure*}
\begin{center}
\includegraphics[angle=90,width=0.475\textwidth]{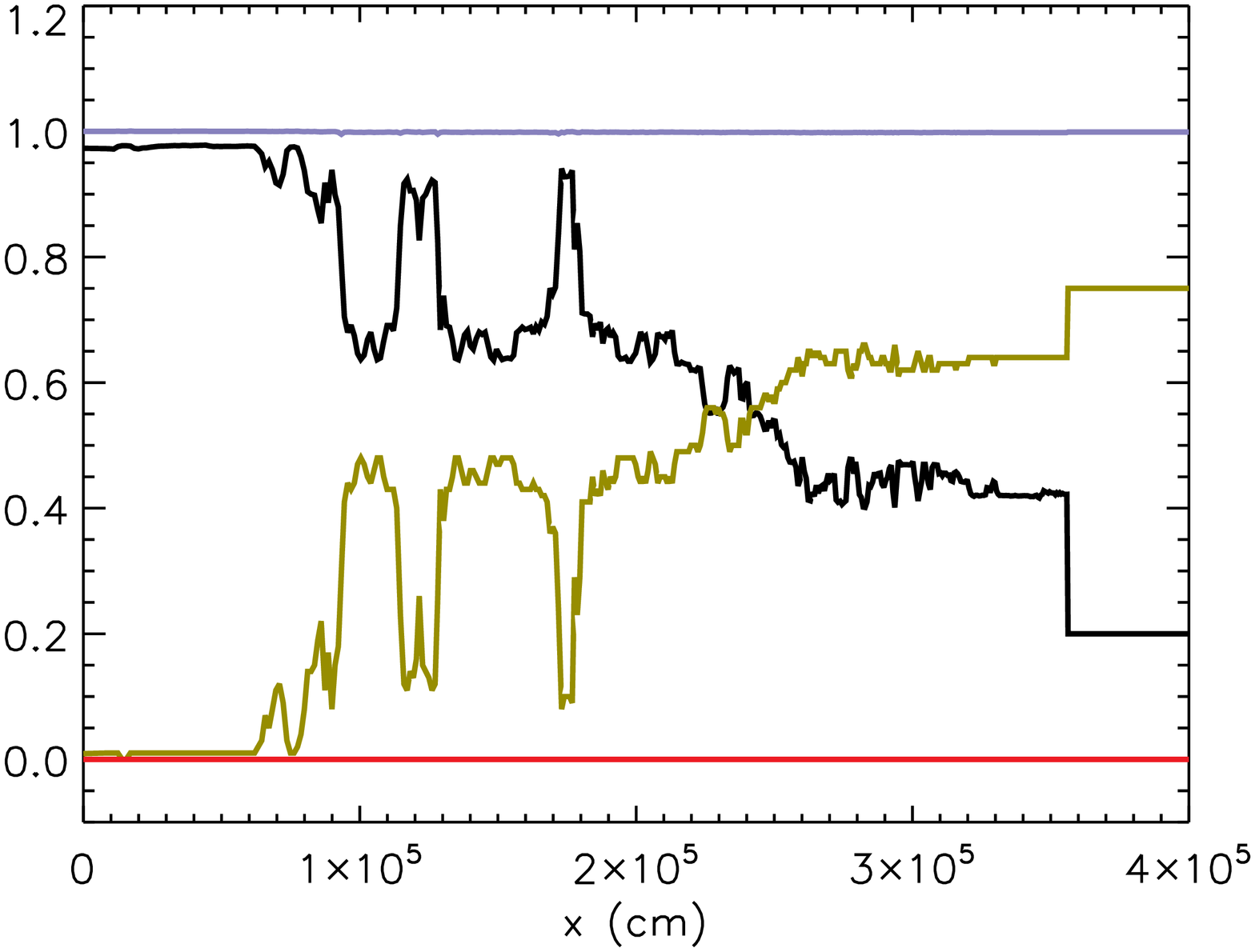}
\hfill
\includegraphics[angle=90,width=0.475\textwidth]{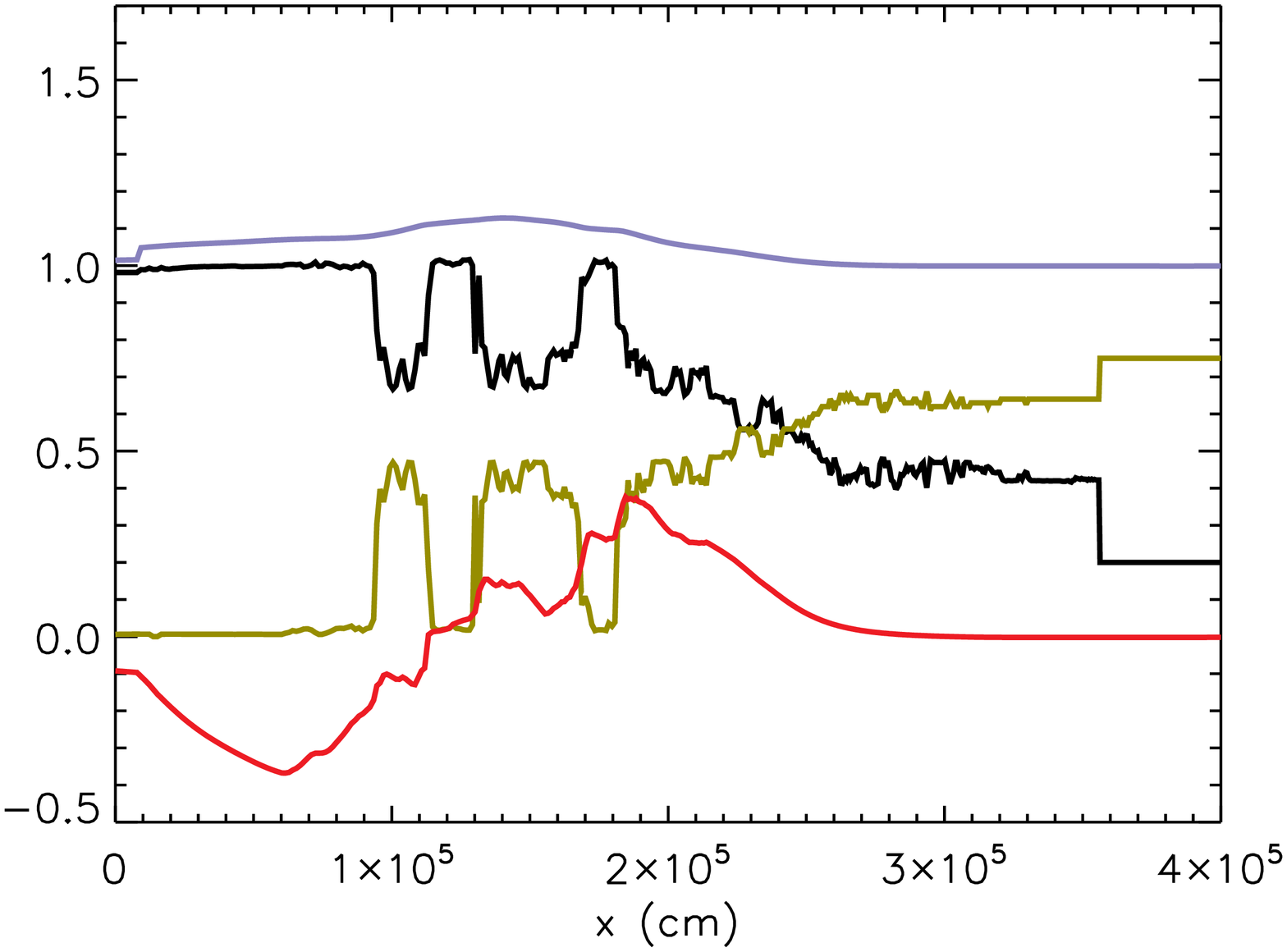}
\includegraphics[angle=90,width=0.475\textwidth]{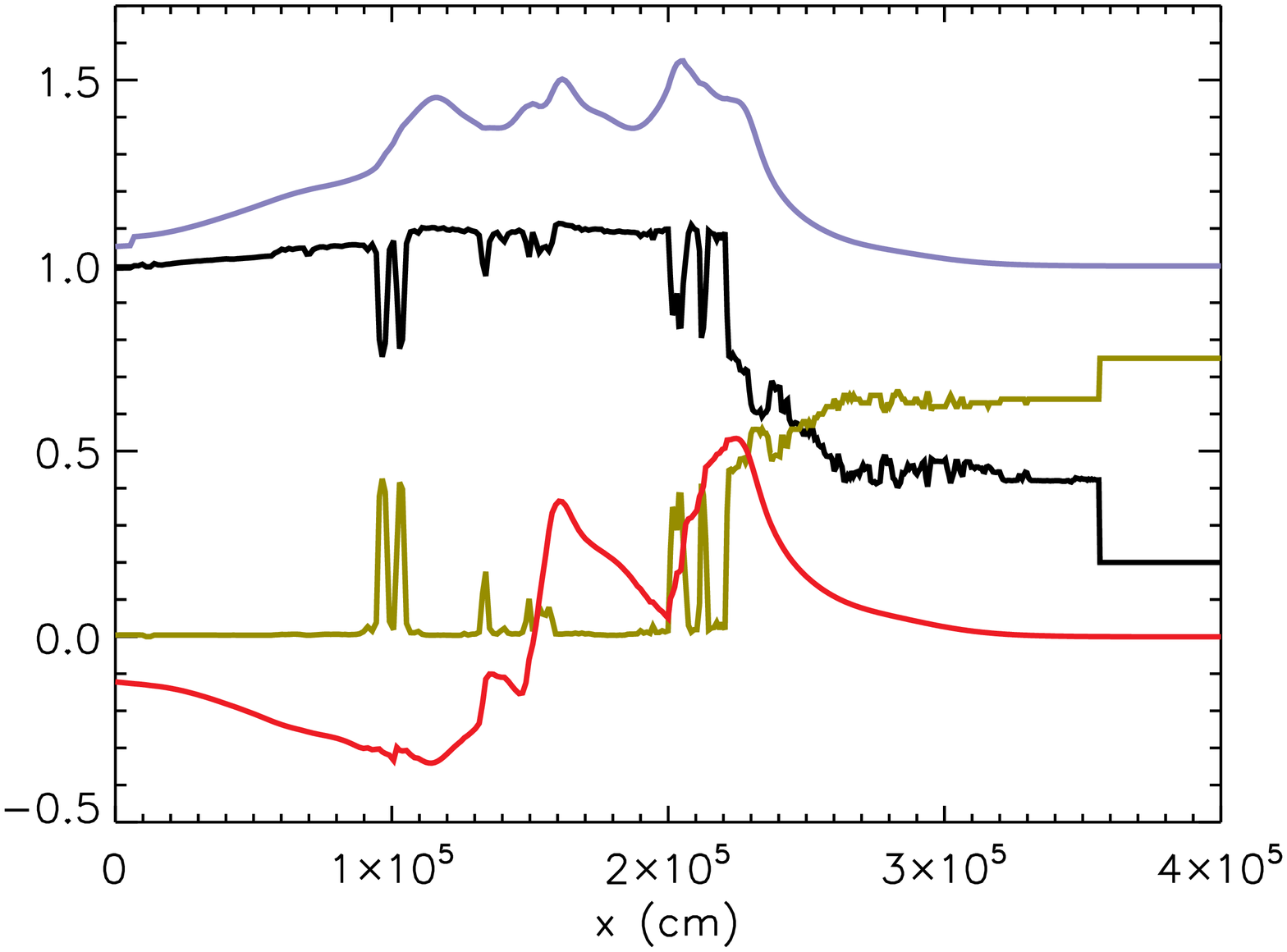}
\hfill
\includegraphics[angle=90,width=0.475\textwidth]{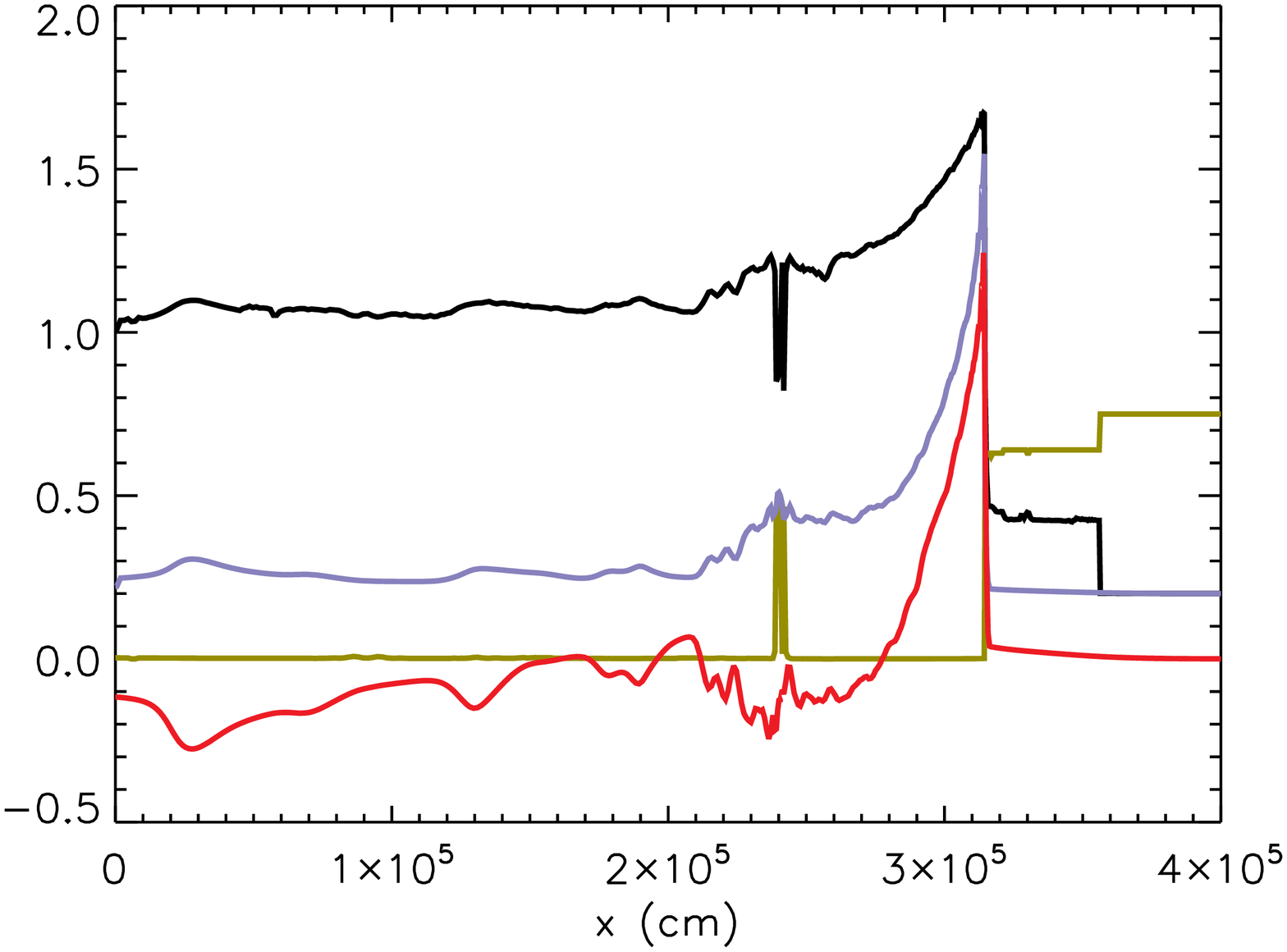}

\caption{The birth of a detonation. A sample mixture calculated using
  LEM for a density $1.0 \times 10^7$ g cm$^{-3}$, v' = 500 km
  s$^{-1}$, and L = 10 km was mapped into a compressible hydrodynamics
  code and its subsequent evolution was followed. The time selected
  was characterized by a very subsonic flame speed but temperature
  gradients that looked ``interesting''. Following the remap, which
  preserved distance scale, ash was on the left and fuel on the
  right. A detonation developed that, barring large barriers of ash,
  would explode the whole star (see text). Shown in the plot are
  carbon mass fraction (gold), pressure (blue), temperature (black),
  and velocity (red) at four different times - 0, 0.15, 0.30, and 0.43
  ms after the mapping. The velocity has been divided by 500 km
  s$^{-1}$ in frames 1 and 2, 2000 km s$^{-1}$ in frame 3 and 5000 km
  s$^{-1}$ in frame 4. The pressure has been scaled by the background
  value, $9.08 \times 10^{23}$ dyne, in frames 1, 2, and 3, and by an
  additional factor of 5 in frame 4. The temperature has been divided
  by $3.0 \times 10^9$ K.  \lFig{detonation}}
\end{center}
\end{figure*}

\section{Conditions for Detonation}
\lSect{ddt}

In order that a detonation occur, three conditions must be satisfied.
First, some region must burn supersonically. The simplest example
would be a hot, isothermal volume in which the nuclear induction time
scale was a constant. As the temperature rises, the time scale
decreases and for a sufficiently large volume there comes a
temperature where the burning time is shorter than the sound crossing
time. The pressure will increase in that region faster than expansion
can damp its growth and, after a maximum temperature is reached (i.e.,
the fuel is gone), the region expands at a speed that is somewhat
higher than the sound speed in the surrounding medium. In reality, the
temperature is never completely isothermal, but, as we shall see, it
suffices to burn only some fraction of the fuel within the volume in a
sound crossing time. A subset of fluid elements within the volume must
have the same temperature to some level of tolerance that depends on
their burning time scale. The smaller the fraction, the smaller the
overpressure, but sonic expansion nevertheless occurs. The colder
matter is an inert dilutant.

Second, the size of the ``detonator'' must exceed some critical value
\citep{Nie97,Dur06}. That critical mass is larger if the mass fraction
of combustible fuel is small or if the fraction of inert cold matter
is large. Because some mixing of ash into the pre-detonation region is
unavoidable and because an appreciable amount of carbon must burn to
reach temperatures where the time scales start to approach sonic, the
critical masses we compute here will be larger than those for mixtures
of just carbon and oxygen with a smooth temperature structure.  For a
given turbulent energy, density must fall to lower values to obtain
these larger structures.

Third, and perhaps most subtly, there must exist, in a significant
fraction of the mass, preferably at its edge, a nearly sonic phase
velocity for the burning \citep{Zel85,Kho97}. This is the ``shock wave
amplification by coherent energy release'' (SWACER) mechanism for
initiating a detonation in an unconfined medium. One requires an
appreciable boundary layer where
\begin{equation}
\frac{d \tau_{\rm nuc}(T)}{dx} \approx \ (c_{\rm sound})^{-1}.
\end{equation}

Two properties of the flame in the SF regime assist in satisfying
these three conditions. First is the unsteady nature of the burning
(\Sect{complex}). As fuel and ash are mixed, burning may briefly
almost go out, only to return with a vengeance after sufficient mixing
and slow burning have occurred. This allows the creation of regions
that, after some delay comparable to a turnover time, can consume fuel
at a rate faster than a single flame moving at the turbulent {\sl rms}
speed on the integral scale. Amplification factors of three are
frequently observed, and larger values are presumably possible in rare
instances.

Second, as discussed in \Sect{ledges}, transient well-mixed
structures, ledges, are a frequent occurrence in the SF
regime. Turbulence does not always lead to heterogeneity on
macroscopic scales. These mixed regions have relatively constant
induction time and they are at least occasionally bounded by regions
in which the temperature decreases (and fuel concentration increases)
fairly smoothly. The characteristic size of these regions is
approximately $\lambda$, which increases with decreasing density.  As
a result, the ratio of mixing time to sound crossing time decreases,
since
\begin{equation}
\frac{\tau_t}{\tau_{\rm sonic}} \ \propto
\ \frac{\lambda^{2/3}}{\lambda} \ = \lambda^{-1/3},
\end{equation}
and this helps detonation happen.  However, as the density decreases,
the critical size required to initiate a detonation also increases
dramatically \citep{Nie97,Woo07}, so mixing a smaller region at higher
density may, in some cases, be better than a larger one at low
density.  Detonation may occur before the characteristic size of the
mixed region becomes equal to the integral scale.

Another important point favoring detonation is that once burning
starts to happen on a time scale approaching sonic, it no longer
occurs at constant pressure. The pressure rises with respect to the
surrounding fuel and is not immediately relieved by expansion. At
constant volume, a given amount of carbon burning raises the
temperature more, thus appreciably shortening the time scale for
additional burning.  That is, for the relevant conditions, the heat
capacity at constant pressure is appreciably larger than the heat
capacity at constant volume. The final temperature from burning all
the fuel is also higher.

\subsection{Detonation Observed for a Mixture Calculated Using LEM}
\lSect{itworks}

In order to verify that some of the mixed regions calculated using LEM
would actually detonate, the burning of select mixtures was followed
using the compressible hydrodynamics code, Kepler
\citep[][\Fig{detonation}]{Wea78,Woo02}. The composition and temperature
structure were taken from an LEM simulation of flame-turbulence
interaction at a density of 1.0 $\times 10^7$ g cm$^{-3}$ for a
characteristic turbulent speed of 500 km s$^{-1}$ on an integral scale
of 10 km (Table 2). Carbon had a mass fraction of 0.75 in the
fuel. Use of 65536 zones gave a characteristic Reynolds number of
80,000 and a resolution not achievable in a multi-dimensional
simulation on this length scale. The LEM calculation was run for 0.10
s, or 5 eddy turnover times on the integral scale, during which 40
dumps were created at equal time intervals. Visual inspection isolated
several cases for further study. One of these was dump number 28, made
70 ms after the beginning of the run. The overall burning rate at this
time was not particularly high, corresponding to an effective speed of
only 250 km s$^{-1}$. However, most of the energy generation was
occurring in regions where the carbon mass fraction was already low
(\Fig{detonation}), so the local {\sl rate} of pressure increase was
quite large, even though the integrated total change in pressure in
the end was small.

Kepler, frequently used for studying stars and supernova explosions,
is an implicit 1D hydrodynamics code with a spherical Lagrangian
mesh. In order to simulate a problem with approximately plane parallel
geometry, the 4 km of interest was mapped on top of a sphere of pure
ash (mostly magnesium and silicon) with a radius 10 km so that LEM
zones of constant thickness had approximately constant mass and
thickness in Kepler. In the figure, that 10 km has been subtracted off
of the x-coordinate. The overall thickness in km of the region of
interest was the same in the Kepler study as in the LEM
calculation. The temperature was by no means isothermal in this
region. In fact, an isothermal runaway would not have initiated a
detonation here, even if the burning were supersonic. However, several
key elements favoring detonation were in place. First, there were some
regions where the temperature was already high because of previous
mixing and burning. The rate of carbon burning was quite high in these
regions with a peak energy generation rate near 10$^{21}$ erg g
s$^{-1}$. These were embedded in an extended region with X$_{12}$
$\approx 0.4$ that was already quite warm due to mixing. Finally that
mixture lay at the base of a region where the carbon mass fraction
increased steadily, if somewhat noisily, in an outwardly direction. It
is important to emphasize that this was the result of an LEM
calculation, and not an artificial construct.

During the first phase, the ``initiator'' burned rapidly producing an
overpressure in the surrounding zones of about 10 - 15\%. This
increase was sufficient to cause the warm material to burn faster, on
a roughly sonic time scale. Their expansion then compressed and
ignited material at successively higher carbon mass fractions in the
gradient. While weak at first, the detonation strengthened and by the
last frame shown was strong enough that its permanent propagation was
guaranteed. It was followed until it left the grid. Among the
interesting implications here is that a region need not be highly
isothermal, nor need all of it initially burn on a supersonic time
scale to provoke a detonation, though near sonic speeds are needed. A
pile up of strong acoustic pulses at around 2.3 km initiates the
runaway.

\begin{figure*}
\begin{center}
\includegraphics[angle=90,width=0.475\textwidth]{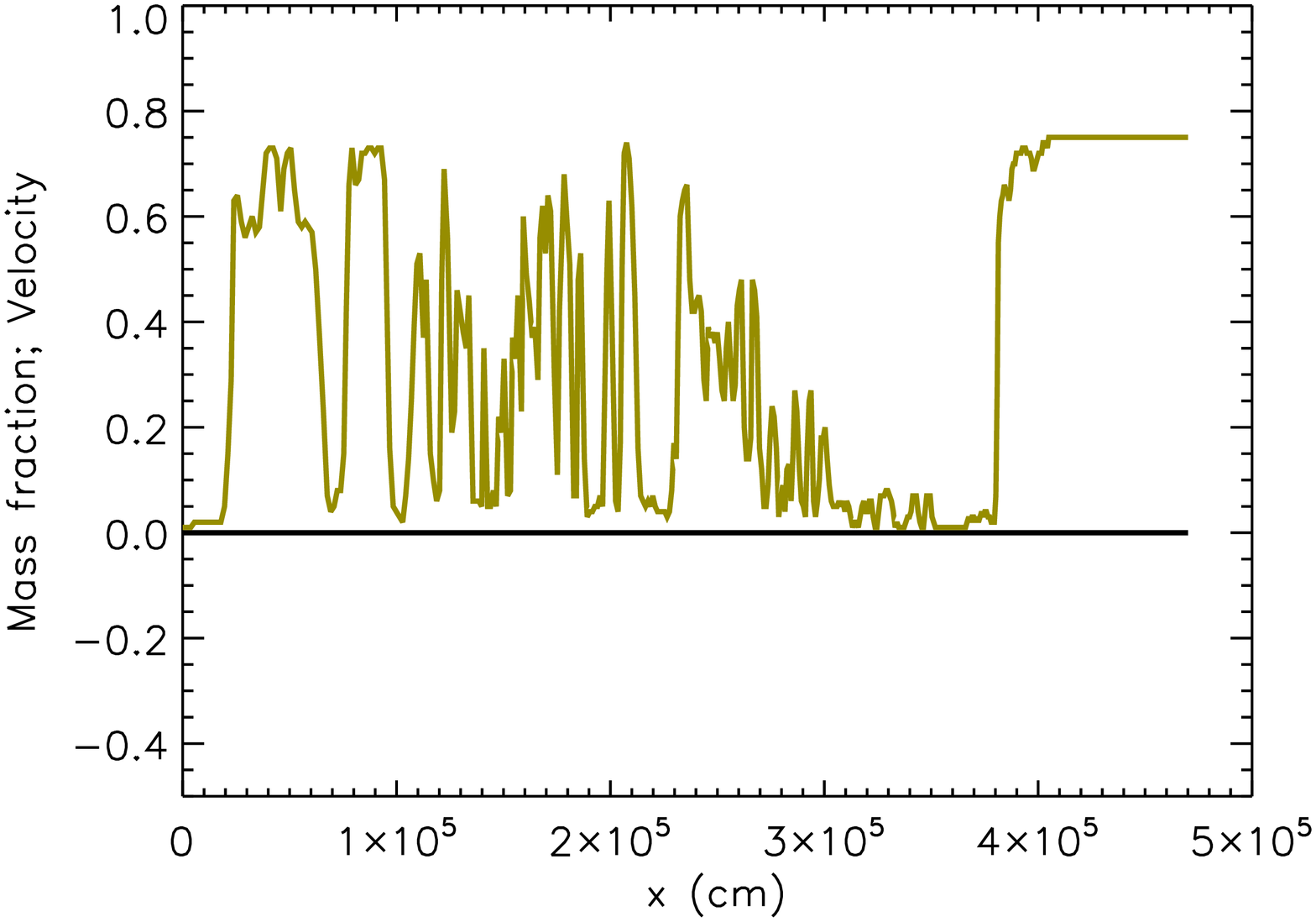} \hfill
\includegraphics[angle=90,width=0.475\textwidth]{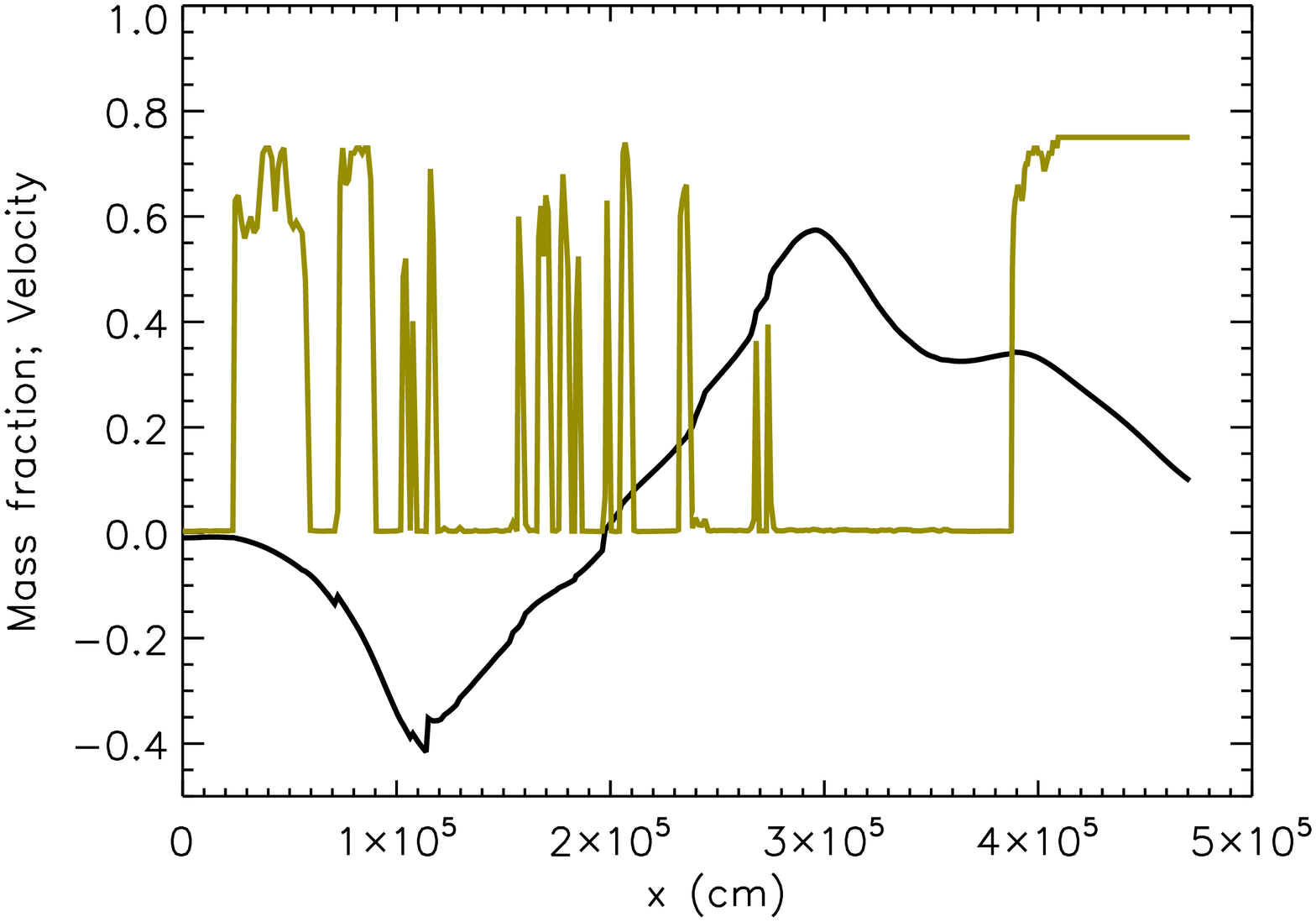} 
\caption{Development of a small scale explosion within a flame front
  in the WSR regime. The highly variable gold line is the carbon mass
  fraction. The dark line that is initially zero is the velocity. The
  velocity scale is in units of 1000 km s$^{-1}$. Unlike in
  \Fig{detonation}, the rapid burning of a mixed region does not
  initiate a detonation here, but it does produce a strong subsonic
  pulse. These may be very common events and would be a mechanism for
  adding turbulent energy at the flame scale.  The two edits here are
  from the same run that produced the detonation in \Fig{detonation}
  sampled $3.9 \times 10^{-4}$ s apart.  \lFig{pulse}}
\end{center}
\end{figure*}

\subsection{Active Turbulent Combustion}
\lSect{ATC}

Almost as interesting were the many other sample mixtures in this and
other runs that, when mapped into Kepler, did {\sl not}
detonate. Instead, the irregular burning that is characteristic of
combustion in the SF regime produces strong pressure waves. A
particularly strong pulse occurs roughly every turnover time on the
integral scale because these big mixing events trigger a lot of
burning. An example from the same run that produced the detonation in
\Fig{detonation} is shown about one turnover time later (dump 39 at 95
ms) in \Fig{pulse}. This time the burning did not produce a
detonation, but substantial burning still occurred on less than
a sonic time scale for the 4 km shown (the sound crossing time for
this region is about 1 ms). This burning produced a strong pulse that
sent matter moving inwards and outwards at $\sim600$ km s$^{-1}$, about the
same value as assumed for turbulence on an integral scale of 10 km in
the study.

In three dimensions, these pulses would be quasi-spherical and would
occur all over the surface where burning and mixing are going on.
Collisions between the fronts would pump additional energy into
turbulence on a scale comparable to the width of the burning region,
i.e., the integral scale, or about 10 km.  As noted, the velocities in
these pulses are frequently larger than expected from the assumed
turbulent energy on that length scale. As discussed by
\citet{Ker96,Nie97}, this sets the stage for a potential runaway.
More turbulence leads to increased mixing which leads to more
violent irregular burning, which creates more turbulence. The
culmination of this runaway could be a detonation.

\begin{figure}
\begin{center}
\includegraphics[angle=90,width=0.475\textwidth]{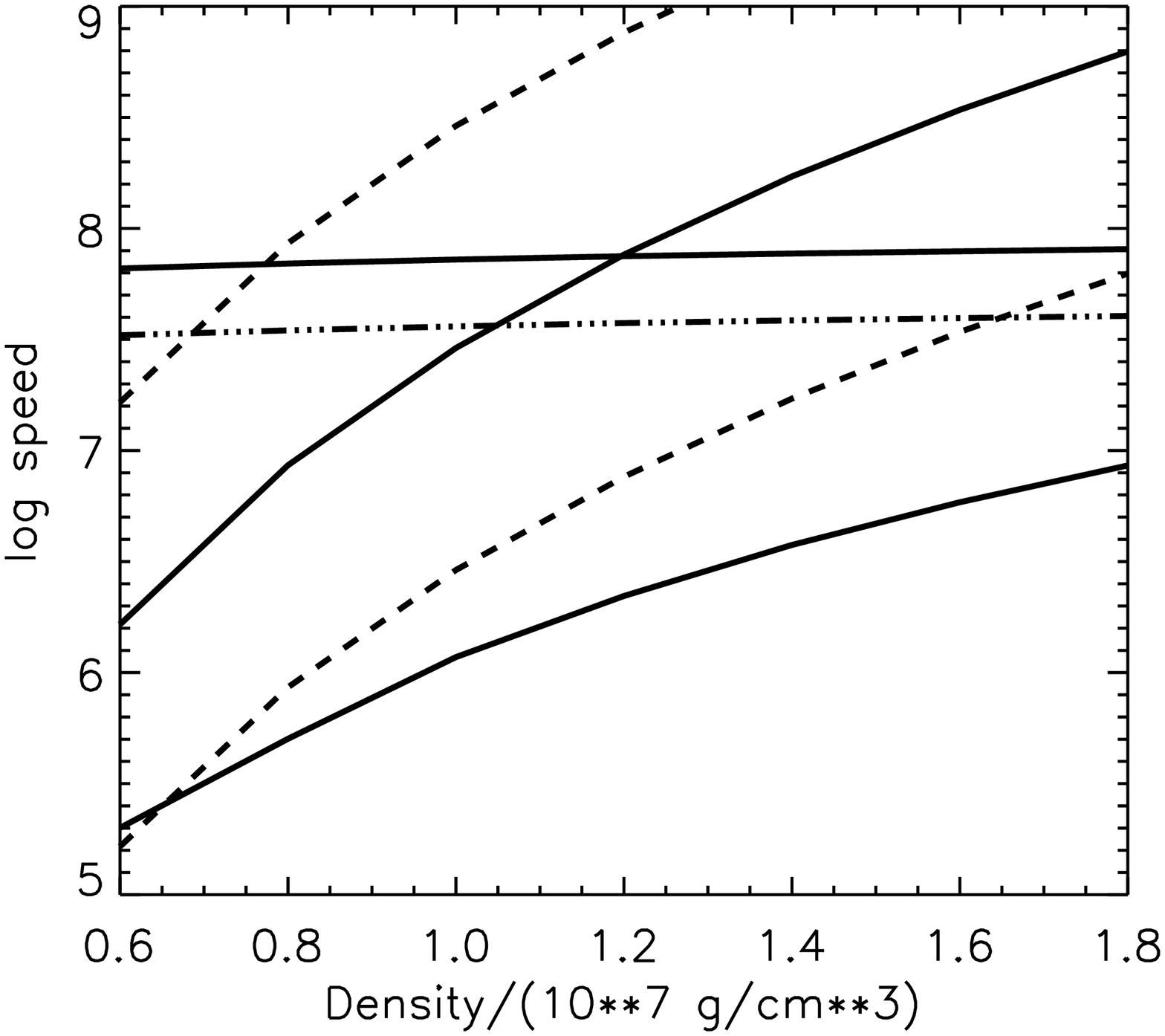} \hfill
\includegraphics[angle=90,width=0.475\textwidth]{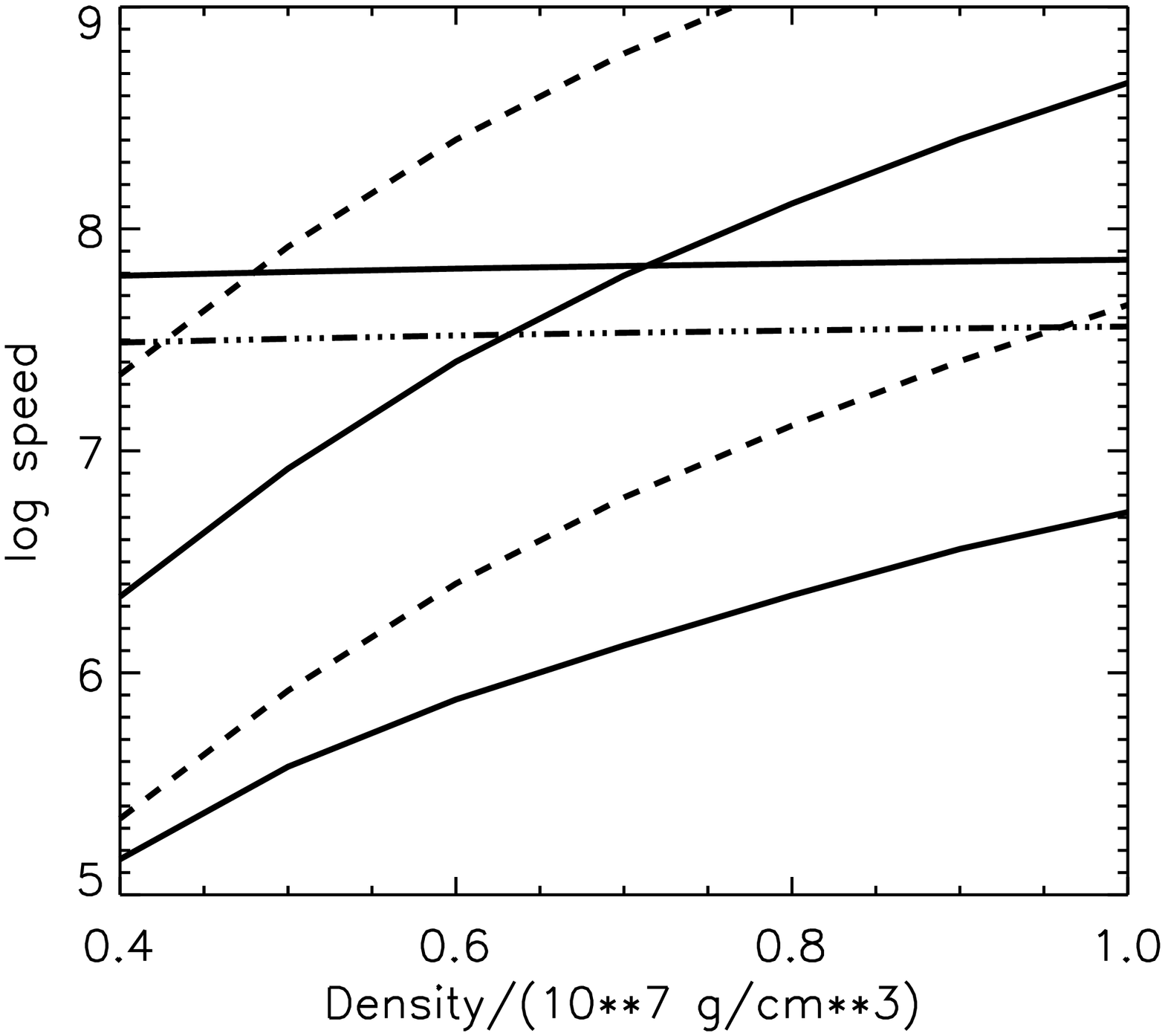} 
\caption{Speeds required on an integral scale of 10 km to establish
  the necessary conditions for detonation. The first panel is for an
  initial carbon mass fraction of 50\%; the second, for 75\%. The solid
  line that increases rapidly from left to right is the condition Da =
  10 and the two dashed line parallel to it are Da = 1 (upper) and 100
  (lower). The nearly horizontal solid line is the sound speed divided
  by 5 and the dash-dotted line beneath is the sound speed divided by
  10. The lowest solid line is the condition Ka = 10. Scaling to other
  values is given in the text. In order to detonate the speed on the
  integral scale must be greater than that given by Ka = 10 and in the
  vicinity of Da = 10. It is also necessary that the turbulent speed
  be at least 1/10 the sound speed and, better still, 1/5 the sound
  speed, but turbulent speeds above 20\% sonic may not be achieved in
  the star. Thus detonation for the assumed conditions and a carbon
  mass fraction of 0.5 probably occurs in the band between 0.8 and 1.6
  $\times 10^7$ g cm$^{-3}$. For 75\% carbon, the likely density range
  is 0.5 to 1.0 $\times 10^7$ g cm$^{-3}$. \lFig{darm}}
\end{center}
\end{figure}

\section{Conclusions}
\lSect{concl}

Three regimes of turbulent flame propagation relevant to nuclear
burning in a Type Ia supernova have been explored. At high density,
for Karlovitz numbers less than about one, burning occurs in multiply
folded laminar flames. The overall progress of burning is governed by
the turbulent energy and has a speed that is independent of the
laminar speed. A similar description of ``laminar flame brushes'' has been
given many times in the literature, but this is the first time it has
been simulated in a supernova (e.g., \Fig{flamelet}).

The average number of flamelets in the flame brush is $U_L/S_{\rm
  lam}$, and initially is not large. Because of this, there will be
considerable variation in the burning rate.  We estimate such
variations to be as large as a factor of three. However, detonation
does not happen so early because the turbulent speed then is very
subsonic and the individual flamelets are thin.  Later, the number of
flamelets becomes very large, reaching hundreds or even thousands, and
variations in the overall burning rate are much smaller. Detonation
remains impossible in the flamelet regime. No critical mass of hot
fuel can be obtained.

As the Karlovitz number increases above about 10, hot ash and cold
fuel can be mixed for the first time and detonation is, in principle,
possible.  Two cases of turbulent burning were explored corresponding
to the well-stirred reactor (WSR; Da $< 1$, \Sect{distrib}) and
stirred flames (SF; Da $> 1$; \Sect{stirred}). In the WSR regime,
turbulent diffusion substitutes for heat diffusion and turbulently
broadened flames result. These flames can be much larger than the
integral scale. Because the integral scale in a supernova is
so large, this limiting case is probably never fully realized, but is
amenable to numerical simulation \citep{Asp08}. We normalized an
uncertain constant in the LEM calculation, $C$, (\Sect{LEM}) to those
simulations. In the stirred flame regime, one again has something like
a flame brush, but with turbulently broadened structures substituting
for the individual flamelets in the flamelet regime. The structure of
burning here is complex and highly variable. Because the burning time
scale is very temperature sensitive, it is possible to mix large
regions of ``warm'' fuel and ash in which are embedded smaller nuggets
on the verge of explosion. Mixed regions of nearly constant temperature
are sometimes observed (\Fig{andy5}, \Sect{ledges}). 

Spontaneous detonation can happen in the SF regime
(\Fig{detonation}, \Sect{itworks}). It does not require complete mixing
on the integral scale, and hence Da $\approx 1$ as discussed in
\citet{Woo07}, but is favored by values of Da that are not large.
The necessary conditions occur infrequently and require a favorable
confluence of several mixed regions including: a) some region, perhaps
not large and with a low mass fraction of carbon, where the burning
occurs sufficiently rapidly to increase the pressure by a small
fraction supersonically; and b) an extended region where the carbon
mass fraction rises gradually and the temperature falls
slowly. Neither region need be particularly homogeneous as long as
some part of the composition burns supersonically for condition a) and
there are neither large barriers of ash or abrupt, sustained
increases in carbon mass fraction for condition b).

In general, the production of these situations require turbulent
speeds that are already a considerable fraction of sonic, certainly
within 10\% and probably 20\%.  Overall the progress of burning in the
SF regime varies frequently by a factor of three up and down, and rare
excursions to larger values probably also occur due to intermittency
\citep{Pan08}. We also find that appreciable turbulence can be put
into the burning region on a scale comparable to the integral scale by
small subsonic explosions of mixtures that fail to detonate
(\Sect{ATC}, \Fig{pulse}). Before the big bang in these stars, there is
a lot of thunder. Though we have not demonstrated it here, we
speculate that this energy input may exceed that put in at the large
scale by the flame instabilities. If so, there is the possibility of a
turbulent runaway in which mixing pumps energy into turbulence which
in turn causes accelerated mixing \citep{Ker96}. The endpoint would be
detonation.

These necessary conditions for detonation are summarized in
\Fig{darm}. Detonation is estimated to occur for a reasonable range of
turbulent energies in the nearly horizontal band between the lines Da
= 10 and Da = 100.  It must be acknowledged that this figure is very
approximate because the nuclear time scale (and hence Damk\"ohler
number) are poorly determined. The effective nuclear time scale in Da
is probably longer than in Table 3 and intermittency essentially
raises the effective turbulent speed making detonation possible at a
higher density. This is why the conditions Da = 10 - 100 in \Fig{darm}
are probably more favored than say 1 to 10. Above Da = 100, the mixed
regions may be too small to initiate a detonation.

Three-dimensional simulations by \citet{Roe07a} show that the necessary
degree of turbulence for detonation, roughly $u'$ = 500 km s$^{-1}$, is
realized in the full star models. 

We also find some dependence of the detonation conditions on initial
carbon mass fraction \citep[see also][]{Woo07}. For lower carbon mass
fractions, Damk\"ohler numbers of order 10 are reached at a
higher density. If detonation does occur at a higher density, the
explosion makes more $^{56}$Ni and a brighter supernova. However,
detonation depends on achieving a significant overpressure on a sonic
time scale. By the time that carbon-poor fuel carbon reaches a
temperature where it burns rapidly (\Fig{taunuc}), the remaining
burning produces too small an overpressure. Detonating carbon-rich fuel
is, in this sense, easier \citep[see also][]{Ume99a,Ume99b}. It is
important to note that the relevant location for detonation is
probably in the outer layers where the density first declines below
10$^7$ g cm$^{-1}$ at the flame front, not near the center. The carbon
abundance is higher in these outer layers.

All in all, our results are supportive of the hypothesis that
some, perhaps even all Chandrasekhar mass white dwarfs explode by a
delayed detonation that occurs shortly after the burning enters the SF
regime. This would then make the location and number of detonation
points, along with the ignition conditions \citep{Kuh06}, the principal
determining factors in the intrinsic properties of a Type Ia
supernova.

Several avenues for future investigation are opened up by this work.
First, our results hinge on the calibration of the 1D LEM model to
direct 3D simulation \citep{Asp08}. The 3D study used for
normalization here was carried out in the WSR regime, but the results
(e.g., the constant C) were assumed to be valid in the SF regime.
Given the subgrid model developed here, it should be possible to carry
out equivalent 3D studies for the SF regime ($Da \gg 1$).

The results for detonation (\Sect{itworks}) and active turbulent
combustion (\Sect{ATC}) were obtained by mapping results from LEM
using a linear grid into a 1D compressible hydro-code with spherical
coordinates. It would be greatly preferable to see both the mixing and
the strong pressure waves in the same, preferably 3D study. Because of
the range in length scales, the need for a large effective Reynolds
number, and the rare, transient nature of the phenomena, this will
require a very major investment of computational resources, but should
be practical in the near future.

Ultimately, of course, one would want to see these results applied to
full scale models of the supernova and its light curve.

\acknowledgements

The authors gratefully acknowledges helpful conversations on the
subject of the paper with Andy Aspden, John Bell, and Martin Lisewski.
This research has been supported by the NASA Theory Program
(NNG05GG08G) and the DOE SciDAC Program (DE-FC02-06ER41438). Work at
Sandia was supported by the US Department of Energy, Office of Basic
Energy Sciences, Division of Chemical Sciences, Geosciences and
Biosciences. Sandia is a multiprogram laboratory operated by Sandia
Corporation, a Lockheed Martin Company, for the US Department of
Energy under contract DE-AC04-94AL85000.

\vskip 0.5 in

\begin{widetext}
\begin{appendices}
\section{LEM flame speed in the flamelet regime}

Properties of the triplet map imply a simple, exact expression for the
turbulent burning velocity $v_{\rm turb}$ in terms of the LEM
parameters $D_{\rm turb}$ (turbulent diffusivity) and $L$ (largest
allowed map size) under the conditions $v_{\rm turb}\gg S_{\rm lam}$ (here
assuming the flamelet regime) and $D_{\rm turb}\gg \nu /\rho$
(implying Re $\gg 1$).  On the LEM domain (coordinate $x$), assume
fuel on the right and ash on the left, so the flame advances rightward
(direction of increasing $x$).  $v_{\rm turb}$ is evaluated by
tracking the forward progress of the rightmost (forwardmost) ash
location.  The only mechanism affecting this location, denoted $r$, is
the triplet map because small $S_{\rm lam}$ implies that the contribution of
laminar burning is negligible.

Any triplet map containing $r$ maps $r$ to three locations, the
rightmost of which, denoted $r'$, exceeds $r$.  Specifically, if the
map interval is $[x_1,x_2]$, then $r'=x_2-\frac{1}{3}(x_2-r)$, i.e.,
the distance from $r'$ to $x_2$ is one third of the pre-map distance.
The advancement of the rightmost ash location is therefore $r'-r
=\frac{2}{3}(x_2-r)$.

Suppose that the interval $[x_1,x_2]$ is chosen arbitrarily.  If it
contains $r$, then $r$ is equally likely to be anywhere within the
interval, and hence is uniformly distributed in the interval.  The
advancement is therefore averaged over $r$ values in the interval.
Because the advancement is linear in $r$, this implies that the
average value $(x_1+x_2)/2$ is substituted for $r$, giving the average
advancement $\langle r'-r\rangle =\frac{1}{3}(x_2-x_1)=\frac{1}{3}l$,
where $l$ is the map size.

Suppose that all maps were the same size $l$ and denote the frequency
of maps containing a given location as $\phi$.  Then the average rate
$v_{\rm turb}$ of advancement of $r$ is $\phi$ times the average
advancement of $r$ per map, giving $v_{\rm turb}=\frac{1}{3}\phi l$.

$\phi$ can be expressed in terms of $D_{\rm turb}$.  From random walk
theory, $D_{\rm turb}$ is one-half of the frequency of events that
displace a point times the mean-square displacement per event.  The
mean-square displacement induced by a size-$l$ triplet map is
$\frac{4}{27}l^2$ cite{part5}, so $D_{\rm turb}=\frac{2}{27}\phi l^2$,
giving $\phi = \frac{27}{2}D_{\rm turb}/l^2$ and thus $v_{\rm
  turb}=\frac{9}{2}D_{\rm turb}/l$.  This illustrates the analysis but
is not the case of interest because inertial-range turbulence is
represented in LEM by a distribution of map sizes $l$.

As explained in \citet{Ker91}, in LEM the map size distribution is
$f(l)=Al^{-8/3}$ for $l$ in the range $[\eta,L]$, where $A$ is a
normalization factor.  The total frequency of maps of all sizes per
unit domain length is denoted $\Lambda$.  The frequency of maps in the
size range $[l,l+dl]$ that contain a given point is then $\Lambda l
f(l) \, dl$, so $D_{\rm turb}=\frac{2}{27} \Lambda \int l^3 f(l)\, dl=
\frac{1}{18} \Lambda A(L^{4/3}-\eta^{4/3})$.  For high Re, the $\eta$
term is negligible and is dropped, giving $D_{\rm turb}= \frac{1}{18}
\Lambda AL^{4/3}$.

To obtain $v_{\rm turb}$, the average advancement $\frac{1}{3}l$ for
map size $l$ is multiplied by $\Lambda l f(l) \, dl$ and integrated
over $l$ to obtain (ignoring the $\eta$ term) $v_{\rm turb} = \Lambda
AL^{1/3}$.  In terms of $D_{\rm turb}$, the result $v_{\rm turb} = 18
D_{\rm turb}/L$ is obtained.  The numerical factor is four times
larger than if all maps were of size $L$.  A heuristic interpretation
of this result is that the typical map size from a flame propagation
viewpoint is $L/4$ when $f(l)$ is based on inertial-range scaling.

\end{appendices}
\end{widetext}

\clearpage

\begin{deluxetable}{cccccc} 
\tablecaption{Properties of Laminar Flames}
\tablehead{
X$_{12}$ & $\rho$ & $S_{\rm lam}$ &$\delta_{\rm lam}(\epsilon)$ & $\delta_{\rm lam}(T)$ & $\delta_{\rm lam}(C)$ \\
        & ($10^7$ g cm$^{-3}$)  & (cm/s) &  (cm)  &  (cm)       &   (cm) }
\startdata
0.50 &  0.6  & 8.62(2) & 11.7   & 15.6  & 19.3   \\
0.50 &  0.8  & 1.77(3) &  4.35  & 6.01  & 7.03   \\
0.50 &  1.0  & 3.23(3) &  2.08  & 3.12  & 3.32   \\
0.50 &  1.2  & 4.99(3) &  1.15  & 1.78  & 1.82   \\
0.50 &  1.4  & 7.19(3) &  0.70  & 1.14  & 1.10   \\
0.50 &  1.6  & 9.66(3) &  0.45  & 0.76  & 0.72   \\
0.50 &  1.8  & 1.26(4) &  0.31  & 0.54  & 0.50   \\
0.50 &  2.0  & 1.58(4) &  0.22  & 0.40  & 0.36   \\        
0.50 &  2.5  & 2.47(4) &  0.12  & 0.23  & 0.19   \\
0.50 &  3.0  & 3.54(4) &  0.067 & 0.14  & 0.11   \\  
0.50 &  3.5  & 4.66(4) &  0.041 & 0.090 & 0.065  \\
     &       &         &        &       &        \\
0.75 &  0.3  & 1.73(2) &   172  & 157   & 325    \\
0.75 &  0.4  & 8.65(2) &  21.4  & 22.4  & 40.0   \\
0.75 &  0.5  & 1.74(3) &  9.80  & 11.4  & 17.7   \\
0.75 &  0.6  & 2.80(3) &  5.02  & 6.12  & 8.94   \\
0.75 &  0.7  & 4.10(3) &  2.94  & 3.52  & 5.29   \\
0.75 &  0.8  & 5.86(3) &  1.80  & 2.27  & 3.21   \\
0.75 &  0.9  & 8.39(3) &  1.25  & 1.73  & 2.20   \\
0.75 &  1.0  & 1.08(4) &  0.85  & 1.21  & 1.49   \\
0.75 &  1.2  & 1.70(4) &  0.43  & 0.66  & 0.77   \\
0.75 &  1.4  & 2.37(4) &  0.25  & 0.40  & 0.45  
\enddata
\end{deluxetable}

\clearpage

\begin{deluxetable}{ccccccccc} 
\tablecaption{Flame Properties at $\rho = 1.0 \times 10^7$ g cm$^{-3}$}
\tablehead{
X$_{12}$ & $l$   & $U_L$   & Zones & $\Delta x$  & Transport & $v_f$ & 
$\delta_f(\epsilon)$  & $\delta_f(T)$ \\
        & (cm)  & (km/s) &       &  (cm)       &           &  (km/s) & cm & cm }
\startdata
0.50 &  -      &   0   &  2048  & 0.0244 & rad   & 0.0323  &  2.1    & 3.1 \\
0.50 &  -      &   0   &   256  & 0.196  & rad   & 0.0314  &  2.1    & 3.1 \\
0.50 & 15      & 2.47  &  2048  & 0.244  & rad.  & 0.149   &  40     & 70  \\
0.50 & 15      & 2.47  &  2048  & 0.244  & SG    & 0.156   &  40     & 70  \\
0.50 & 120     & 4.93  &  2048  & 0.977  & SG    & 0.61    &  160    & 300 \\
0.50 & 960     & 9.86  &  2048  & 9.77   & SG    & 2.3     & $\sim$700 &
 $\sim$1200 \\
0.50 & 960     & 9.86  & 16384  & 1.22   & SG    & 2.3     & $\sim$700 & 
 $\sim$1200 \\
0.50 & 7680    & 19.7  &  8192  & 9.77   & SG    & 8.2     &  fluc. & fluc. \\
0.50 & 6.14(4) & 39.4  & 32768  & 9.16   & SG    & 26      &  fluc. & fluc.\\
0.50 & 4.92(5) & 78.9  & 65536  & 49.9   & SG    & 110     &  fluc. & fluc.\\
0.50 & 3.93(6) & 158   & 65536  & 360    & SG    & 280     &  fluc. & fluc.\\
0.50 & 3.93(6) & 340   & 65536  & 360    & SG    & 440 - 470? & fluc. & fluc.\\
     &         &       &        &        &       &         &          &    \\
0.75 &  -      &  0    &  2048  & 0.0391 & rad   & 0.113   &  0.85 & 1.2 \\
0.75 & 10      & 2.16  &  2048  & 0.0977 & rad   & 0.415   &  10   & 20   \\
0.75 & 40      & 3.42  &  4096  & 0.0977 & rad   & 0.922   &  $\sim$25 & 
 $\sim$60\\
0.75 & 80      & 4.31  &  8192  & 0.0977 & rad   & 1.46    &  $\sim$40 & 
 $\sim$90 \\
0.75 & 80      & 4.31  &  8192  & 0.0977 & SG    & 1.54    &  $\sim$40 &
 $\sim$90 \\
0.75 & 320     & 6.84  & 32768  & 0.0977 & SG    & 3.2     &  fluc.  & fluc.\\
0.75 & 2560    & 13.7  & 32768  & 0.391  & SG    & 12      &  fluc.  & fluc.\\
0.75 & 2.05(4) & 27.4  & 32768  & 3.05   & SG    & 40      &  fluc.  & fluc.\\
0.75 & 1.00(4) & 215   & 32768  & 2.50   & SG    & 96      &  fluc. & fluc. \\
0.75 & 1.00(6) & 500   & 32768  & 152    & SG    & 1020    &  fluc. & fluc. \\
0.75 & 1.00(6) & 500   & 65536  & 76.3   & SG    & 870     &  fluc. & fluc. \\
\enddata
\end{deluxetable}

\clearpage

\begin{deluxetable}{cccccc} 
\tablecaption{Characteristic Scales in the WSR and ST Regimes for $U_L = 100$ km s$^{-1}$ and L = 10 km}
\tablehead{
X$_{12}$ & $\rho$ & $\tau_{\rm nuc}$ &$\lambda$ & $d$ & Ka \\
        & ($10^7$ g cm$^{-3}$)  & (sec) &  (cm)  &  (cm)  &   }
\startdata
0.50 & 0.6  & 6.1(-2) &  4.7(5) & 1.0(2) & 3.5(3)  \\
0.50 & 0.8  & 1.2(-2) &  4.0(4) & 1.8(2) & 8.9(2)  \\
0.50 & 1.0  & 3.5(-3) &  6.4(3) & 3.2(2) & 2.5(2)  \\
0.50 & 1.2  & 1.3(-3) &  1.5(3) & 5.0(2) & 9.6(1)  \\
0.50 & 1.4  & 5.8(-4) &  4.5(2) & 7.2(2) & 4.3(1)  \\
0.50 & 1.6  & 2.9(-4) &  1.6(2) & 9.7(2) & 2.2(1)  \\
0.50 & 1.8  & 1.6(-4) &  6.4(1) & 1.3(3) & 1.3(1)  \\
0.50 & 2.0  & 9.5(-5) &  2.9(1) & 1.6(3) & 7.5(0)  \\        
0.50 & 2.5  & 3.2(-5) &  5.7(0) & 2.5(3) & 2.8(0)  \\
0.50 & 3.0  & 1.3(-5) &  1.5(0) & 3.5(3) & 1.2(0)  \\ 
0.50 & 3.5  & 6.4(-6) &  5.1(-1)& 4.7(3) & 6.4(-1) \\              \\
0.75 & 0.3  & 3.6(-1) &  6.8(6) & 1.7(1) & 1.8(5)  \\  
0.75 & 0.4  & 6.0(-2) &  4.6(5) & 8.7(1) & 5.8(3)  \\
0.75 & 0.5  & 1.6(-2) &  6.4(4) & 1.7(2) & 1.4(3)  \\
0.75 & 0.6  & 5.3(-3) &  1.2(4) & 2.8(2) & 4.8(2)  \\
0.75 & 0.7  & 2.1(-3) &  3.0(3) & 4.1(2) & 2.1(2)  \\
0.75 & 0.8  & 1.0(-3) &  9.9(2) & 5.9(2) & 9.5(1)  \\
0.75 & 0.9  & 5.0(-4) &  3.5(2) & 8.4(2) & 4.6(1)  \\
0.75 & 1.0  & 2.8(-4) &  1.4(2) & 1.1(3) & 2.6(1)  \\
0.75 & 1.2  & 1.0(-4) &  3.3(1) & 1.7(3) & 9.4(0)  \\
0.75 & 1.4  & 4.5(-5) &  9.6(0) & 2.4(3) & 4.3(0)
\enddata
\end{deluxetable}

\end{document}